\overfullrule=0pt
%%%%%%%%%%%%%%%%%%  tex macros for preprints, cm version %%%%%%%%%%%%%%
%         (P. Ginsparg <ginsparg@lanl.gov>, last updated 7/94)
%                if confused, type `b' in response to query 
%           hypertex extensions (still provisional), 7/26/94
%
%---------------------------------------------------------------------%
%\input hyperbasics %comment out this line to restore non-hyper functionality
%
%% site dependent options:
%% \unredoffs and \redoffs define horizontal and vertical offsets
%% respectively for unreduced and reduced modes. \speclscape defines
%% the \special{} call that sets printer to landscape (sideways) mode.
%% from standard set below, leave uncommented as appropriate or redefine
%
%%% next 400dpi
\def\unredoffs{} \def\redoffs{\voffset=-.31truein\hoffset=-.48truein}
\def\speclscape{}
%\def\speclscape{\special{papersize=11in,8.5in}}
%
%%% apple lw
%\def\unredoffs{} \def\redoffs{\voffset=-.31truein\hoffset=-.59truein}
%\def\speclscape{\special{ps: landscape}}
%
%%% qms lasergrafix:
%\def\unredoffs{} \def\redoffs{\voffset=-.4truein\hoffset=.125truein}
%\def\speclscape{\special{qms: landscape}}
%
%%% saclay A4 paper:
%\def\unredoffs{\hoffset-.14truein\voffset-.2truein}
%\def\redoffs{\voffset=-.45truein\hoffset=-.21truein}
%\def\speclscape{\special{landscape}}
%
%---------------------------------------------------------------------%
%
\newbox\leftpage \newdimen\fullhsize \newdimen\hstitle \newdimen\hsbody
\tolerance=1000\hfuzz=2pt
\catcode`\@=11 % This allows us to modify PLAIN macros.
\ifx\hyperdef\UNd@FiNeD\def\hyperdef#1#2#3#4{#4}\def\hyperref#1#2#3#4{#4}\fi
\def\bigans{b }
\def\answ{b }
%\message{ big or little (b/l)? }\read-1 to\answ
%
\ifx\answ\bigans\message{(This will come out unreduced.}
\magnification=1200\unredoffs\baselineskip=16pt plus 2pt minus 1pt
\hsbody=\hsize \hstitle=\hsize %take default values for unreduced format
\else\message{(This will be reduced.} \let\l@r=L
\magnification=1000\baselineskip=16pt plus 2pt minus 1pt \vsize=7truein
\redoffs \hstitle=8truein\hsbody=4.75truein\fullhsize=10truein\hsize=\hsbody
\output={\ifnum\pageno=0 %%% This is the HUTP version
  \shipout\vbox{\speclscape{\hsize\fullhsize\makeheadline}
    \hbox to \fullhsize{\hfill\pagebody\hfill}}\advancepageno
  \else
  \almostshipout{\leftline{\vbox{\pagebody\makefootline}}}\advancepageno
  \fi}
\def\almostshipout#1{\if L\l@r \count1=1 \message{[\the\count0.\the\count1]}
      \global\setbox\leftpage=#1 \global\let\l@r=R
 \else \count1=2
  \shipout\vbox{\speclscape{\hsize\fullhsize\makeheadline}
      \hbox to\fullhsize{\box\leftpage\hfil#1}}  \global\let\l@r=L\fi}
\fi
%---------------------------------------------------------------------
%
\newcount\yearltd\yearltd=\year\advance\yearltd by -1900

\def\Title#1#2{\nopagenumbers\abstractfont\hsize=\hstitle\rightline{#1}%
\vskip 1in\centerline{\titlefont #2}\abstractfont\vskip .5in\pageno=0}
\def\Date#1{\vfill\leftline{#1}\tenpoint\supereject\global\hsize=\hsbody%
\footline={\hss\tenrm\hyperdef\hypernoname{page}\folio\folio\hss}}%
% (restores pagenumbers)
%
%       use following instead of \Date on the preliminary draft,
%       puts date/time on each page in big mode, writes labels in margins

\def\draftmode{\message{ DRAFTMODE }\def\draftdate{{\rm preliminary draft:
\number\month/\number\day/\number\yearltd\ \ \hourmin}}%
\headline={\hfil\draftdate}\writelabels\baselineskip=20pt plus 2pt minus 2pt
 {\count255=\time\divide\count255 by 60 \xdef\hourmin{\number\count255}
  \multiply\count255 by-60\advance\count255 by\time
  \xdef\hourmin{\hourmin:\ifnum\count255<10 0\fi\the\count255}}}
%       use \nolabels to get rid of eqn, ref, and fig labels in draft mode
\def\nolabels{\def\wrlabeL##1{}\def\eqlabeL##1{}\def\reflabeL##1{}}
\def\writelabels{\def\wrlabeL##1{\leavevmode\vadjust{\rlap{\smash%
{\line{{\escapechar=` \hfill\rlap{\sevenrm\hskip.03in\string##1}}}}}}}%
\def\eqlabeL##1{{\escapechar-1\rlap{\sevenrm\hskip.05in\string##1}}}%
\def\reflabeL##1{\noexpand\llap{\noexpand\sevenrm\string\string\string##1}}}
\nolabels
%
% tagged sec numbers
\global\newcount\secno \global\secno=0
\global\newcount\meqno \global\meqno=1
\def\s@csym{}
\def\newsec#1{\global\advance\secno by1%
{\toks0{#1}\message{(\the\secno. \the\toks0)}}%
%\ifx\answ\bigans \vfill\eject \else \bigbreak\bigskip \fi  %if desired
\global\subsecno=0\eqnres@t\let\s@csym\secsym\xdef\secn@m{\the\secno}\noindent
{\bf\hyperdef\hypernoname{section}{\the\secno}{\the\secno.} #1}%
\writetoca{{\string\hyperref{}{section}{\the\secno}{\vskip2pt \bf \the\secno\quad}} {\bf #1}}%
\par\nobreak\medskip\nobreak}
\def\eqnres@t{\xdef\secsym{\the\secno.}\global\meqno=1\bigbreak\bigskip}
\def\sequentialequations{\def\eqnres@t{\bigbreak}}\xdef\secsym{}
\global\newcount\subsecno \global\subsecno=0
\def\subsec#1{\global\advance\subsecno by1%
{\toks0{#1}\message{(\s@csym\the\subsecno. \the\toks0)}}%
\ifnum\lastpenalty>9000\else\bigbreak\fi
\noindent{\it\hyperdef\hypernoname{subsection}{\secn@m.\the\subsecno}%
{\secn@m.\the\subsecno.} #1}\writetoca{\string\hskip1.45cm
{\string\hyperref{}{subsection}{\secn@m.\the\subsecno}{\secn@m.\the\subsecno.}}
{#1}}\par\nobreak\medskip\nobreak}
\def\appendix#1#2{\global\meqno=1\global\subsecno=0\xdef\secsym{\hbox{#1.}}%
\bigbreak\bigskip\noindent{\bf Appendix \hyperdef\hypernoname{appendix}{#1}%
{#1.} #2}{\toks0{(#1. #2)}\message{\the\toks0}}%
\xdef\s@csym{#1.}\xdef\secn@m{#1}%
\writetoca{{\string\hyperref{}{appendix}{#1}{\vskip2pt \bf {#1}\quad}} {\bf #2}}%
\par\nobreak\medskip\nobreak}
%
%       \eqn\label{a+b=c}	gives displayed equation, numbered
%				consecutively within sections.
%     \eqnn and \eqna define labels in advance (of eqalign?)
%
\def\checkm@de#1#2{\ifmmode{\def\f@rst##1{##1}\hyperdef\hypernoname{equation}%
{#1}{#2}}\else\hyperref{}{equation}{#1}{#2}\fi}
\def\eqnn#1{\DefWarn#1\xdef #1{(\noexpand\relax\noexpand\checkm@de%
{\s@csym\the\meqno}{\secsym\the\meqno})}%
\wrlabeL#1\writedef{#1\leftbracket#1}\global\advance\meqno by1}
\def\f@rst#1{\c@t#1a\em@ark}\def\c@t#1#2\em@ark{#1}
\def\eqna#1{\DefWarn#1\wrlabeL{#1$\{\}$}%
\xdef #1##1{(\noexpand\relax\noexpand\checkm@de%
{\s@csym\the\meqno\noexpand\f@rst{##1}}{\hbox{$\secsym\the\meqno##1$}})}
\writedef{#1\numbersign1\leftbracket#1{\numbersign1}}\global\advance\meqno by1}
\def\eqn#1#2{\DefWarn#1%
\xdef #1{(\noexpand\hyperref{}{equation}{\s@csym\the\meqno}%
{\secsym\the\meqno})}$$#2\eqno(\hyperdef\hypernoname{equation}%
{\s@csym\the\meqno}{\secsym\the\meqno})\eqlabeL#1$$%
\writedef{#1\leftbracket#1}\global\advance\meqno by1}
\def\xeqn{\expandafter\xe@n}\def\xe@n(#1){#1}
\def\xeqna#1{\expandafter\xe@n#1}
\def\eqns#1{(\e@ns #1{\hbox{}})}
\def\e@ns#1{\ifx\UNd@FiNeD#1\message{eqnlabel \string#1 is undefined.}%
\xdef#1{(?.?)}\fi{\let\hyperref=\relax\xdef\next{#1}}%
\ifx\next\em@rk\def\next{}\else%
\ifx\next#1\xeqn#1\else\def\n@xt{#1}\ifx\n@xt\next#1\else\xeqna#1\fi
\fi\let\next=\e@ns\fi\next}

\def\DefWarn#1{\ifx\UNd@FiNeD#1\else
\immediate\write16{*** WARNING: the label \string#1 is already defined ***}\fi}
%
%			 footnotes
\newskip\footskip\footskip14pt plus 1pt minus 1pt %sets footnote baselineskip
\def\footnotefont{\ninepoint}\def\f@t#1{\footnotefont #1\@foot}
\def\f@@t{\baselineskip\footskip\bgroup\footnotefont\aftergroup\@foot\let\next}
\setbox\strutbox=\hbox{\vrule height9.5pt depth4.5pt width0pt}
\global\newcount\ftno \global\ftno=0
\def\foot{\global\advance\ftno by1\def\foot@rg{\hyperref{}{footnote}%
{\the\ftno}{\the\ftno}\xdef\foot@rg{\noexpand\hyperdef\noexpand\hypernoname%
{footnote}{\the\ftno}{\the\ftno}}}\footnote{$^{\foot@rg}$}}
%
%say \footend to put footnotes at end
%will cause problems if \ref used inside \foot, instead use \nref before
\newwrite\ftfile
\def\footend{\def\foot{\global\advance\ftno by1\chardef\wfile=\ftfile
%%$^{\the\ftno}$\ifnum\ftno=1\immediate\openout\ftfile=\jobname.fts\fi%
\hyperref{}{footnote}{\the\ftno}{$^{\the\ftno}$}%
\ifnum\ftno=1\immediate\openout\ftfile=\jobname.fts\fi%
\immediate\write\ftfile{\noexpand\smallskip%
%%\noexpand\item{f\the\ftno:\ }\pctsign}\findarg}%
\noexpand\item{\noexpand\hyperdef\noexpand\hypernoname{footnote}
{\the\ftno}{f\the\ftno}:\ }\pctsign}\findarg}%
\def\footatend{\vfill\eject\immediate\closeout\ftfile{\parindent=20pt
\centerline{\bf Footnotes}\nobreak\bigskip\input \jobname.fts }}}
\def\footatend{}
%
%     \ref\label{text}
% generates a number, assigns it to \label, generates an entry.
% To list the refs on a separate page,  \listrefs
%
\global\newcount\refno \global\refno=1
\newwrite\rfile
\def\ref{[\hyperref{}{reference}{\the\refno}{\the\refno}]\nref}
\def\nref#1{\DefWarn#1%
\xdef#1{[\noexpand\hyperref{}{reference}{\the\refno}{\the\refno}]}%
\writedef{#1\leftbracket#1}%
\ifnum\refno=1\immediate\openout\rfile=\jobname.refs\fi
\chardef\wfile=\rfile\immediate\write\rfile{\noexpand\item{[\noexpand\hyperdef%
\noexpand\hypernoname{reference}{\the\refno}{\the\refno}]\ }%
\reflabeL{#1\hskip.31in}\pctsign}\global\advance\refno by1\findarg}
%	horrible hack to sidestep tex \write limitation
\def\findarg#1#{\begingroup\obeylines\newlinechar=`\^^M\pass@rg}
{\obeylines\gdef\pass@rg#1{\writ@line\relax #1^^M\hbox{}^^M}%
\gdef\writ@line#1^^M{\expandafter\toks0\expandafter{\striprel@x #1}%
\edef\next{\the\toks0}\ifx\next\em@rk\let\next=\endgroup\else\ifx\next\empty%
\else\immediate\write\wfile{\the\toks0}\fi\let\next=\writ@line\fi\next\relax}}
\def\striprel@x#1{} \def\em@rk{\hbox{}}
\def\lref{\begingroup\obeylines\lr@f}
\def\lr@f#1#2{\DefWarn#1\gdef#1{\let#1=\UNd@FiNeD\ref#1{#2}}\endgroup\unskip}
\def\semi{;\hfil\break}
\def\addref#1{\immediate\write\rfile{\noexpand\item{}#1}} %now unnecessary
\def\listrefs{\footatend\vfill\supereject\immediate\closeout\rfile\writestoppt
\baselineskip=\footskip\centerline{{\bf References}}\bigskip{\parindent=20pt%
\frenchspacing\escapechar=` \input \jobname.refs\vfill\eject}\nonfrenchspacing}
\def\startrefs#1{\immediate\openout\rfile=\jobname.refs\refno=#1}
\def\xref{\expandafter\xr@f}\def\xr@f[#1]{#1}
\def\refs#1{\count255=1[\r@fs #1{\hbox{}}]}
\def\r@fs#1{\ifx\UNd@FiNeD#1\message{reflabel \string#1 is undefined.}%
\nref#1{need to supply reference \string#1.}\fi%
\vphantom{\hphantom{#1}}{\let\hyperref=\relax\xdef\next{#1}}%
\ifx\next\em@rk\def\next{}%
\else\ifx\next#1\ifodd\count255\relax\xref#1\count255=0\fi%
\else#1\count255=1\fi\let\next=\r@fs\fi\next}
%

%
% this is ugly, but moore insists
\newwrite\ffile\global\newcount\figno \global\figno=1
\def\fig{fig.~\hyperref{}{figure}{\the\figno}{\the\figno}\nfig}
\def\nfig#1{\DefWarn#1%
\xdef#1{fig.~\noexpand\hyperref{}{figure}{\the\figno}{\the\figno}}%
\writedef{#1\leftbracket fig.\noexpand~\xfig#1}%
\ifnum\figno=1\immediate\openout\ffile=\jobname.figs\fi\chardef\wfile=\ffile%
{\let\hyperref=\relax
\immediate\write\ffile{\noexpand\medskip\noexpand\item{Fig.\ %
\noexpand\hyperdef\noexpand\hypernoname{figure}{\the\figno}{\the\figno}. }
\reflabeL{#1\hskip.55in}\pctsign}}\global\advance\figno by1\findarg}
\def\listfigs{\vfill\eject\immediate\closeout\ffile{\parindent40pt
\baselineskip14pt\centerline{{\bf Figure Captions}}\nobreak\medskip
\escapechar=` \input \jobname.figs\vfill\eject}}
\def\xfig{\expandafter\xf@g}\def\xf@g fig.\penalty\@M\ {}
\def\figs#1{figs.~\f@gs #1{\hbox{}}}
\def\f@gs#1{{\let\hyperref=\relax\xdef\next{#1}}\ifx\next\em@rk\def\next{}\else
\ifx\next#1\xfig #1\else#1\fi\let\next=\f@gs\fi\next}
\def\figin{\epsfcheck\figin}\def\figins{\epsfcheck\figins}
\def\epsfcheck{\ifx\epsfbox\UNd@FiNeD
\message{(NO epsf.tex, FIGURES WILL BE IGNORED)}
\gdef\figin##1{\vskip2in}\gdef\figins##1{\hskip.5in}% blank space instead
\else\message{(FIGURES WILL BE INCLUDED)}%
\gdef\figin##1{##1}\gdef\figins##1{##1}\fi}
\def\DefWarn#1{}
\def\figinsert{\goodbreak\midinsert}
\def\ifig#1#2#3{\DefWarn#1\xdef#1{fig.~\noexpand\hyperref{}{figure}%
{\the\figno}{\the\figno}}\writedef{#1\leftbracket fig.\noexpand~\xfig#1}%
\figinsert\figin{\centerline{#3}}\medskip\centerline{\vbox{\baselineskip12pt
\advance\hsize by -1truein\noindent\wrlabeL{#1=#1}\footnotefont%
{\bf Fig.~\hyperdef\hypernoname{figure}{\the\figno}{\the\figno}:} #2}}
\bigskip\endinsert\global\advance\figno by1}
\newwrite\lfile
{\escapechar-1\xdef\pctsign{\string\%}\xdef\leftbracket{\string\{}
\xdef\rightbracket{\string\}}\xdef\numbersign{\string\#}}
\def\writedefs{\immediate\openout\lfile=\jobname.defs \def\writedef##1{%
{\let\hyperref=\relax\let\hyperdef=\relax\let\hypernoname=\relax
 \immediate\write\lfile{\string\def\string##1\rightbracket}}}}%
\def\writestop{\def\writestoppt{\immediate\write\lfile{\string\pageno
 \the\pageno\string\startrefs\leftbracket\the\refno\rightbracket
 \string\def\string\secsym\leftbracket\secsym\rightbracket
 \string\secno\the\secno\string\meqno\the\meqno}\immediate\closeout\lfile}}
\def\writestoppt{}\def\writedef#1{}
\def\seclab#1{\DefWarn#1%
\xdef #1{\noexpand\hyperref{}{section}{\the\secno}{\the\secno}}%
\writedef{#1\leftbracket#1}\wrlabeL{#1=#1}}
\def\subseclab#1{\DefWarn#1%
\xdef #1{\noexpand\hyperref{}{subsection}{\secn@m.\the\subsecno}%
{\secn@m.\the\subsecno}}\writedef{#1\leftbracket#1}\wrlabeL{#1=#1}}
\def\applab#1{\DefWarn#1%
\xdef #1{\noexpand\hyperref{}{appendix}{\secn@m}{\secn@m}}%
\writedef{#1\leftbracket#1}\wrlabeL{#1=#1}}
\newwrite\tfile \def\writetoca#1{}
\def\leaderfill{\leaders\hbox to 1em{\hss.\hss}\hfill}
%	use this to write file with table of contents
\def\writetoc{\immediate\openout\tfile=\jobname.toc
   \def\writetoca##1{{\edef\next{\write\tfile{\noindent ##1
   \string\leaderfill  {\string\hyperref{}{page}{\noexpand\number\pageno}%
                       {\noexpand\number\pageno}} \par}}\next}}}
%       and this lists table of contents on second pass
\newread\ch@ckfile
\def\listtoc{\immediate\closeout\tfile\immediate\openin\ch@ckfile=\jobname.toc
\ifeof\ch@ckfile\message{no file \jobname.toc, no table of contents this pass}%
\else\closein\ch@ckfile\centerline{\bf Contents}\nobreak\medskip%
{\baselineskip=16pt\footnotefont\parskip=0pt\catcode`\@=11\input\jobname.toc
\catcode`\@=12\bigbreak\bigskip}\fi}
\catcode`\@=12 % at signs are no longer letters
%
%	Unpleasantness in calling in abstract and title fonts
\edef\tfontsize{\ifx\answ\bigans scaled\magstep3\else scaled\magstep4\fi}
\font\titlerm=cmr10 \tfontsize \font\titlerms=cmr7 \tfontsize
\font\titlermss=cmr5 \tfontsize \font\titlei=cmmi10 \tfontsize
\font\titleis=cmmi7 \tfontsize \font\titleiss=cmmi5 \tfontsize
\font\titlesy=cmsy10 \tfontsize \font\titlesys=cmsy7 \tfontsize
\font\titlesyss=cmsy5 \tfontsize \font\titleit=cmti10 \tfontsize
\skewchar\titlei='177 \skewchar\titleis='177 \skewchar\titleiss='177
\skewchar\titlesy='60 \skewchar\titlesys='60 \skewchar\titlesyss='60
\def\titlefont{\def\rm{\fam0\titlerm}% switch to title font
\textfont0=\titlerm \scriptfont0=\titlerms \scriptscriptfont0=\titlermss
\textfont1=\titlei \scriptfont1=\titleis \scriptscriptfont1=\titleiss
\textfont2=\titlesy \scriptfont2=\titlesys \scriptscriptfont2=\titlesyss
\textfont\itfam=\titleit \def\it{\fam\itfam\titleit}\rm}
 \ifx\answ\bigans\else scaled\magstep1\fi
\ifx\answ\bigans\def\abstractfont{\tenpoint}\else
\font\absit=cmti10 scaled \magstep1
\font\abssl=cmsl10 scaled \magstep1
\font\absrm=cmr10 scaled\magstep1 \font\absrms=cmr7 scaled\magstep1
\font\absrmss=cmr5 scaled\magstep1 \font\absi=cmmi10 scaled\magstep1
\font\absis=cmmi7 scaled\magstep1 \font\absiss=cmmi5 scaled\magstep1
\font\abssy=cmsy10 scaled\magstep1 \font\abssys=cmsy7 scaled\magstep1
\font\abssyss=cmsy5 scaled\magstep1 \font\absbf=cmbx10 scaled\magstep1
\skewchar\absi='177 \skewchar\absis='177 \skewchar\absiss='177
\skewchar\abssy='60 \skewchar\abssys='60 \skewchar\abssyss='60
\def\abstractfont{\def\rm{\fam0\absrm}% switch to abstract font
\textfont0=\absrm \scriptfont0=\absrms \scriptscriptfont0=\absrmss
\textfont1=\absi \scriptfont1=\absis \scriptscriptfont1=\absiss
\textfont2=\abssy \scriptfont2=\abssys \scriptscriptfont2=\abssyss
\textfont\itfam=\absit \def\it{\fam\itfam\absit}\def\footnotefont{\tenpoint}%
\textfont\slfam=\abssl \def\sl{\fam\slfam\abssl}%
\textfont\bffam=\absbf \def\bf{\fam\bffam\absbf}\rm}\fi
\def\tenpoint{\def\rm{\fam0\tenrm}% switch back to 10-point type
\textfont0=\tenrm \scriptfont0=\sevenrm \scriptscriptfont0=\fiverm
\textfont1=\teni  \scriptfont1=\seveni  \scriptscriptfont1=\fivei
\textfont2=\tensy \scriptfont2=\sevensy \scriptscriptfont2=\fivesy
\textfont\itfam=\tenit \def\it{\fam\itfam\tenit}\def\footnotefont{\ninepoint}%
\textfont\bffam=\tenbf \def\bf{\fam\bffam\tenbf}\def\sl{\fam\slfam\tensl}\rm}
\font\ninerm=cmr9 \font\sixrm=cmr6 \font\ninei=cmmi9 \font\sixi=cmmi6
\font\ninesy=cmsy9 \font\sixsy=cmsy6 \font\ninebf=cmbx9
\font\nineit=cmti9 \font\ninesl=cmsl9 \skewchar\ninei='177
\skewchar\sixi='177 \skewchar\ninesy='60 \skewchar\sixsy='60
\def\ninepoint{\def\rm{\fam0\ninerm}% switch to footnote font
\textfont0=\ninerm \scriptfont0=\sixrm \scriptscriptfont0=\fiverm
\textfont1=\ninei \scriptfont1=\sixi \scriptscriptfont1=\fivei
\textfont2=\ninesy \scriptfont2=\sixsy \scriptscriptfont2=\fivesy
\textfont\itfam=\ninei \def\it{\fam\itfam\nineit}\def\sl{\fam\slfam\ninesl}%
\textfont\bffam=\ninebf \def\bf{\fam\bffam\ninebf}\rm}
%
%---------------------------------------------------------------------
%

\hyphenation{anom-aly anom-alies coun-ter-term coun-ter-terms}
\def\inv{^{\raise.15ex\hbox{${\scriptscriptstyle -}$}\kern-.05em 1}}

\def\Dsl{\,\raise.15ex\hbox{/}\mkern-13.5mu D} %this one can be subscripted
\def\dsl{\raise.15ex\hbox{/}\kern-.57em\partial}

 %pound sterling
\def\lspace{\ifx\answ\bigans{}\else\qquad\fi}
\def\lbspace{\ifx\answ\bigans{}\else\hskip-.2in\fi} % $$\lbspace...$$
\def\boxeqn#1{\vcenter{\vbox{\hrule\hbox{\vrule\kern3pt\vbox{\kern3pt
	\hbox{${\displaystyle #1}$}\kern3pt}\kern3pt\vrule}\hrule}}}
\def\mbox#1#2{\vcenter{\hrule \hbox{\vrule height#2in
		\kern#1in \vrule} \hrule}}  %e.g. \mbox{.1}{.1}
%	matters of taste
%\def\tilde{\widetilde} \def\bar{\overline} \def\hat{\widehat}
%
% some sample definitions
  %     curly letters

\def\darr#1{\raise1.5ex\hbox{$\leftrightarrow$}\mkern-16.5mu #1}
 %pound sterling

\def\half{{\textstyle{1\over2}}} %puts a small half in a displayed eqn
\def\roughly#1{\raise.3ex\hbox{$#1$\kern-.75em\lower1ex\hbox{$\sim$}}}

\input tikz.tex
% TeX wont stop to complain about errors
%\nonstopmode

\def\a{{\alpha}}
\def\l{{\lambda}}
\def\lb{{\overline\lambda}}

\def\b{{\beta}}
\def\g{{\gamma}}

\def\d{{\delta}}

\def\s{{\sigma}}

\def\half{{1\over 2}}
\def\p{{\partial}}
\def\pb{{\overline\partial}}
\def\t{{\theta}}
\def\tb{{\overline\theta}}
\def\bar{\overline}
\def\({\left(}
\def\){\right)}
\def\cF{{\cal F}}

\def\Tt{{\tilde T}}
\def\frac#1#2{{#1 \over #2}}

%\newread\instream \openin\instream= \jobname.defs
%\ifeof\instream \message{No labels in advance yet. Wait till next pass.}
%\else \closein\instream \input \jobname.defs
%\fi
%\writedefs

\input label.defs

%%%%%%%%%%%%%%

% Caption for inline tikzpictures
\def\DefWarn#1{}
\def\tikzcaption#1#2{\DefWarn#1\xdef#1{Fig.~\the\figno}
\writedef{#1\leftbracket Fig.\noexpand~\the\figno}%
{
\smallskip
\leftskip=20pt \rightskip=20pt \baselineskip12pt\noindent
{{\bf Fig.~\the\figno}\ \ninepoint #2}
\bigskip
\global\advance\figno by1 \par}}

% convert numbers [1-9] to upper case letters [A-I]
\def\ntoalpha#1{%
\ifcase#1%
@%
\or A\or B\or C\or D\or E\or F\or G\or H\or I
\fi
}

% Appendix label
\global\newcount\appno \global\appno=1
\def\applab#1{\xdef #1{\ntoalpha\appno}\writedef{#1\leftbracket#1}\wrlabeL{#1=#1}
\global\advance\appno by1}

% small caps

\lref\MSSTFT{
C.R.~Mafra, O.~Schlotterer, S.~Stieberger and D.~Tsimpis,
``A recursive formula for N-point SYM tree amplitudes,''
  arXiv:1012.3981 [hep-th], to appear in Phys.\ Rev.\  D.
  %%CITATION = ARXIV:1012.3981;%%
}

\lref\PARTTWO{
C.R.~Mafra, O.~Schlotterer and S.~Stieberger,
``Complete $N$--Point Superstring Disk Amplitude. II. Amplitude and Hypergeometric Function Structure,''
  AEI--2011--35, MPP-2011--65.
  }

\Title{\vbox{\vskip -2cm \rightline{AEI--2011--34}\rightline{MPP--2011--47}}}{
	       	\vbox{\vskip-1.5cm
		\centerline{Complete $N$--Point Superstring Disk Amplitude}\vskip6pt
		\centerline{I. Pure Spinor Computation}
		}
	}
\vskip-12pt

\centerline{Carlos R. Mafra$^{a,b,}$\foot{e-mail: crmafra@aei.mpg.de}, 
Oliver Schlotterer$^{b,c,}$\foot{e-mail: olivers@mppmu.mpg.de}, 
and Stephan Stieberger$^{b,c,}$\foot{e-mail: stephan.stieberger@mpp.mpg.de}}

\bigskip\medskip
\centerline{\it $^a$ Max--Planck--Institut f\"ur Gravitationsphysik} 
\centerline{\it Albert--Einstein--Institut, 14476 Potsdam, Germany}
\vskip4pt
\centerline{\it $^b$ Kavli Institute for Theoretical Physics} 
\centerline{\it University of California, Santa Barbara, CA 93106, USA}
\vskip4pt
\centerline{\it $^c$ Max--Planck--Institut f\"ur Physik}
\centerline{\it Werner--Heisenberg--Institut, 80805 M\"unchen, Germany}
\bigskip\medskip\medskip
\noindent
In this paper the pure spinor formalism is used to obtain a compact expression for the superstring 
$N$--point disk amplitude. The color ordered string amplitude is given by a sum over $(N-3)!$ 
super Yang--Mills subamplitudes multiplied by multiple Gaussian hypergeometric functions. In 
order to obtain this result, the cohomology structure of the 
pure spinor superspace is exploited to generalize the Berends--Giele method of computing
super Yang--Mills amplitudes. The method was briefly presented in \MSSTFT, and this  
paper elaborates on the details and contains higher-rank examples of building blocks
and associated cohomology objects.
But the main achievement of this work is to identify these field-theory structures in the 
pure spinor computation of the superstring amplitude. In particular, the associated 
set of basis worldsheet integrals is constructively obtained here and thoroughly investigated together with the structure and properties of the amplitude in \PARTTWO.

\Date{}
\goodbreak

\lref\StieSusy{
  S.~Stieberger and T.R.~Taylor,
  ``Supersymmetry Relations and MHV Amplitudes in Superstring Theory,''
  Nucl.\ Phys.\  B {\bf 793}, 83 (2008)
  [arXiv:0708.0574 [hep-th]].
  %%CITATION = NUPHA,B793,83;%%
}

\lref\StieOpr{
  D.~Oprisa and S.~Stieberger,
  ``Six gluon open superstring disk amplitude, multiple hypergeometric  series
  and Euler-Zagier sums,''
  arXiv:hep-th/0509042
  %%CITATION = HEP-TH/0509042;%%
  \semi
  S.~Stieberger, T.R.~Taylor,
  ``Amplitude for N-Gluon Superstring Scattering,''
Phys.\ Rev.\ Lett.\  {\bf 97}, 211601 (2006).
[hep-th/0607184];
%%CITATION = hep-th/0607184%%

% S.~Stieberger, T.R.~Taylor,
  ``Multi-Gluon Scattering in Open Superstring Theory,''
Phys.\ Rev.\  {\bf D74}, 126007 (2006).
[hep-th/0609175];
%%CITATION = hep-th/0609175%%

% S.~Stieberger, T.R.~Taylor,
  ``Complete Six-Gluon Disk Amplitude in Superstring Theory,''
Nucl.\ Phys.\  {\bf B801}, 128-152 (2008).
[arXiv:0711.4354 [hep-th]].
%%CITATION = arXiv:0711.4354%%
}
\lref\Medinas{
  R.~Medina, F.T.~Brandt and F.R.~Machado,
  ``The open superstring 5-point amplitude revisited,''
  JHEP {\bf 0207}, 071 (2002)
  [arXiv:hep-th/0208121]
  %%CITATION = JHEPA,0207,071;%%
\semi
  L.A.~Barreiro and R.~Medina,
  ``5-field terms in the open superstring effective action,''
  JHEP {\bf 0503}, 055 (2005)
  [arXiv:hep-th/0503182].
  %%CITATION = JHEPA,0503,055;%%
}

\lref\FTAmps{
  C.R.~Mafra,
  ``Towards Field Theory Amplitudes From the Cohomology of Pure Spinor
  Superspace,''
  JHEP {\bf 1011}, 096 (2010)
  [arXiv:1007.3639 [hep-th]].
  %%CITATION = JHEPA,1011,096;%%
}
\lref\wittentwistor{
  E.~Witten,
  ``Twistor-Like Transform In Ten-Dimensions,''
  Nucl.\ Phys.\  B {\bf 266}, 245 (1986).
  %%CITATION = NUPHA,B266,245;%%
}
\lref\psf{
  N.~Berkovits,
  ``Super-Poincare covariant quantization of the superstring,''
  JHEP {\bf 0004}, 018 (2000)
  [arXiv:hep-th/0001035].
  %%CITATION = JHEPA,0004,018;%%
}
\lref\thetaSYM{
  	J.P.~Harnad and S.~Shnider,
	``Constraints And Field Equations For Ten-Dimensional Superyang-Mills
  	Theory,''
  	Commun.\ Math.\ Phys.\  {\bf 106}, 183 (1986)
  	%%CITATION = CMPHA,106,183;%%
\semi
	P.A.~Grassi and L.~Tamassia,
        ``Vertex operators for closed superstrings,''
        JHEP {\bf 0407}, 071 (2004)
        [arXiv:hep-th/0405072].
        %%CITATION = HEP-TH 0405072;%%
}
\lref\tsimpis{
  G.~Policastro and D.~Tsimpis,
``$R^4$, purified,''
  Class.\ Quant.\ Grav.\  {\bf 23}, 4753 (2006)
  [arXiv:hep-th/0603165].
  %%CITATION = CQGRD,23,4753;%%
}
\lref\ictp{
  N.~Berkovits,
  ``ICTP lectures on covariant quantization of the superstring,''
[hep-th/0209059].
%%CITATION = hep-th/0209059%%
}
\lref\FivePt{
  C.R.~Mafra,
``Simplifying the Tree-level Superstring Massless Five-point Amplitude,''
  JHEP {\bf 1001}, 007 (2010)
  [arXiv:0909.5206 [hep-th]].
  %%CITATION = JHEPA,1001,007;%%
}
\lref\mafraids{
  C.R.~Mafra,
  ``Pure Spinor Superspace Identities for Massless Four-point Kinematic Factors,''
JHEP {\bf 0804}, 093 (2008).
[arXiv:0801.0580 [hep-th]].
%%CITATION = arXiv:0801.0580%%
}
\lref\humberto{
  H.~Gomez and C.R.~Mafra,
  ``The Overall Coefficient of the Two-loop Superstring Amplitude Using Pure
  Spinors,''
  JHEP {\bf 1005}, 017 (2010)
  [arXiv:1003.0678 [hep-th]].
  %%CITATION = JHEPA,1005,017;%%
}

\lref\MSST{
  C.R.~Mafra, O.~Schlotterer, S.~Stieberger and D.~Tsimpis,
  ``Six Open String Disk Amplitude in Pure Spinor Superspace,''
  Nucl.\ Phys.\  B {\bf 846}, 359 (2011)
  [arXiv:1011.0994 [hep-th]].
  %%CITATION = NUPHA,B846,359;%%
}
\lref\dhokerRev{
  E.~D'Hoker, D.H.~Phong,
  ``The Geometry of String Perturbation Theory,''
Rev.\ Mod.\ Phys.\  {\bf 60}, 917 (1988).
%%CITATION = PUPT-1039%%
}
\lref\siegel{
  W.~Siegel,
  ``Classical Superstring Mechanics,''
Nucl.\ Phys.\  {\bf B263}, 93 (1986).
%%CITATION = UCB-PTH-85-23%%
}
\lref\PSS{
  C.R.~Mafra,
  ``PSS: A FORM Program to Evaluate Pure Spinor Superspace Expressions,''
[arXiv:1007.4999 [hep-th]].
%%CITATION = arXiv:1007.4999%%
}
\lref\BCJ{
  Z.~Bern, J.J.M.~Carrasco, H.~Johansson,
  ``New Relations for Gauge-Theory Amplitudes,''
Phys.\ Rev.\  {\bf D78}, 085011 (2008).
[arXiv:0805.3993 [hep-ph]].
%%CITATION = arXiv:0805.3993%%
}
\lref\anomaly{
  N.~Berkovits and C.R.~Mafra,
  ``Some superstring amplitude computations with the non-minimal pure spinor
  formalism,''
  JHEP {\bf 0611}, 079 (2006)
  [arXiv:hep-th/0607187].
  %%CITATION = JHEPA,0611,079;%%
}
\lref\BG{
  F.A.~Berends, W.T.~Giele,
  ``Recursive Calculations for Processes with n Gluons,''
Nucl.\ Phys.\  {\bf B306}, 759 (1988).
%%CITATION = Print-88-0100 (LEIDEN)%%
}
\lref\FORM{
  J.A.M.~Vermaseren,
  ``New features of FORM,''
  arXiv:math-ph/0010025.
  %%CITATION = MATH-PH/0010025;%%
\semi
  M.~Tentyukov and J.A.M.~Vermaseren,
  ``The multithreaded version of FORM,''
  arXiv:hep-ph/0702279.
  %%CITATION = HEP-PH/0702279;%%
}
\lref\KK{
  R.~Kleiss, H.~Kuijf,
  ``Multi - Gluon Cross-sections And Five Jet Production At Hadron Colliders,''
Nucl.\ Phys.\  {\bf B312}, 616 (1989).
%%CITATION = Print-88-0425 (LEIDEN)%%
}
\lref\Mangano{
  M.L.~Mangano, S.J.~Parke,
  ``Multiparton amplitudes in gauge theories,''
Phys.\ Rept.\  {\bf 200}, 301-367 (1991).
[hep-th/0509223].
%%CITATION = hep-th/0509223%%
}
\lref\monodVanhove{
  N.E.J.~Bjerrum-Bohr, P.H.~Damgaard, P.~Vanhove,
  ``Minimal Basis for Gauge Theory Amplitudes,''
Phys.\ Rev.\ Lett.\  {\bf 103}, 161602 (2009).
[arXiv:0907.1425 [hep-th]].
%%CITATION = arXiv:0907.1425%%
}
\lref\monodStie{
  S.~Stieberger,
  ``Open \& Closed vs. Pure Open String Disk Amplitudes,''
[arXiv:0907.2211 [hep-th]].
%%CITATION = arXiv:0907.2211%%
}
\lref\KITP{
  C.R.~Mafra, O.~Schlotterer, S.~Stieberger,
  ``Explicit BCJ Numerators from Pure Spinors,''
  [arXiv:1104.5224 [hep-th]].
  %%CITATION = arXiv:1104.5224%%
}
\lref\HoweMF{
  P.S.~Howe,
  ``Pure spinors lines in superspace and ten-dimensional supersymmetric theories,''
Phys.\ Lett.\  {\bf B258}, 141-144 (1991).
}

\lref\GS{
  M.B.~Green, J.H.~Schwarz,
  ``Covariant Description of Superstrings,''
  Phys.\ Lett.\  {\bf B136}, 367-370 (1984).
  %%CITATION = QMC-83-7%%
}
\lref\expPSS{
  N.~Berkovits,
  ``Explaining Pure Spinor Superspace,''
  [hep-th/0612021].
  %%CITATION = hep-th/0612021%%
}
\lref\cohoSO{
  N.~Berkovits,
  ``Cohomology in the pure spinor formalism for the superstring,''
JHEP {\bf 0009}, 046 (2000).
[hep-th/0006003].
%%CITATION = hep-th/0006003%%
}

\lref\twolooptwo{
  N.~Berkovits, C.R.~Mafra,
  ``Equivalence of two-loop superstring amplitudes in the pure spinor and RNS formalisms,''
Phys.\ Rev.\ Lett.\  {\bf 96}, 011602 (2006).
[hep-th/0509234].
%%CITATION = hep-th/0509234%%
}
\lref\stahn{
  C.~Stahn,
  ``Fermionic superstring loop amplitudes in the pure spinor formalism,''
  JHEP {\bf 0705}, 034 (2007).
  [arXiv:0704.0015 [hep-th]].
  %%CITATION = arXiv:0704.0015%%
}
\lref\breno{
  N.~Berkovits, B.C.~Vallilo,
  ``Consistency of superPoincare covariant superstring tree amplitudes,''
JHEP {\bf 0007}, 015 (2000).
[hep-th/0004171].
%%CITATION = hep-th/0004171%%
}
\lref\GSWI{
  M.B.~Green, J.H.~Schwarz, E.~Witten,
  ``Superstring Theory. Vol. 1: Introduction,''
Cambridge, Uk: Univ. Pr. (1987) 469 P. (Cambridge Monographs On Mathematical Physics).
}
\lref\SchwarzJN{
M.B.~Green, J.H.~Schwarz,
``Supersymmetrical Dual String Theory. 2. Vertices and Trees,''
Nucl.\ Phys.\  {\bf B198}, 252-268 (1982)\semi
%%CITATION = CALT-68-872%%
J.H.~Schwarz,
``Superstring Theory,''
Phys.\ Rept.\  {\bf 89}, 223-322 (1982)\semi
%%CITATION = CALT-68-911%%
A.A.~Tseytlin,
``Vector Field Effective Action in the Open Superstring Theory,''
Nucl.\ Phys.\  {\bf B276}, 391 (1986).
%%CITATION = LEBEDEV-86-06%%
}

\listtoc
\writetoc
\break

%********************
\newsec{Introduction}
%********************

The computation of tree-level superstring scattering amplitudes is an important problem since
the birth of string theory (see e.g. \GSWI). But despite being already four decades old, explicit
results for tree amplitudes with more than four external legs \SchwarzJN\ have only recently been completed
using the Ramond--Neveu--Schwarz (RNS) formalism at five points \Medinas, at six points \StieOpr\ and partially up to seven points \StieSusy.
In addition to conceptual issues about higher-point worldsheet integrals, the huge amount of algebraic 
manipulations required to complete these calculations has proven to be a major obstacle to further 
developments. When written in terms of ten-dimensional momenta and polarizations, the
amplitudes simply become too big.

However, since the year 2000 a new formalism for the superstring which can be used 
to compute manifestly super-Poincar\'e invariant scattering amplitudes in superspace is available \psf.
A general proof that the disk amplitudes in the pure spinor formalism 
for an arbitrary number of bosonic and for up to four fermionic external state agree
with the standard RNS prescription was given in \breno; and the supersymmetric four-, five- and six-point tree amplitudes
have been explicitly computed in \refs{\tsimpis\mafraids\FivePt-\MSST}.

In this paper the general problem will be solved; i.e.\ the complete solution for {\it all\/} $N$--point
superstring color-ordered disk amplitude ${\cal A}_N \equiv {\cal A}(1,2,\ldots,N)$ is given by
\eqn\fullsol{
{\cal A}_N = \int \limits_{z_i<z_{i+1}} \prod_{i<j} |z_{ij}|^{-s_{ij}} \bigg[ \prod_{k=2}^{N-2}  
\sum_{m=1}^{k-1} {s_{mk} \over z_{mk}}\; {\cal A}_{YM}(1,2,\ldots,N)  +  {\cal P}(2,\dots,N-2) \bigg],
}
where ${\cal A}_{YM}(1,2, \ldots,N)$ is the color-ordered $N$--point super Yang--Mills subamplitude in ten dimensions, ${\cal P}(2, \ldots,N-2)$ means 
the summation over all $(N-3)!$ permutations of the labels $(2, \ldots,N-2)$
inside the brackets, and the color ordering of the superstring subamplitude is defined by 
the integration region
$\int \limits_{z_i<z_{i+1}} \kern-0.3cm \equiv \prod_{j=2}^{N-2} \int ^1 _{z_{j-1}} dz_j$.

It is straightforward to obtain subamplitudes associated with different color orderings 
$(1,2,\ldots,N) \mapsto (1_\sigma,2_\sigma,\ldots ,(N-1)_\sigma,N)$ for $\sigma \in S_{N-1}$ 
and $i_\sigma \equiv \sigma(i)$ from \fullsol. The worldsheet integrand with its $(N-3)!$ 
kinematic ${\cal A}_{YM}$ packages stay the same, only the integration region has to be adapted to
$$
I_\sigma \equiv \{ z_i \in  {\bf R}, \ 0= z_{1_{\sigma}} \leq z_{2_{\sigma}} \leq \ldots \leq z_{(N-2)_{\sigma}}  \leq z_{(N-1)_{\sigma}} = 1 \},
$$
according to the $\sigma \in S_{N-1}$ permutation in question,
$$
{\cal A}(1_\sigma,2_\sigma,\ldots ,(N-1)_\sigma,N) = \int _{I_\sigma} \prod_{l=2}^{N-2} dz_{l_\sigma} \prod_{i<j} |z_{ij}|^{-s_{ij}} \hskip4cm
$$
\vskip-0.5cm
\eqn\fullsolSIGMA{
 \times \bigg[ \prod_{k=2}^{N-2}  
\sum_{m=1}^{k-1} {s_{mk} \over z_{mk}}\; {\cal A}_{YM}(1,2,\ldots,N)  +  {\cal P}(2,\dots,N-2) \bigg].
}
\noindent
By taking the $\a'\to 0$ field-theory limit of \fullsolSIGMA\ (in particular of the integrals involved using the methods presented in  \PARTTWO), 
it follows that all color-ordered field theory amplitudes
can be written in terms of the $(N-3)!$ dimensional basis $\{ {\cal A}_{YM}(1,2_\sigma, \ldots,(N-2)_\sigma, N-1, N) \mid \s \in S_{N-3}\}$,
a result which was proposed in \BCJ\ and later proved in \refs{\monodVanhove,\monodStie} using monodromy relations in string theory.
Furthermore, plugging in the explicit field-theory limits of the integrals appearing in \fullsol\ (using the method described in
\PARTTWO), one derives the BCJ relations among different color-ordered subamplitudes discussed in \BCJ.

This paper is organized as follows. In section~2 a brief review of the pure spinor formalism is given; with special emphasis
to the elements necessary for the scattering amplitude computations in the following sections.
In section~3 the BRST building blocks which encode the information of the pure spinor CFT correlator will be defined
and their BRST properties studied at length. In particular, a diagrammatic method which associates arbitrary cubic graphs
to certain building block combinations is fully presented (partial results have already been shown in \MSSTFT).
In section~4 a pure spinor generalization of the recursive method of
Berends--Giele \BG\ to compute super Yang--Mills in ten-dimensions is developed which extends the previous results
of \MSSTFT. In section~5 the general $N$--point CFT correlator of the superstring amplitude involved in
the pure spinor prescription is obtained in a compact form using the BRST cohomology objects of the previous sections.
Finally, using a mixture of pure spinor superspace manipulations together with total derivative relations for the
superstring integrals, the superstring $N$--point amplitude is rewritten in terms of the field-theory subamplitudes
as in the result \fullsol\ presented above.
In the appendix~A, the calculations involving the explicit derivation of the building block $T_{12345}$ in
terms of super Yang--Mills superfields (which were omitted from the main text due to its lenghty nature) are
presented in full detail. In appendix~B, the explicit expressions for the pure spinor Berends--Giele currents
$M_{123 \ldots p}$ are written down in terms of BRST building blocks 
for up to and including $M_{1234567}$. Finally, in appendix~C the cubic
graphs which were used to find the expressions of appendix~B are depicted up to $M_{123456}$ (the
132 graphs used to derive $M_{1234567}$ would occupy too much space and were omitted).

%**********************************************************
\newsec{The pure spinor formalism}
%**********************************************************

In the pure spinor formalism \psf, the worldsheet action for the type IIB superstring is
\eqn\action{
S = {1\over 2\pi} \int d^2 z \; \( {1\over 2}\p X^m \pb X_m + p_\a \pb\t^\a 
+ {\bar p}_\a \p\t^\a - \omega_\a \pb\l^\a - {\bar \omega}_\a\p\l^\a\),
}
where $[X^m(z,{\bar z}),\t^\a(z), p_\a(z); \tb^\a({\bar z}), {\bar p}_\a({\bar z})]$ 
and $[\l^\a(z), \omega_\a(z); \lb^\a({\bar z}), {\bar \omega}_\a({\bar z})]$ 
are the Green-Schwarz-Siegel matter variables \refs{\GS,\siegel} and the Berkovits ghosts.
The bosonic pure spinor $\l^\a$ satisfies
\eqn\psconst{
\l^\a \g^m_{\a\b} \l^\b = 0,\qquad m=0, \ldots,9 \quad \a,\b = 1, \ldots,16
}
where $\g^m_{\a\b}$ are the symmetric $16\times 16$ Pauli matrices in $D=10$.
The right-moving fields have
opposite chirality for the type IIA, for the heterotic superstring they are the same as
in the RNS formalism, and for the open superstring the boundary conditions relate the two sectors.
This paper only considers the open superstring, so the right-moving fields will
be ignored.

The supersymmetric momentum and Green-Schwarz constraint are given by
\eqn\susymom{
\Pi^m(z) = \p X^m + \half (\t\g^m \p\t), \quad d_\a(z) = p_\a -\half(\g^m\t)_{\a}\p X_m 
-{1\over 8}(\g^m\t)_{\a}(\t\g_m\p\t),
}
while the ghost contribution to the Lorentz currents is denoted by $N^{mn}(z) = \half (\l\g^{mn}w)$. 
Furthermore, the energy-momentum tensor $T$ with vanishing central charge and
the ghost-number current $J$ are given by 
\eqn\ENMOM{
T(z) = -{1\over 2}\Pi^m\Pi_m - d_\a \p\t^\a + \omega_\a \p\l^\a, \qquad J = \omega_\a \l^\a .
}
Finally, the physical spectrum is obtained from the cohomology of the BRST charge \psf
\eqn\BRSTc{
Q = \oint \l^\a(z)d_\a(z).
}
One can show that these operators satisfy the following relations \refs{\psf,\siegel,\ictp}
\eqnn\OPEs
$$\displaylines{
 d_\a(z)d_\b(w) \rightarrow   -  {\g^m_{\a\b}\Pi_m\over z- w},\quad
 \Pi^m(z)\Pi^n(w) \rightarrow   -  {\eta^{mn}\over (z - w)^2},\quad
 d_\a(z)\,\t^\b(w)\, \rightarrow   {\d^\b_\a\over (z- w)} \cr
  N^{mn}(z)N_{pq}(w) \rightarrow  {4 \over z - w}N^{[m}_{\phantom{m}[p}\d^{n]}_{q]}
- {6 \over (z - w)^2}\d^n_{[p}\d^m_{q]}, \quad
N^{mn}(z)\l^\a(w) \rightarrow - \half {(\l\g^{mn})^\a\over z- w}\cr
\ \ \ \ \ d_\a(z) \Pi^m(w) \rightarrow  {(\g^m\p\t)_\a\over z- w},\quad
 \Pi^m(z)X^n(w) \rightarrow - {\eta^{mn}\over z-w}, \quad J(z)\l^\a(w) \rightarrow {\l^\a \over z-w} \hfill \OPEs \cr
}$$
where the antisymmetrization bracket $[\ldots ]$ encompassing $N$ indices is defined to 
contain an overall factor of $1/N!$. Furthermore, if $f(X,\t)$ is a superfield containing 
only the zero modes of $\t$ and  $D_\a = \p_\a + \half (\g^m\t)_\a \p_m$ is the supersymmetric covariant derivative,
$$
   d_\a(z) f(X(w),\t(w)) \rightarrow  {D_\a f(X(w),\t(w)) \over z- w},\quad
 \Pi^m(z) f(X(w),\t(w)) \rightarrow    -  {k^m f(X(w),\t(w))\over z- w}.
$$
Hence, the action of the BRST operator on superfields is $Qf = \lambda^\alpha D_{\alpha}f$. It is easy 
to show using the OPEs of \OPEs\ and the pure spinor constraint \psconst\ that the BRST charge
indeed satisfies $Q^2=0$. So, the pure spinor formalism can be covariantly quantized, 
is manifestly space-time supersymmetric and contains no worldsheet spinor fields; avoiding from the outset
the issues which make the computation of scattering amplitudes with
the RNS and GS formalisms a difficult task.

Throughout this paper $k^{12\ldots  n}_m$ stands for $k^{1}_m + k^{2}_m + \cdots + k^{n}_m$,
the dimensionless (generalized) Mandelstam invariants are given by
\eqn\genMan{
s_{12\ldots  n} = \a' (k^{1} + k^{2} + \cdots +k^{n})^2,
}
and whenever an $\a'$ is not explicitly written down the convention $2\a' = 1$ has been used.

%*****************************************************
\subsec{Massless vertex operators and SYM superfields}
%*****************************************************

For the open superstring, the vertex operators for the massless states in unintegrated and integrated forms 
are given by
\eqn\vertices{
V^i = \l^\a A^i_\a(x,\t), \qquad 
U^i = \p\t^\a A^i_\a + \Pi^m A^i_m + d_\a W^\a_i + \half {\cal F}_{mn}^i N^{mn},
}
where $i$~denotes the label of the string whose massless modes are described by the 
ten-dimensional super Yang--Mills (SYM) superfields $[A_\a,A_m,W^\a,{\cal F}_{mn}]$ 
satisfying \refs{\ictp, \wittentwistor}
\eqnn\SYM
$$\displaylines{
\hfill D_\a A_\b + D_\b A_\a = \g^m_{\a\b} A_m, \qquad D_\a A_m = (\g_m W)_\a + k_m A_\a  \hfill\phantom{(1.1)}\cr
\hfill D_\a{\cal F}_{mn} = 2k_{[m} (\g_{n]} W)_\a, \qquad  D_\a W^{\b} = {1\over 4}(\g^{mn})^{\phantom{m}\b}_\a{\cal F}_{mn}.  \hfill\SYM\cr
}$$
Their $\t$--expansions can be computed using the gauge $\t^\a A_\a =0$ \refs{\tsimpis, \thetaSYM},
\eqnn\expansions
$$\eqalignno{
A_{\a}(x,\t)& ={1\over 2}a_m(\g^m\t)_\a -{1\over 3}(\xi\g_m\t)(\g^m\t)_\a
-{1\over 32}F_{mn}(\g_p\t)_\a (\t\g^{mnp}\t) + \cdots \cr
A_{m}(x,\t) &= a_m - (\xi\g_m\t) - {1\over 8}(\t\g_m\g^{pq}\t)F_{pq}
         + {1\over 12}(\t\g_m\g^{pq}\t)(\p_p\xi\g_q\t) + \cdots & \expansions \cr
W^{\a}(x,\t) &= \xi^{\a}\kern-0.15cm - {1\over 4}(\g^{mn}\t)^{\a} F_{mn}
           \kern-0.05cm + {1\over 4}(\g^{mn}\t)^{\a}(\p_m\xi\g_n\t)
	   \kern-0.05cm + {1\over 48}(\g^{mn}\t)^{\a}(\t\g_n\g^{pq}\t)\p_m F_{pq} 
	   + \cdots \cr
\cF_{mn}(x,\t) &= F_{mn} - 2(\p_{[m}\xi\g_{n]}\t) + {1\over
4}(\t\g_{[m}\g^{pq}\t)\p_{n]}F_{pq} 
+ {1\over 6}\p_{[m}(\t\g_{n]}^{\phantom{m}pq}\t)(\xi\g_q\t)\p_p + \cdots
}$$
where $a_m(X) = e_m {\rm e}^{ik\cdot X}$, $\xi^{\a}(X) =\chi^\a {\rm e}^{ik\cdot X}$ are the bosonic
and fermionic polarizations and $F_{mn} = 2\p_{[m} a_{n]}$ is the field-strength.
Using the OPEs \OPEs\ and equations of motion \SYM\ one can show that
\eqn\conve{
(\l\g^m W^i)(z_i)U^j(z_j) \rightarrow {1\over z_j - z_i}\big[
(\l\g^n W^j)\cF^i_{mn} - (\l\g^m W^i)(k^i\cdot A^j) + Q(W^i\g^m W^j)
\big],
}
which will be frequently used in the computations below.

As shown by Howe in 1991 \HoweMF, the use of a pure spinor field simplifies
the description of ten-dimensional super Yang--Mills, and this
is naturally incorporated in the pure spinor formalism. For example,
it can be shown that $QV=0$ is equivalent to putting the SYM superfields 
on-shell and it also implies that the BRST variation of the integrated
vertex $U$ is given by $QU = \p V$ \ictp, and many simplifications occur due to this
compact description.
In fact, it has recently been shown how the cohomology of pure spinor superspace \refs{\expPSS,\cohoSO}
is enough to fix all $N$--point scattering amplitudes of $D=10$ SYM \refs{\FTAmps,\MSSTFT}.
So unless otherwise stated, all superfield manipulations in the next sections are done 
on-shell; where both $Q V = 0$ and $QU = \p V$ are satisfied.

%****************************************
\subsec{Tree-level scattering amplitudes}
%****************************************

The prescription to compute a tree-level open-string scattering
amplitude with the pure spinor formalism is given by \psf\ (see also \breno),
\eqn\treepresc{
{\cal A}_N = \left\langle V^1(0)\,V^{(N-1)}(1)\,V^N(\infty) \int dz_2 \, U^2(z_2) 
\;\cdots \int dz_{(N-2)}\, U^{(N-2)}\(z_{(N-2)}\)\right\rangle,
}
where $V^i$ and $U^i$ are the massless
vertex operators of \vertices\ and the $SL(2, R)$ invariance of the disk worldsheet has already been used to fix
three vertex positions to the convenient values $(z_1,z_{N-1},z_N) =( 0,1,\infty)$. The pure spinor bracket $\langle \ldots  \rangle$ appearing in \treepresc\
denotes a zero-mode integration prescription for the variables $\l^\a$ and $\t^\a$, which are the only ones 
among $[d_\a,\Pi^m,N^{mn},\t^\a,\p\t^\a, \l^\a, w_\a]$ to contain zero modes on the disk because they have
conformal weight zero \dhokerRev. Furthermore, the integration regions of \treepresc\ encode the
different color orderings of the external states.
For example, the ordering $A_N(1,2,3,\dots,N)$ is
computed when the integration region is $0=z_1\le z_2 \le \cdots \le z_{N-2} \le z_{N-1} = 1$.

After integrating out the conformal weight-one variables $[d_\a,\Pi^m,N^{mn},\p\t^\a]$ from the 
tree-level amplitude \treepresc\ using the OPEs of \OPEs\ and evaluating the world-sheet integrals,
one is left with a generic pure spinor superspace expression
containing the zero modes of $\l^\a$ and $\t^\a$
\eqn\endres{
{\cal A}_N = \langle \l^\a\l^\b\l^\g f_{\a\b\g}^{i_1 \ldots i_n}(\t,\a') \rangle.
}
In \endres, $f^{i_1\ldots  i_n}_{\a\b\g}(\t,\a')$ is both a composite superfield in the labels $[i_1,\dots, i_n]$ of the external states 
and a function of the string scale $\a'$ satisfying $\l^\a\l^\b\l^\g\l^\d D_\d f_{\a\b\g}^{i_1 \ldots i_n}(\t,\a')=0 $. 
Its specific form in terms of the super Yang--Mills superfields $[A^i_\a,A^i_m,W^\a_i,{\cal F}_{mn}^i]$
follows from the OPE contractions discussed above while its functional dependence on $\a'$
is determined by the momentum expansion of n-point hypergeometric integrals \refs{\Medinas,\StieOpr,\StieSusy}.
As explained in \psf, the zero-mode integration of $\langle \ldots \rangle$ selects from the $\t-$expansion of the enclosed
superfields the unique
element in the cohomology of the pure spinor BRST operator at ghost-number three;
$(\l\g^m\t)(\l\g^n\t)(\l\g^p\t)(\t\g_{mnp}\t)$. Its tree-level normalization can be chosen as 
\eqn\psnorm{
\langle (\l\g^m\t)(\l\g^n\t)(\l\g^p\t)(\t\g_{mnp}\t)\rangle = 1,
}
and although \psnorm\ involves only five $\t^\a$ out of sixteen, it can be shown to
be supersymmetric \psf. Furthermore, given the fact that there is only one scalar in the
decomposition of $(\l^3 \t^5)$ it is possible to compute any correlator using symmetry arguments
and the normalization condition \psnorm\ \refs{\twolooptwo, \stahn}.

%************************************************************
\subsec{Component expansions of amplitudes: a simple example}
%*************************************************************

Given a pure spinor superspace expression like in \endres\ it is
straightforward to perform the $\t$-expansion of the SYM superfields and select the terms
according to \psnorm\ to obtain the supersymmetric result of the scattering amplitude in terms of
the more familiar gluon and gluino polarizations $[e^i_m,\chi_i^\a]$ and their momenta $k^m_i$.
For example, let us obtain the 3-gluon scattering from the component expansion of the 3-point amplitude \psf,
\eqn\three{
{\cal A}_3 =\langle (\l A^1)(\l A^2)(\l A^3)\rangle.
}
Plugging in the $\t$-expansions \expansions\ and selecting the terms with a total of five $\t$'s which contain
only gluon fields results in,
\eqn\threegluon{
{\cal A}_3 =
- {1\over 64}\(
k^3_m e^1_{r}e^2_{s} e^3_{n} -
k^2_m e^1_{r}e^2_{n} e^3_{s} +
k^1_m e^1_{n}e^2_{r} e^3_{s}
\)
\langle (\l \g^r\t) (\l \g^s\t)(\l \g_p \t)(\t\gamma^{pmn}\t)\rangle.
}
In the appendix of \anomaly\ one finds a catalog of the most common pure spinor correlators
and, in particular,
$
\langle (\l \g^r\t) (\l \g^s\t)(\l \g_p \t)(\t\gamma^{pmn}\t)\rangle 
= {1\over 120}\d^{rsp}_{pmn} = {1\over 45}\d^{rs}_{mn}.
$
Therefore the 3-gluon amplitude \threegluon\ is given by
\eqn\tgfim{
{\cal A}_3 = - {1\over 2880}\( 
  (e^1\cdot e^2)(k^2 \cdot e^3) 
+ (e^1\cdot e^3)(k^1 \cdot e^2)
+ (e^2\cdot e^3)(k^3 \cdot e^1)\).
}
Performing the above steps becomes a tedious task 
when higher-point calculations are involved. Fortunately, this procedure
is suitable for an automated handling \refs{\PSS,\FORM}.

%****************************
\newsec{BRST building blocks}
%****************************

Only terms which are in the cohomology of the pure spinor BRST charge \BRSTc\ contribute to the 
$n$-point scattering amplitude \endres.
Therefore it will be convenient to foresee the BRST properties of the objects which naturally appear
in the tree-level calculation of \treepresc. With this intent in mind, in this
section the OPEs among the massless vertex operators \vertices\ are used 
to define composite superfields $L_{2131\ldots  p1}$ and their BRST properties are studied in detail.
It will be found that these superfields transform covariantly under the BRST charge and
generically contain BRST-exact parts. A prescription to consistently remove these parts
will then be given and that will 
define the so-called {\it BRST building blocks\/}: $T_{123\ldots  p}$.

In a later section these building blocks will be used to define other composite
superfields $M_{123\ldots  p}$ and $E_{123\ldots  p}$ with well-defined BRST cohomology
properties. They will turn out to be the natural objects with which to write the superstring
scattering amplitudes. In the course of doing that, several general structures of the string tree amplitudes
will become apparent -- like the fact that they can be written using a $(N-3)!$ dimensional basis
of integrals as conjectured some years ago in \StieOpr.

%****************************************
\subsec{OPE residues of vertex operators}
\subseclab\residues
%****************************************

\noindent
Motivated by the computations one needs to perform when computing tree-level higher-point
amplitudes \refs{\mafraids,\FivePt,\MSST} it is convenient to define composite superfields $L_{2131...p1}$ as
\eqn\Ldefs{
\lim_{z_2\to z_1} V^1(z_1)U^2(z_2) \rightarrow {L_{21}\over z_{21}}, \quad
\lim_{z_p \to z_1} L_{2131...(p-1)1}(z_1) U^p(z_p) \rightarrow {L_{2131...(p-1)1p1}\over z_{p1} },
}
which transform covariantly under the action of the pure spinor BRST charge \FTAmps.
To see this one uses $QV=0$ and $QU=\p V$ to obtain
\eqn\Qrecursd{
QL_{2131\ldots  p1} = \lim_{z_p\to z_1}z_{p1}\big[\( QL_{2131\ldots  (p-1)1}\)(z_1) U^p(z_p) - L_{2131\ldots  (p-1)1}(z_1) \p V^p(z_p)\big].
}
The OPE in the first term of \Qrecursd\ can be computed using the definition \Ldefs\ recursively while
the second term evaluates to $\sum_{j=1}^{p-1} s_{jp} L_{2131\ldots  (p-1)1}V_p$;
as one can easily show by using $\p V^i = (\p \l^\a)A^i_\a + \Pi^m k_m V^i + \p\t^\a D_\a V^i$ and the
OPEs of \OPEs. Therefore,
\eqnn\QLs
$$\eqalignno{
QL_{21} = &\, s_{12} V_1 V_2, \cr
QL_{2131} 
= &\, (s_{13}+s_{23})L_{21}V_3 + s_{12}(L_{31}V_2 + V_1 L_{32}) , \cr
QL_{213141} = & \, (s_{14} + s_{24} + s_{34}) L_{2131} V_4 + (s_{13} +s_{23})( L_{21} L_{43} + L_{2141} V_3 )\cr
& + s_{12}(  L_{3141} V_2 + L_{31} L_{42}  +  L_{41} L_{32} +  V_1 L_{3242}),\cr
QL_{21314151} = &\, (s_{15} + s_{25} + s_{35} + s_{45}) L_{213141} V_{5} + (s_{14} + s_{24} + s_{34}) (L_{213151}V_{4} + L_{2131}L_{54}) \cr
 & + (s_{13} + s_{23}) (L_{214151} V_{3} + L_{2141} L_{53} + L_{2151} L_{43} + L_{21} L_{4353}) \cr
 & + s_{12} (L_{314151} V_{2} + V_{1} L_{324252} + L_{3141} L_{52} + L_{3151} L_{42} + L_{4151} L_{32} \cr
 & \qquad + L_{31} L_{4252} + L_{41} L_{3252} + L_{51} L_{3242}), & \QLs\cr
}$$
while $QL_{2131 \ldots p1}$ for $p\ge 6$ can be also be easily obtained (the general BRST variation of a object related to $L_{2131 \ldots p1}$ will
be written down in the next subsection).

The expressions for $L_{2131\ldots p1}$ in terms of SYM superfields can be obtained using the
OPEs of \OPEs\ in the definition \Ldefs. For example, 
\eqn\Ldo{
L_{21} \equiv \lim_{z_2\to z_1}z_{21}V^1(z_1)U^2(z_2) =
- A^1_m (\l\g^m W^2) - V^1(k^1\cdot A^2) + Q(A^1W^2).
}
Similar calculations yield the expressions for  $L_{2131\ldots  p1}$ and one can show that (discarding
BRST-exact quantities for reasons to be explained in later sections) they are given by:
\eqnn\Ltwo
$$\eqalignno{
L_{21} =&  - A^1_m (\l\g^m W^2) - V^1(k^1\cdot A^2), \cr
L_{2131} = & - L_{21}(k^{12}\cdot A^3) -\big [ \(L_{31} + V^1(k^1\cdot A^3)\)(k^1\cdot A^2) - (1\leftrightarrow 2)\big] \cr
& - (\l\g^m W^3) \( (W^1\g_m W^2) - k^2_m (A^1\cdot A^2)\) \cr
L_{213141} = & - L_{2131} (k^{123}\cdot A^4) - (L_{2141} + L_{21}(k^{12}\cdot A^4)) (k^{12}\cdot A^3) \cr
& - \Big[ (L_{3141} + L_{31}  (k^{13}\cdot A^4))(k^1\cdot A^2) + (L_{41} + V^1 (k^1\cdot A^4)) (k^1\cdot A^3)(k^1\cdot A^2)\cr
& - {1\over 4} (\l\g^m W^4)(W^2\g^{pq}\g_m W^3){\cal F}^1_{pq} - (1\leftrightarrow 2) \Big]\cr
& + (\l\g^m W^4) \( (W^1\g^n W^2) - k_2^n (A^1\cdot A^2) \){\cal F}^3_{mn} & \Ltwo \cr
L_{21314151} = & - L_{213141} (k^{1234}\cdot A^5) - \(L_{213151} + L_{2131}  (k^{123}\cdot A^5)\) (k^{123}\cdot A^4)\cr 
& - \big[ L_{214151} + L_{2141}  (k^{124}\cdot A^5)  + \(L_{2151} + L_{21} (k^{12}\cdot A^5)\) (k^{12}\cdot A^4)\big](k^{12}\cdot A^3) \cr
& - \Big[ \big[ L_{314151} + L_{3141}(k^{134}\cdot A^5) + (L_{3151} + L_{31} (k^{13}\cdot A^5))(k^{13}\cdot A^4)  \cr
& + \(L_{4151} + L_{41}(k^{14}\cdot A^5) + (L_{51} + V^1(k^1\cdot A^5))(k^1\cdot A^4)\)(k^1\cdot A^3)\big] (k^1\cdot A^2)\cr
& + (\l\g^m W^5)\big[ {1\over 4}(W^1\g^{pq}\g^n W^3)\cF^2_{pq}\cF^4_{mn}
+  {1\over 16}(W^4\g^m\g^{pq}\g^{rs}W^1)\cF^2_{rs}\cF^3_{pq}\big] - (1\leftrightarrow 2)\Big] \cr
& + (\l\g^m W^5)\Big[ (W^1\g^n W^2) ( \cF^4_{mp}\cF^3_{np}
-   (W^3\g_m W^4)k^3_n - \half (W^4\g_m\g_n\g^p W^3)k^{12}_p ) \cr
& - \half (W^3\g^{pq}\g_m W^4)\cF^1_{pa}\cF^2_{qa}
+ (A^1\cdot A^2)\( \cF^3_{pq}\cF^4_{mp} k^2_q + (W^3\g_m W^4) (k^2\cdot k^3)\)\Big],
}$$
and can be checked to satisfy the BRST identities \QLs. 

Due to the recursive definition of $L_{2131 \ldots p1}$ care must be taken when discarding BRST-exact terms
when evaluating the OPEs for the next $p+1$ step.
For example, if the BRST-exact term in $L_{21}$ is kept then it follows that \FivePt\
$$
L_{2131} = 
\big[ A^1_m (\l\g^{m}W^2) + V^1 (k^1\cdot A^2)\big](k^{12}\cdot A^3)
$$
$$ 
+  (\l\g^{m}W^3)\big[  A^1_m(k^1\cdot A^2) +  A^{1\,n}{\cal F}^2_{mn}
       -  (W^1\g_{m}W^2)\big]
$$
\eqn\lastl{
+ s_{12}\big[ (A^1W^3)V^2 - (A^2W^3)V^1\big] + (s_{13}+s_{23})(A^1W^2)V^3.
}
Equation \QLs\ implies that after discarding $Q(A^i W^j)$ from $L_{ji}$ 
the last line of \lastl\ must be discarded as well, in order for
$QL_{2131} =s_{12}(L_{31}V_2 + V_1 L_{32}) + (s_{13}+s_{23})L_{21}V_3 $
continue to hold.
Equivalently, one can consider the expressions in \Ltwo\
as an explicit representation for composite superfields $L_{2131\ldots  p1}$ which
satisfy the BRST identities of \QLs. 

It is worth mentioning that the BRST-exact terms
dropped from $L_{ji}$, $L_{jiki}$ and $L_{jikili}$
were observed to cancel out in the final superspace expressions for the five- and six-point computations
of \refs{\FivePt,\MSST}. This seems natural in view of the requirement that the overall amplitude should 
live in the BRST cohomology like its basic ingredients, the vertex operators. This will be the main idea 
to be exploited in the next subsection.

Furthermore, the energy-momentum tensor and the ghost-number current of \ENMOM\ can be used together
with the OPEs of \OPEs\ to show that the conformal weight $h$ of $L_{2131 \ldots p1}$ and its ghost number are given by,
\eqn\confweight{
h\(L_{2131 \ldots p1}\) = (k^1+ \cdots + k^n)^2 \neq 0, \qquad {\rm ghost\ \#}(L_{2131 \ldots p1}) = + 1.
}
This will prove essential to argue that the BRST cohomology for composite superfields it generically empty.

%*************************************************************
\subsec{Definition of BRST building blocks $T_{123\ldots  p }$}
\subseclab\BBB
%*************************************************************

The definition of a rank-$q$ BRST building block $T_{123\ldots  q}$ follows from two steps
\eqn\steps{
L_{2131\ldots q1 } \buildrel{(i)}\over\longrightarrow {\tilde T}_{123\ldots  q} \buildrel{(ii)}\over\longrightarrow T_{123\ldots  q}
}
which are designed to remove BRST-exact terms in $L_{2131\ldots  q1}$ and in ${\tilde T}_{123\ldots  q}$
while still preserving the fundamental BRST variation identities \QLs\ when the
combined redefinition $L_{2131\ldots  q1}\longrightarrow T_{123\ldots  q}$ is used in both
sides of \QLs. 

The first step $(i)$ of \steps\ to obtain ${\tilde T}_{123 \ldots q1}$ from the
composite superfield $L_{2131\ldots  q1}$
depends on all the previous redefinitions of $L_{2131 \ldots p1}$ with $p < q$ which were made
to get the BRST building blocks $T_{123 \ldots p}$. Its purpose is to absorb the extra terms (in the left-hand side)
when the substitutions
$L_{2131\ldots  p1} \rightarrow T_{123 \ldots p}$ are made in the right-hand side of
the BRST variation identity for $QL_{2131 \ldots q1}$.
Therefore the first step $(i)$ ensures
that $Q{\tilde T}_{123 \ldots q}$ is written in terms of $T_{123 \ldots p}$ rather than
$L_{2131 \ldots  p1}$,
\eqnn\QTtildes
$$\eqalignno{
Q{\tilde T}_{123}  =\; &s_{12}(T_{13}V_2 + V_1 T_{23}) + (s_{13}+s_{23})T_{12}V_3\cr
Q{\tilde T}_{1234} =\; &  (s_{14} + s_{24} + s_{34}) T_{123} V_4 + (s_{13} +s_{23})( T_{12} T_{34} + T_{124} V_3 )\cr
& + s_{12}(  T_{134} V_2 + T_{13} T_{24}  +  T_{14} T_{23} +  V_1 T_{234}), &\QTtildes \cr
}$$
and similarly for $\Tt_{123 \ldots q}$ with $q\ge 5$.

One can check using \QTtildes\ that there are certain specific combinations of $\Tt$'s which are BRST-closed, like for
example $Q( \Tt_{123} + \Tt_{231} + \Tt_{312}) = 0$. Furthermore, it was shown in \confweight\ that
the composite superfields $L_{2131 \ldots p1}$ (and therefore also $\Tt_{123 \ldots p}$) have conformal weights $h\neq 0$,
so those combinations must also be BRST-exact -- because the cohomology of $Q$ at ghost-number $+1$ is non-trivial only at
zero conformal weight\foot{We thank Nathan Berkovits for illuminating discussions on this point.}.

So the second step $(ii)$ of \steps\  will involve searching for sums of ${\tilde T}_{123\ldots  q}$ which are BRST-closed in
order to subtract the corresponding BRST-exact parts from ${\tilde T}_{123\ldots  q}$. In principle these
sums can be found by a brute-force analysis of the identities in \QTtildes, but in subsection \secgraphsym\ 
a simple diagrammatic method to find all those sums will be presented. That in turn allows one to obtain
the explicit expressions for all $q-1$ BRST-exact parts $R^{(I)}_{123\ldots  q}$ of $\Tt_{123 \ldots q}$;
\eqn\expsec{
\sum {\tilde T}_{123\ldots  q} = QR^{(I)}_{123\ldots  q}, \quad I = 1,2,3,\dots, q-1,
}
where the $q-1$ different sums will involve different label permutations of $\Tt_{123\ldots  q}$ with $\pm$~signs, see 
subsection~\secgraphsym\ for their precise forms.

The prescription to remove the BRST-exact parts from $\Tt_{123\ldots  q}$ --  which completes
the second step $(ii)$ of \steps\ -- will be explained in
subsection~\explBBBs. After doing that, the previous BRST-closed sums of $\Tt_{123 \ldots q}$ become 
BRST-symmetries of the building blocks $T_{123 \ldots q}$, i.e.,
\eqn\manifest{
\sum  T_{123\ldots  q} = 0.
}
In summary, the two steps in \steps\ are:
\item{$(i)$} Redefine $L_{2131\ldots  q1}\rightarrow {\tilde T}_{123\ldots  q}$ such that $Q{\tilde T}_{123\ldots  q}$ is expressed
in terms of building blocks $T_{123\ldots  p}$ of lower-level $p<q$.
\item{$(ii)$} Remove the BRST-exact parts of ${\tilde T}_{123\ldots  q}$ given by \expsec\ such that $T_{123\ldots  q}$ 
satisfies the symmetry properties \manifest.

\noindent The composite superfields $T_{123\ldots  q}$ defined in this way are the BRST building blocks and obey the
following identities,
\eqnn\QTs
$$\eqalignno{
QT_{12} = & \, s_{12} V_1 V_2, \cr
QT_{123} 
= &\, (s_{13}+s_{23})T_{12}V_3 + s_{12}(T_{13}V_2 + V_1 T_{23}) , \cr
QT_{1234} = & \, (s_{14} + s_{24} + s_{34}) T_{123} V_4 + (s_{13} +s_{23})( T_{12} T_{34} + T_{124} V_3 )\cr
& + s_{12}(  T_{134} V_2 + T_{13} T_{24}  +  T_{14} T_{23} +  V_1 T_{234}), \cr
QT_{12345} = & \, (s_{15} + s_{25} + s_{35} + s_{45}) T_{1234} V_{5} + (s_{14} + s_{24} + s_{34}) (T_{1235} V_{4} + T_{123}T_{45}) \cr
 & + (s_{13} + s_{23}) (T_{1245} V_{3} + T_{124} T_{35} + T_{125} T_{34} + T_{12} T_{345}) \cr
 & + s_{12} (T_{1345} V_{2} + V_{1} T_{2345} + T_{134} T_{25} + T_{135} T_{24} + T_{145} T_{23} \cr
 & + T_{13} T_{245} + T_{14} T_{235} + T_{15} T_{234}) & \QTs
}$$
and so forth. The relations \QTs\ can be generalized as follows,
\eqn\powerQTs{
QT_{12{\dots} n} = \sum_{j=2}^{n} \sum_{\a \in P(\beta_j)} \( s_{1j} + s_{2j} + \cdots + s_{j-1,j}\) 
T_{12\ldots  j-1,\{\a\}}\; T_{j, \{\beta_j \backslash \a\}},
}
where $\b_j = \{j+1,\dots, n\}$, $P(\b_j)$ is the powerset of $\b_j$ and $V_i \equiv T_i$.
Furthermore, the first few BRST symmetries of \manifest\ are given by
\eqn\BRSTexact{\eqalign{
0& = T_{12} + T_{21},\cr
0& = T_{123} + T_{231} + T_{312},\cr
0& = T_{1234} - T_{1243} + T_{3412} - T_{3421},\cr
0& = T_{12345}  - T_{12354} + T_{12543} - T_{12453} + T_{45321} - T_{45312},
}}
where each higher-order building block $T_{123\ldots  q}$ inherits
all the lower-order identities in its first $q-1$ labels (this can be seen 
from the recursive definition of $L_{2131\ldots  p1}$ in \Ldefs). For example, $T_{1234}$ not 
only satisfies the third equation of \BRSTexact\ but also the previous two in the form of
$T_{1234} + T_{2134} = T_{1234} + T_{2314} + T_{3124} = 0.$
Using the diagrammatic method explained below, the following general BRST symmetries for building blocks
will be derived, 
\eqn\TrankNu{\eqalign{
p=2n+1 &: \ \ \ T_{12\ldots n+1[n+2[\ldots [2n-1[2n,2n+1]] \ldots ]]} - 2 T_{2n+1\ldots n+2[n+1[\ldots [3[21]] \ldots ]]}  = 0 \cr
p=2n &: \ \ \ T_{12\ldots n[n+1[\ldots [2n-2[2n-1,2n]] \ldots ]]} + T_{2n\ldots n+1[n[\ldots [3[21]] \ldots ]]}  = 0.
}}
The notation $[i[jk]]$ means consecutive antisymmetrization of pairs of labels starting from the outermost
label, e.g. $[i[jk]] = 1/2(i[jk] - [jk]i) = 1/4(ijk - ikj - jki + kji)$

%**************************************************************************
\subsec{Diagrammatic interpretation of $T_{123\ldots  p}$ building blocks}
\subseclab\secgraphs
%**************************************************************************

As discussed in \BCJ, every color-ordered tree-level field theory amplitude can be
arranged into a form which manifests the kinematic poles that appear,
\eqn\treepoles{
A_{ YM}(1,2,\dots, N) = \sum_i {n_i \over \prod_{\a_i} p^2_{\a_i}}
}
where the sum is over the set of $(2N-4)!/((N-1)!(N-2)!)$ diagrams with only cubic vertices,
$n_i$ represent some kinematic numerator factor and $p^2_{\a_i}$ are the 
propagators of each diagram.
Using this representation for the $N$--point amplitudes it was suggested in \FTAmps\ that
the BRST cohomology of the pure spinor formalism 
might be enough to fix
the ten-dimensional SYM amplitudes, bypassing the need to perform the $\a'\rightarrow 0$
limit of their corresponding open superstring amplitudes. 
%%%%%%%%% COPY INTO THESIS FROM HERE
To that end it is useful to require that the numerator factors $n_i$ have BRST transformations
which are proportional to the Mandelstam invariants associated to their poles,
$Qn_i = \sum_j p^2_{\a_j} m_j$
for some $m_j$. This makes sure that each term in $Qn_i$ cancel one of the poles
and different terms can be concocted to yield an overall BRST-closed amplitude.
So in order for the empirical cohomology method of \FTAmps\ to work, one needs to have
explicit mappings between cubic diagrams and ghost-number {\it three\/} pure spinor
superspace expressions. Although some lower-order examples were presented in \FTAmps,
a general solution was still missing.
But as it became clear later, it is better to have mappings
between cubic diagrams and ghost-number {\it one\/} composite superfields; the BRST building
blocks. This realization
led to the discovery in \MSSTFT\ of a general recursive method to construct expressions in the cohomology of the
BRST charge with the correct properties of $N$--point SYM amplitudes.
So in this section we describe in detail the solution of \MSSTFT\ to find the general dictionary between
cubic-vertex diagrams and ghost-number one pure spinor building blocks.

The idea to obtain the dictionary is to find the precise sums of
building blocks whose BRST variation contains the same set of Mandelstam variables associated
to a particular cubic diagram. And this problem can be solved
by understanding the patterns present in the BRST variation identities of \powerQTs.

\topinsert
\centerline{\hskip\parindent
\tikzpicture [line width=0.30mm]
\scope[xshift=-4.2cm]
\draw (0,0) -- (-1,1) node[above]{$i_2$};
\draw (0,0) -- (-1,-1) node[below]{$i_1$};
\draw (0,0) -- (5,0);
\draw (0.5,-0.2) node{$s_{i_1i_2}$};
\draw (1,0) -- (1,1) node[above]{$i_3$};
\draw (1.6,-0.2) node{$s_{i_1i_2i_3}$};
\draw (2.2,0) -- (2.2,1) node[above]{$i_4$};;
\draw (3, 0.5) node{$\ldots$};
\draw (4,0) -- (4,1) node[above]{$i_n$};;
\draw (4.7,-0.2) node{$s_{i_1 ... i_n}$};
\draw (5.3,0) node{$\ldots$};
\draw (7.0,0) node{$\quad \longleftrightarrow \quad T_{i_1i_2i_3\ldots  i_n\ldots  }$};
\draw (2.4, -1.4) node{(a)};
\endscope
\scope[xshift=-7.0cm,yshift=-4.4cm]
\draw node[left]{$\ldots$} (0,0) -- (1,0) node[right]{$\ldots$};
\draw (0.5, 0.0) -- (0.5, 0.5);
\draw (0.5, 0.5) -- (0, 1) node[above]{$i_1$};
\draw (0.5, 0.5) -- (1, 1) node[above]{$i_2$};
\draw (0.5, -0.3) node[below]{$T_{\ldots  [i_1i_2]\ldots  }$};
\endscope
\scope[xshift=-3.9cm,yshift=-4.4cm]
\draw node[left]{$\ldots$} (0,0) -- (1,0) node[right]{$\ldots$};
\draw (0.5, 0.0) -- (0.5, 1.0);
\draw (0.5, 1.0) -- (0.0, 1.5) node[above]{$i_1$};
\draw (0.5, 1.0) -- (1.0, 1.5) node[above]{$i_2$};
\draw (0.5, 0.5) -- (1.0, 0.5) node[right]{$i_3$};
\draw (0.5, -0.3) node[below]{$T_{\ldots  [[i_1i_2]i_3]\ldots  }$};
\draw (2.1, -1.3) node{(b)};
\endscope
 \scope[xshift=-0.60cm,yshift=-4.4cm]
 \draw node[left]{$\ldots$} (0,0) -- (1,0) node[right]{$\ldots$};
 \draw (0.5, 0.0) -- (0.5, 1.5);
 \draw (0.5, 1.5) -- (0.0, 2.0) node[above]{$i_1$};
 \draw (0.5, 1.5) -- (1.0, 2.0) node[above]{$i_2$};
 \draw (0.5, 1.0) -- (1.0, 1.0) node[right]{$i_3$};
 \draw (0.5, 0.5) -- (1.0, 0.5) node[right]{$i_4$};
 \draw (0.5, -0.3) node[below]{$T_{\ldots  [[[i_1i_2]i_3]i_4]\ldots  }$};
 \endscope
\scope[xshift=2.8cm,yshift=-4.4cm]
\draw node[left]{$\ldots$} (0,0) -- (1,0) node[right]{$\ldots$};
\draw (0.5, 0.0) -- (0.5, 2.0);
\draw (0.5, 2.0) -- (0.0, 2.5) node[above]{$i_1$};
\draw (0.5, 2.0) -- (1.0, 2.5) node[above]{$i_2$};
\draw (0.5, 1.5) -- (1.0, 1.5) node[right]{$i_3$};
\draw (1.3, 1.1) node{$\vdots$};
\draw (0.5, 0.5) -- (1.0, 0.5) node[right]{$i_n$};
\draw (0.5, -0.3) node[below]{$T_{\ldots  [[\ldots  [ [i_1i_2] i_3]\ldots  ]i_n]\ldots  }$};
\endscope
\endtikzpicture
}
\tikzcaption\figdiags{
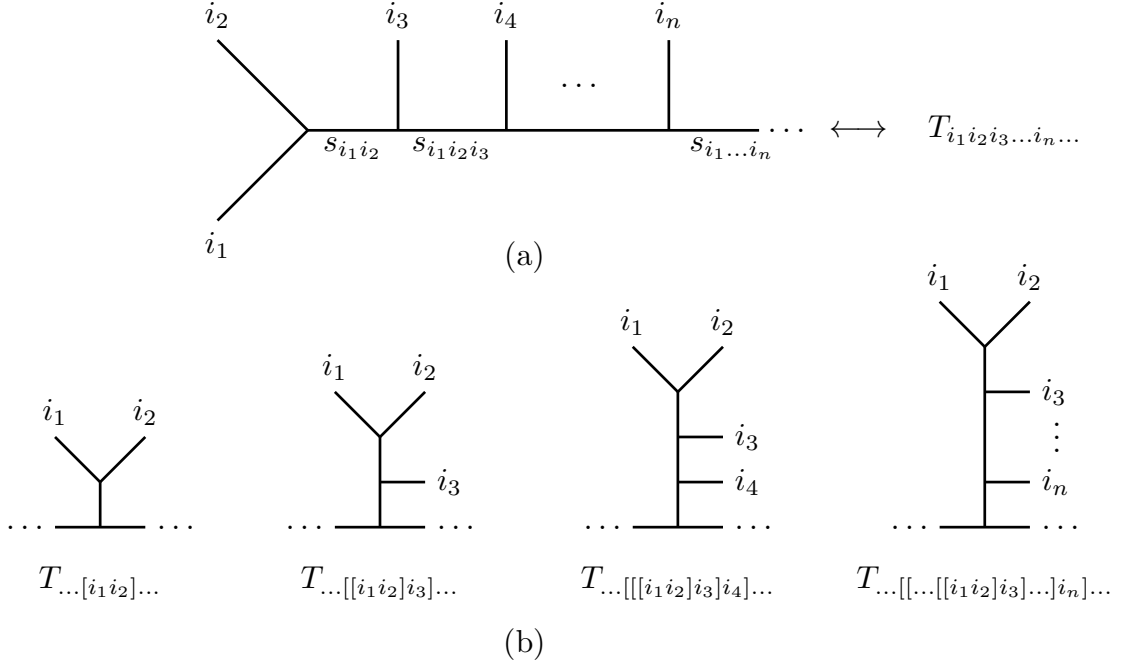
 (a) A tail-end cubic diagram with kinematic poles $\{ s_{i_1i_2},\dots, s_{i_1i_2\ldots i_n}\}$
% and its associated
corresponds to the
building block $T_{i_1i_2\ldots  i_n\ldots  }\,$.
 (b) Branches of cubic diagrams and their associated building blocks.
The motivation behind this dictionary lies on the fact that all kinematic invariants specified by the cubic graphs
are present in the BRST variation of their corresponding building blocks.}
\endinsert

To see this consider the diagram (a) of \figdiags\ where one leg has been removed
and which contains the set of kinematic poles $\{ s_{i_1i_2}, s_{i_1i_2i_3},\dots, s_{i_1...i_n}\}$.
From equation \powerQTs\ one checks that all terms in the BRST variation of $T_{i_1i_2i_3\ldots  i_n\ldots  }$
contain at least one of those Mandelstam variables without exception, schematically
\eqn\Qtailend{
QT_{i_1i_2i_3\ldots  i_n\ldots  } \longrightarrow \left\{ s_{i_1i_2},s_{i_1i_2i_3},\dots, s_{i_1i_2i_3\ldots  i_n\ldots  }\right\}
}
where the trailing dots on the labels of the building block correspond to the amputated part of the diagram.
Given this match, we associate the building block of \Qtailend\ to the cubic graph of \figdiags~(a).

\noindent 
To find the appropriate BRST building blocks which can be associated with the {\it branches\/} containing two amputated legs
in \figdiags~(b), note the pattern that certain sums
of $T_{123\ldots  p}$ with different label orderings have a different set of Mandelstam invariants 
in their BRST variation. As seen on \Qtailend, the BRST variation of $T_{i_1 i_2\ldots i_n}$ 
contains all elements of the set $\{s_{i_1i_2},s_{i_1i_2i_3},\dots, s_{i_1\ldots i_n}\}$
but antisymmetrization in certain labels replaces some elements by others, e.g.
\eqnn\sbranches
$$\eqalignno{
\quad\qquad Q T_{i_1\ldots i_p [jk ] r_1\ldots r_q}
\longrightarrow &\; s_{jk} \hbox{ instead of } s_{i_1 i_2 \ldots i_p j}\cr
\quad\qquad Q T_{i_1 \ldots i_p [j [kl]] r_1 \ldots r_q}
\longrightarrow &\;  s_{kl} , s_{jkl} \hbox{ instead of } s_{i_1 \ldots i_p j}, s_{i_1 \ldots i_p jk} & \sbranches\cr
\quad\qquad Q T_{i_1 \ldots i_p [j [k [lm]]] r_1 \ldots r_q}
\longrightarrow &\; s_{lm},  s_{klm},  s_{jklm} \hbox{ instead of }  s_{i_1 \ldots i_p j},
s_{i_1 \ldots i_p jk},  s_{i_1 \ldots i_p jkl}\ ,
}$$
where the two sets of dots in the building blocks correspond to the amputated parts of the graphs (b) in \figdiags.
The patterns shown in \sbranches\ therefore justify the general dictionary given in \figdiags (b).

%******************************************
\subsec{BRST symmetries of building blocks}
\subseclab\secgraphsym
%******************************************

It is not difficult to use the BRST variations of ${\tilde T}_{123\ldots  q}$ in \QTtildes\
to find their BRST-closed sums for small $q$ by trial and error. Since the cohomology at conformal weight $h \neq 0$ is empty,
these same BRST-closed combinations of ${\tilde T}$'s are also BRST-exact. As explained in the previous subsection,
the removal of the BRST-exact parts of ${\tilde T}_{123\ldots  q}$ gives rise to the definition of the
building block $T_{123\ldots  q}$ and at the same time the BRST-closed sum of ${\tilde T}$'s translates
into a symmetry of the associated $T_{12\ldots  n}$ (see equation \manifest).
Therefore it is imperative to find the general BRST-closed sums of ${\tilde T}$'s, or equivalently,
the general symmetries of $T$'s.

So in this subsection we use the diagrammatic interpretation of building blocks to predict the symmetry 
properties of $T_{12\ldots n}$ which in turn allow the BRST-exact parts of ${\tilde T}_{123\ldots  n}$ to
be found (see subsection \explBBBs).

\topinsert
\centerline{\hskip\parindent
\tikzpicture[line width=0.30mm]
\draw (0,0.5) node[above]{$2$} -- (0.5,0);
\draw (0,-0.5) node[below]{$1$}  -- (0.5,0);
\draw (1,0.5) node[above]{$3$} -- (1,0);
\draw (0.5,0.0) -- (1.5,0.0) node[right]{$\ldots$};
\draw (3.6,0) node{$\displaystyle = \; \left\{ \eqalign{ &T_{123} \cr &T_{321} \ - \ T_{312} } \right.$};
\endtikzpicture
}
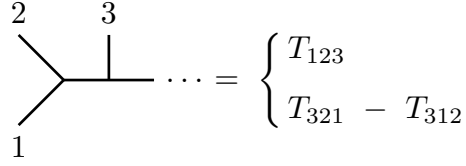
\tikzcaption\figTtsym{Two different ways to interpret the same diagram give rise to an identity for $T_{ijk}$.
In the first expression it is viewed as a tail-end graph, while in the second it is interpreted as a branch.}
\endinsert
\noindent
As a first example, consider the diagram of \figTtsym.
In the first expression the diagram is interpreted as a tail-end graph like the one
depicted in (a) of \figdiags\ and  is associated with the building block $T_{123}$. However, in the second expression
the diagram is viewed as a branch like the first graph of (b) in \figdiags, where
one of the ``missing'' legs now contains the label $3$ and it is therefore associated with
$2T_{3[21]}= T_{321} - T_{312}$.
The fact that both interpretations have to agree implies
the symmetry identity \BRSTexact\ for $T_{ijk}$,
$$
0 = T_{123} - T_{321} + T_{312} = T_{123} + T_{231} + T_{312}.
$$
The relative sign between the two viewpoints is fixed by the fact that diagram associated with 
$T_{12\ldots n}$ catch a $(-1)^{n-1}$ sign under inversion $(1,2,3,\dots, n-1, n) \leftrightarrow (n, n-1,\dots, 1)$. 
Hence, we have to make sure that the sign of $T_{123\ldots n}$ relative to $T_{n,n-1,\ldots 21}$ is $(-1)^n$ 
in \manifest, e.g. $T_{123} + (-1)^{3} T_{321} + \cdots = 0$.

This same idea can be used to obtain the BRST symmetries for higher-order building blocks. For example, the 
symmetries of $T_{123\ldots  n}$ for $n=4$, $5$, $6$, $7$, $8$ are obtained from the diagrams of \figYsix,
\eqn\BRSTexacttwo{\eqalign{
0&= 2T_{12[34]} + 2T_{43[21]} ,\cr 0 &= 2T_{123[45]} - 4 T_{54[3[21]]},\cr
0& = 4T_{123[4[56]]} + 4T_{654[3[21]]},\cr 0 &= 4T_{1234[5[67]]} - 8 T_{765[4[3[21]]]} ,\cr
0& =8 T_{1234[5[6[78]]]} + 8 T_{8765[4[3[21]]]}.
}}

\topinsert
\centerline{\hskip\parindent
\tikzpicture[scale=0.8,line width=0.30mm]
\scope[xshift=-2.5cm]
\draw (-0.5,0) node[left]{$\ldots$} -- (0,0);
\draw (0,0) -- (1,1) node[right]{$1$};
\draw (0,0) -- (1,-1) node[right]{$4$};
\draw (0.5,0.5) -- (1,0.5) node[right]{$2$};
\draw (0.5,-0.5) -- (1,-0.5) node[right]{$3$};
\draw (0.2,-2.1) node{$ \displaystyle =
\left\{ \eqalign{ 2& T_{12[34]} \cr  -2 &T_{43[21]} } \right.$};
\endscope
%%%%%%%%%%%%%%%
\scope[xshift=-1.1cm]
\draw (2.5,0) node[left]{$\ldots$} -- (3,0);
\draw (3,0) -- (4,1) node[right]{$1$};
\draw (3,0) -- (4,-1) node[right]{$5$};
\draw (3.6,0.6) -- (4,0.6) node[right]{$2$};
\draw (3.2,0.2) -- (4,0.2) node[right]{$3$};
\draw (3.5,-0.5) -- (4,-0.5) node[right]{$4$};
\draw (3.2,-2.1) node{$ \displaystyle = 
\left\{ \eqalign{ &2 T_{123[45]}  \cr & 4 T_{54[3[21]]}  } \right.$};
\endscope
%%%%%%%%%%%%%%
\scope[xshift=0.0cm]
\draw (5.5,0) node[left]{$\ldots$} -- (6,0);
\draw (6,0) -- (7,1) node[right]{$1$};
\draw (6,0) -- (7,-1) node[right]{$6$};
\draw (6.6,0.6) -- (7,0.6) node[right]{$2$};
\draw (6.2,0.2) -- (7,0.2) node[right]{$3$};
\draw (6.6,-0.6) -- (7,-0.6) node[right]{$5$};
\draw (6.2,-0.2) -- (7,-0.2) node[right]{$4$};
\draw (6.6,-2.1) node{$\displaystyle =
\left\{ \eqalign{ 4&T_{123[4[56]]} \cr -4&T_{654[3[21]]} } \right.$};
\endscope
\endtikzpicture
}
\bigskip
\centerline{\hskip\parindent
\tikzpicture[scale=0.8, line width=0.30mm]
\scope[xshift=-5.7cm]
\draw (-0.5,0) node[left]{$\ldots$} -- (0,0);
\draw (0,0) -- (0.5,1);
\draw (0,0) -- (0.5,-1);
\draw (0.5,1) -- (2,1);
\draw (1,0.5) node[below]{$4$} -- (1,1);
\draw (1.5,0.5) node[below]{$3$} -- (1.5,1);
\draw (2,1) -- (2.5,1.5) node[right]{$1$};
\draw (2,1) -- (2.5,0.5) node[right]{$2$};
\draw (0.5,-1) -- (2,-1);
\draw (1.5,-0.5) node[above]{$5$} -- (1.5,-1);
\draw (2,-1) -- (2.5,-1.5) node[right]{$7$};
\draw (2,-1) -- (2.5,-0.5) node[right]{$6$};
\draw (4.8,0) node{$ \displaystyle\ \; =
\left\{ \eqalign{ &4T_{1234[5[67]]} \cr & 8T_{765[4[3[21]]]} }\right.$};
\endscope
%%%%%%%%%%%%%%%%%%%%
\scope[xshift=-1.7cm]
\draw (3.6,0) node{$\ldots$};
\draw (4,0) -- (4.5,0);
\draw (4.5,0) -- (5,1);
\draw (4.5,0) -- (5,-1);
\draw (5,1) -- (6.5,1);
\draw (5.5,0.5) node[below]{$4$} -- (5.5,1);
\draw (6,0.5) node[below]{$3$} -- (6,1);
\draw (6.5,1) -- (7,1.5) node[right]{$1$};
\draw (6.5,1) -- (7,0.5) node[right]{$2$};
\draw (5,-1) -- (6.5,-1);
\draw (5.5,-0.5) node[above]{$5$} -- (5.5,-1);
\draw (6,-0.5) node[above]{$6$} -- (6,-1);
\draw (6.5,-1) -- (7,-1.5) node[right]{$8$};
\draw (6.5,-1) -- (7,-0.5) node[right]{$7$};;
\draw (9.5,0) node{$ \displaystyle\ \; =
\left\{ \eqalign{ 8&T_{1234[5[6[78]]]} \cr  - 8&T_{8765[4[3[21]]]} } \right.$};
\endscope
\endtikzpicture
}
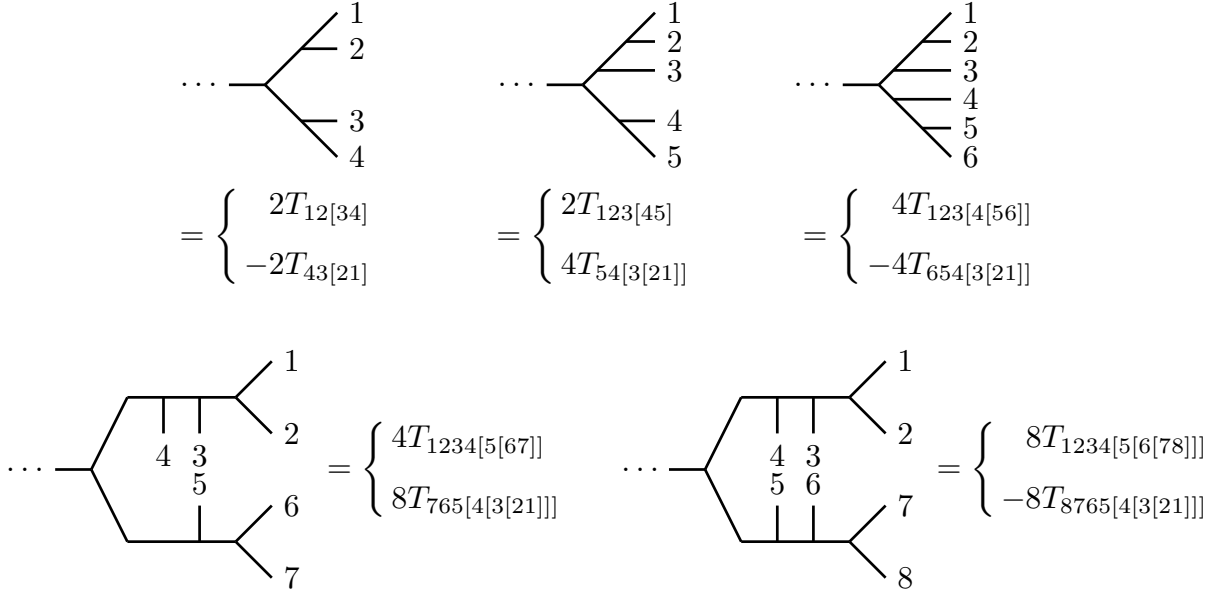
\tikzcaption\figYsix{Diagrammatic derivation of the BRST symmetries of higher-order building blocks. The top (bottom) line corresponds
to the building block association which follow from reading the diagram in a counter-clockwise (clockwise) direction.}
\endinsert

Using the BRST variations \QTs\ we checked up to $T_{12345678}$ that these relations are indeed BRST-closed 
and obtained their explicit BRST-exact parts for up to ${\tilde T}_{12345}$.
The latter was made using the explicit expressions of ${\tilde T}_{123\ldots  p}$ in terms of super Yang--Mills superfields to
find the explicit solutions $R^{(n)}_{123\ldots  p}$ of equation \expsec, and that will be presented in the
next section. 

To write down the generalization of \BRSTexacttwo\ to higher $p >8$, let us distinguish between odd and even ranks for ease of notation:
\eqn \TrankN { \eqalign{
p=2n+1 &: \ \ \ T_{12\ldots n+1[n+2[\ldots [2n-1[2n,2n+1]] \ldots ]]} - 2 T_{2n+1\ldots n+2[n+1[\ldots [3[21]] \ldots ]]}  = 0 \cr
p=2n &: \ \ \ T_{12\ldots n[n+1[\ldots [2n-2[2n-1,2n]] \ldots ]]} + T_{2n\ldots n+1[n[\ldots [3[21]] \ldots ]]}  = 0
}}
The relations for $p=2n+1$ and $p=2n$ involve $3 \cdot 2^{n-1}$ and $2^{n}$ terms, respectively.

We should emphasize again that the lower rank identities for $T_{12\ldots q}$ carry over to $T_{12\ldots p}$ with $p>q$.
The last labels $q+1,\dots,p$ are then simply left untouched, e.g. $0 = T_{(12)345} = T_{[123]45} = T_{12[34]5} + T_{43[21]5}$ 
at rank $p=5$. By applying the $p-1$ symmetries available at rank $p$, one can successively move a particular label 
to the first position, i.e. express $T_{i_1i_2\ldots i_p}$ as a combination of $T_{1j_1j_2\ldots j_{p-1}}$. 
Hence, there are $(p-1)!$ independent rank-$p$ building blocks $T_{i_1i_2\ldots i_p}$.

%***********************************************************
\subsec{Explicit construction of $T_{12\ldots  p}$ }
\subseclab\explBBBs
%***********************************************************

The definition of the first BRST building block $T_{12}$ requires only the step $(ii)$ in \steps,
as there are no lower-order redefinitions to take into account in the first step $(i)$; that is $\Tt_{12} \equiv L_{21}$. 
From the BRST variation of $\Tt_{12}$ in \QLs\ together with the equations
of motion \SYM\ one sees that its symmetric part is BRST-closed;
$Q(\Tt_{21} + \Tt_{12}) = s_{12}(V_1V_2 + V_2V_1) = 0$,
and also BRST-exact \FTAmps
\eqn\Ltwoexact{
\Tt_{21} + \Tt_{12} = - Q(A^1\cdot A^2) \equiv - QD_{12}.
}
As discussed in \manifest, the definition of the BRST building block $T_{12}$ 
must be made to satisfy $T_{12}+ T_{21} = 0$. This is accomplished by
\eqn\Ttwodef{
T_{12} = \Tt_{[21]} = \Tt_{21} + {1\over 2}QD_{12}.
}

The definition of the building block $T_{123}$ now proceeds using
both steps of \steps. The first redefinition $L_{2131}\buildrel{(i)}\over\rightarrow {\tilde T}_{123}$ is found
by substituting $L_{ji} = \Tt_{ij} = T_{ij} - \half QD_{ij}$ in the right-hand side of $QL_{2131}$ in \QLs, which leads to:
$$\displaylines{
Q\(L_{2131} + \half s_{12}\big[D_{13}V_2 - D_{23}V_1\big] + \half (s_{13} + s_{23})D_{12}V_3\) \cr
= s_{12}(T_{13}V_2 + V_1 T_{23}) + (s_{13}+s_{23})T_{12}V_3.\cr
}$$
Therefore by defining
\eqn\Ttthreedef{
{\tilde T}_{123} = L_{2131} + \half s_{12}\big[
D_{13}V_2 - D_{23}V_1\big] + \half (s_{13} + s_{23})D_{12}V_3,
}
one obtains the desired identity $Q{\tilde T}_{123} =s_{12}(T_{13}V_2 + V_1 T_{23}) + (s_{13}+s_{23})T_{12}V_3$.

Two BRST-closed combinations of $\tilde T_{ijk}$ are easily identified,
\eqn\QclosedTtthree{
Q({\tilde T}_{123} + {\tilde T}_{213}) = 0, \quad 
Q({\tilde T}_{123} + {\tilde T}_{312} + {\tilde T}_{231} ) = 0,
}
and one can show using SYM equations of motion \SYM\ that they originate as the BRST variation of 
ghost number zero superfields $R^{(1)}_{123}$, $R^{(2)}_{123}$ \refs{\MSST,\MSSTFT}
\eqn\exactthree{
{\tilde T}_{123} + {\tilde T}_{213} = Q R^{(1)}_{123},
\quad
{\tilde T}_{123} + {\tilde T}_{312} + {\tilde T}_{231} = Q R^{(2)}_{123},
}
where $R^{(1)}_{123}= D_{12}(k^{12}\cdot A^3)$,\quad $R^{(2)}_{123} = D_{12}(k^2\cdot A^3) + {\rm cyclic}(123)$.
The BRST building block  $T_{123}$ is obtained by removing these BRST-exact pieces
\eqn\Tthreedef{
T_{123} = {\tilde T}_{123} - QS^{(1)}_{123}, \quad S^{(1)}_{123} = \half R^{(1)}_{123} + {1\over 3} R^{(2)}_{[12]3},
}
which implies the following BRST symmetries for $T_{ijk}$:
\eqn\Ttman{
T_{123} + T_{213} = T_{123} + T_{312} + T_{231} = 0.
}
%To see that \Tthreedef\ leads to the identities \Ttman\ one uses $R^{(1)}_{213} = R^{(1)}_{123}$ and
%$$
%R^{(1)}_{123} + R^{(1)}_{312} + R^{(1)}_{231} = R^{(2)}_{123} + R^{(2)}_{213}, \quad
%R^{(2)}_{[12]3} + R^{(2)}_{[31]2} + R^{(2)}_{[23]1} = 3R^{(2)}_{[12]3}
%$$
%to obtain
%\eqn\Soneids{
%S^{(1)}_{123} + S^{(1)}_{213} = R^{(1)}_{123}, \quad S^{(1)}_{123} + S^{(1)}_{312} + S^{(1)}_{231} = R^{(2)}_{123},
%}
%which implies \Ttman\ via \exactthree\ and \Tthreedef.
The definition of $T_{1234}$ is done similarly and uses the information from the lower-order redefinitions of
$L_{21}$ and $L_{2131}$. First one rewrites $L_{ji}$ and $L_{jiki}$ in terms of $T_{ij}$ and $T_{ijk}$ in the
RHS of the identity for $QL_{213141}$ given in \QLs.
After some algebra one finds 
\eqnn\Ttildefour
$$\eqalignno{
{\tilde T}_{1234} & = \; L_{213141} - {1\over 4}\big[
(s_{13}+s_{23})D_{12}QD_{34} + s_{12}\( D_{13}QD_{24}  + D_{14}QD_{23} \)
\big] & \Ttildefour \cr
&+ {1\over 2}\big[(s_{13}+s_{23})\(D_{12}T_{34}-D_{34}T_{12}\) 
+ s_{12}\(D_{13}T_{24}+D_{14}T_{23}-D_{23}T_{14}-D_{24}T_{13}\)\big]\cr
&- (s_{14}+s_{24}+s_{34})S^{(1)}_{123}V_4 - (s_{13}+s_{23})S^{(1)}_{124}V_3 
+ s_{12}(S^{(1)}_{234}V_1-S^{(1)}_{134}V_2)\cr
}$$
which satisfies the required property of
\eqnn\Qtildefo
$$\eqalignno{
Q{\tilde T}_{1234} =&\;\, s_{12}(  T_{134} V_2 + T_{13}T_{24}  + T_{14}T_{23} +  V_1 T_{234}) & \Qtildefo\cr
&+  (s_{13} +s_{23})( T_{12}T_{34} + T_{124} V_3 ) +  (s_{14} + s_{24} + s_{34}) T_{123} V_4. \cr
}$$
Using \Qtildefo\ it is easy to check that the lower-order identities of ${\tilde T}_{123}$ given by
\QclosedTtthree\ are inherited by the first three labels of ${\tilde T}_{1234}$ 
and that there is one additional BRST identity involving the fourth label,
$$
Q\({\tilde T}_{1234} + {\tilde T}_{2134}\) =
Q\({\tilde T}_{1234} + {\tilde T}_{3124} + {\tilde T}_{2314}\) =
Q\({\tilde T}_{1234} - {\tilde T}_{1243} + {\tilde T}_{3412} - {\tilde T}_{3421}\)
= 0,
$$
in accord with the discussions of section \secgraphsym.
Using the SYM equations of motion in a long sequence of calculations shows that 
these combinations are indeed BRST-exact,
\eqn\trivialone{\eqalign{
{\tilde T}_{1234} + {\tilde T}_{2134} &= QR^{(1)}_{1234}\cr
{\tilde T}_{1234} + {\tilde T}_{3124} + {\tilde T}_{2314} &= QR^{(2)}_{1234}\cr
{\tilde T}_{1234} - {\tilde T}_{1243} + {\tilde T}_{3412} - {\tilde T}_{3421} &= QR^{(3)}_{1234},
}}
where
\eqnn\trivials
$$\eqalignno{
R^{(1)}_{1234} = &- R^{(1)}_{123} (k^{123}\cdot A^4) - {1\over 4}s_{12}\big[
D_{13}D_{24} + D_{14}D_{23}\big], & \trivials \cr
%%%
R^{(2)}_{1234} = &- R^{(2)}_{123}(k^{123}\cdot A^4)
       - {1\over 4}\big[ 
         s_{12}D_{23}D_{14}
       + s_{23}D_{24}D_{13}
       + s_{13}D_{34}D_{12}
       \big],\cr
%%%
R^{(3)}_{1234} = &\; (k^1\cdot A^2)\big[
  D_{14}(k^4\cdot A^3)
- D_{13}(k^3\cdot A^4)
\big]
- (k^2\cdot A^1)\big[
  D_{24}(k^4\cdot A^3)
- D_{23}(k^3\cdot A^4)
\big]\cr
&+ {1\over 4}D_{12}D_{34}(s_{14} + s_{23} - s_{13} - s_{24}) 
+ D_{12}\big[(k^4\cdot A^3)(k^2\cdot A^4) - (k^3\cdot A^4)(k^2\cdot A^3)\big]\cr
& + D_{34}\big[(k^2\cdot A^1)(k^4\cdot A^2) - (k^1\cdot A^2)(k^4\cdot A^1)\big] + (W^1 \g^m W^2)(W^3 \g_m W^4).  \cr
}$$
Removing these BRST-exact parts leads to the rank-four BRST building block---which is accomplished with the second redefinition
${\tilde T}_{1234} \buildrel{(ii)}\over\longrightarrow T_{1234}$, 
\eqn\Tfourdef{
T_{1234} = {\tilde T}_{1234} - QS_{1234}^{(2)},
}
where $S^{(2)}_{1234}$ is defined recursively by
\eqnn\Stwodef
$$\eqalignno{
S^{(2)}_{1234} &= {3\over 4}S^{(1)}_{1234} + {1\over 4}( S^{(1)}_{1243} - S^{(1)}_{3412} + S^{(1)}_{3421}) + {1\over 4}R^{(3)}_{1234} \cr
S^{(1)}_{1234} &= {1\over 2}R^{(1)}_{1234} + {1\over 3}R^{(2)}_{[12]34}. & \Stwodef\cr
}$$
To see that \Tfourdef\ and \Stwodef\ imply the BRST symmetries of
\eqn\Tfoman{
T_{1234}+T_{2134}=
T_{1234} + T_{3124} +T_{2314}
=T_{1234} - T_{1243} + T_{3412} - T_{3421}=0,
}
it suffices to check that the following identities hold,
\eqnn\Esses
$$\eqalignno{
S^{(2)}_{1234}+S^{(2)}_{2134} &= R^{(1)}_{1234}\cr
S^{(2)}_{1234}+S^{(2)}_{3124} + S^{(2)}_{2314} &= R^{(2)}_{1234} & \Esses \cr
S^{(2)}_{1234} - S^{(2)}_{1243} + S^{(2)}_{3412} - S^{(2)}_{3421} &= R^{(3)}_{1234}.\cr
}$$

Following this same procedure for $L_{21314151}$ is straightforward but somewhat tedious, therefore
the calculations leading to the explicit superfield expression for the building block $T_{12345}$ 
will be deferred to the Appendix~\AppendixTs.

As will be explained in subsection \secBRSTbyparts, the explicit superfield expressions
for $T_{ij}$, $T_{ijk}$, $T_{ijkl}$ and $T_{ijklm}$ allows one to obtain the expansions of
any superstring or field-theory amplitudes up to $N=11$ legs in terms of momenta and polarization \PSS.

%************************************************
\newsec{Supersymmetric Berends--Giele recursions}
%************************************************

In subsection \secgraphs\ we have given a superfield representation in terms of $T_{i_1\ldots i_p}$ for each 
color ordered diagram made of cubic vertices with $p$ on-shell external leg and one off-shell leg. 
In this section, we combine these diagrams to $p+1$ point field theory amplitudes with one off-shell leg. 
These objects were firstly considered in \BG\ in order to derive recursion relations for gluon scattering 
at tree-level and were referred to as ``currents''. The pure spinor supersymmetric analogue of the $p$-point Berends--Giele 
current $J_p$ will be referred to as $M_{12\ldots p}$.

These $M_{12\ldots p}$ allow for a compact representation of the ten-dimensional $N$--point SYM amplitude ${\cal A}_{YM}(1,\ldots,N)$ 
which nicely exhibits its factorization channels. The recursive nature of the Berends--Giele currents is 
inherited by the amplitudes and leads to the recursive method to compute higher-point SYM amplitudes presented
below.

%*****************************************************************
\subsec{Construction of Berends--Giele currents $M_{123\ldots  p}$}
\subseclab\secBGJs
%*****************************************************************
  
The Berends--Giele currents $M_{123\ldots p}$ are written in terms of building blocks $T_{123\ldots p}$ and Mandelstam invariants
$\{s_{12},s_{123},\ldots ,s_{123\ldots  p}\}$ and follow from the recursive definition
\eqnn\recBG
$$
\eqalignno{
E_{123\ldots p} &\equiv \sum_{j=1}^{p-1} M_{12\ldots j} M_{j+1\ldots p} \cr
QM_{123\ldots p} &\equiv E_{123\ldots p}, & \recBG \cr
}
$$
where $M_1 = V_1$. Although the defining system \recBG\ is purely algebraic, it can be
conveniently solved with the recourse of a diagrammatic interpretation for $M_{123\ldots p}$. To see this,
the current $M_{123\ldots  p}$ is first associated to 
the sum of $(2p-2)! / (p! (p-1)!)$ cubic graphs which enter the $p+1$ amplitude where the leg $p+1$ is put off-shell.
Using the dictionary of subsection \secgraphs\ each one of these cubic graphs can be written in terms of
building blocks $T_{123\ldots p}$ and their relative signs are fixed by requiring the system \recBG\ to be
satisfied. For example, using the cubic graphs for the three- and four-point amplitudes
the currents $M_{12}$ and $M_{123}$ are interpreted as
\smallskip
\centerline{\hskip\parindent
\tikzpicture [scale=0.6,line width=0.30mm]
\scope[xshift=-6.0cm]
\draw (-1.9,0) node{$M_{12} \ \ = $};
\draw (0,0) -- (-1,1) node[above]{$2$};
\draw (0,0) -- (-1,-1) node[below]{$1$};
\draw (0,0) -- (0.8,0);
\draw (0.5,-0.35) node{$s_{12}$};
\draw (1.2, 0) node{$\ldots $};
\endscope
\scope[xshift=1.4cm]
\draw (-2,0) node{$M_{123} \ \ =\ \ $};
\draw (0,0) -- (-1,1) node[above]{$2$};
\draw (0,0) -- (-1,-1) node[below]{$1$};
\draw (0,0) -- (1.8,0);
\draw (0.5,0.35) node{$s_{12}$};
\draw (1,0) -- (1,1) node[above]{$3$};
\draw (1.6,-0.35) node{$s_{123}$};
\draw (2.65, 0.1) node{$\ldots  \  \ + $};
\endscope
\scope[xshift=0.65cm]
\draw (5.5,0) -- (4.5,1) node[above]{$3$};
\draw (5.5,0) -- (4.5,-1) node[below]{$2$};
\draw (5.5,0) -- (7.3,0);
\draw (6,0.35) node{$s_{23}$};
\draw (6.5,0) -- (6.5,-1) node[below]{$1$};
\draw (7.1,-0.35) node{$s_{123}$};
\draw (7.7, 0) node{$\ldots  $};
\endscope
\endtikzpicture
}
%%%%%%%%%%%%%%%%%%

\noindent
while $M_{1234}$ is associated to the graphs of the color-ordered five-point amplitude shown in \figMs.
Under the dictionary of subsection \secgraphs\ these graphs correspond to the following 
expressions in terms of building blocks
\eqnn\Mgraphs
$$\displaylines{
\hfill M_{12} = {T_{12}\over s_{12}}\, , \qquad M_{123} = {1 \over s_{123}}  \( {T_{123} \over s_{12} } + { T_{321} \over s_{23}} \),\hfill \Mgraphs \cr
M_{1234} = {1 \over s_{1234}} \( {T_{1234}\over s_{12}s_{123} } 
+ {T_{3214}\over s_{23}s_{123} }+ {T_{3421} \over s_{34}s_{234} } + {T_{3241} \over s_{23}s_{234} }
+ {2T_{12[34]}\over s_{12}s_{34} }\), \cr
}$$

\topinsert
%%%%%%%%%%%%%%%%%%
%%%%%%%%%%%%%%%%%%
% M_1234
\centerline{\hskip\parindent
\tikzpicture [scale=1.3,line width=0.30mm]
\draw (-1.5,2) node {$M_{1234} \quad = $};
\scope[yshift=2cm]
\draw (0,0) -- (-0.5,0.5) node[above]{$2$};
\draw (0,0) -- (-0.5,-0.5) node[below]{$1$};
\draw (0,0) -- (1.3,0);
\draw (0.25,-0.2) node{$s_{12}$};
\draw (0.5,0) -- (0.5,0.5) node[above]{$3$};
\draw (0.75,-0.2) node{$s_{123}$};
\draw (1,0) -- (1,0.5) node[above]{$4$};
\draw (1.40,-0.2) node{$s_{1234}$};
\draw (1.5, 0) node{$\ldots  $};
\endscope
\scope[xshift=2.8cm]
\draw (0,2) -- (-0.5,2.5) node[above]{$3$};
\draw (0,2) -- (-0.5,1.5) node[below]{$2$};
\draw (-0.75,2) node{$+$};
\draw (0,2) -- (1.3,2);
\draw (0.25,1.8) node{$s_{23}$};
\draw (0.5,2) -- (0.5,1.5) node[below]{$1$};
\draw (0.70,2.2) node{$s_{123}$};
\draw (1,2) -- (1,2.5) node[above]{$4$};
\draw (1.35,1.8) node{$s_{1234}$};
\draw (1.5,2) node{$\ldots  $};
\endscope
\scope[xshift=1.4cm]
\draw (4.2, 2) -- (3.7, 2.5) node[above]{$4$};
\draw (4.2, 2) -- (3.7, 1.5) node[below]{$3$};
\draw (3.45,2) node{$+$};
\draw (4.2, 2) -- (5.5, 2);
\draw (4.45, 1.8) node{$s_{34}$};
\draw (4.7, 2) -- (4.7, 1.5) node[below]{$2$};
\draw (4.95, 2.2) node{$s_{234}$};
\draw (5.2, 2) -- (5.2, 1.5) node[below]{$1$};
\draw (5.55, 1.8) node{$s_{1234}$};
\draw (5.7, 2) node{$\ldots  $};
\endscope
\scope[xshift=-4.2cm]
\draw (4.2, 0) -- (3.7, 0.5) node[above]{$3$};
\draw (4.2, 0) -- (3.7, -0.5) node[below]{$2$};
\draw (3.45,0) node{$+$};
\draw (4.2, 0) -- (5.5, 0);
\draw (4.45,-0.2) node{$s_{23}$};
\draw (4.7, 0) -- (4.7, 0.5) node[above]{$4$};
\draw (5.05, 0.2) node{$s_{234}$};
\draw (5.2, 0) -- (5.2, -0.5) node[below]{$1$};
\draw (5.60, -0.2) node{$s_{1234}$};
\draw (5.7, 0) node{$\ldots  $};
\endscope
\scope[xshift=1.8cm,yshift=1.5cm]
  \draw (1, -1.5) -- (4.0, -1.5);
  \draw (1, -1.5) -- (0.5, -1.0) node[above]{$2$};
  \draw (1, -1.5) -- (0.5, -2.0) node[below]{$1$};
  \draw (0.25,-1.5) node{$+$};
  \draw (4.0, -1.5) -- (4.5, -1.0) node[above]{$3$};
  \draw (4.0, -1.5) -- (4.5, -2.0) node[below]{$4$};
  %off-shell leg
  \draw (2.5, -1.5) -- (2.5, -1.75);
  \draw (2.5, -1.85) node{$\vdots$};
  \draw (1.75, -1.3) node{$s_{12}$};
  \draw (3.25, -1.3) node{$s_{34}$};
  \draw (2.9, -1.7) node{$s_{1234}$};
\endscope
%%%%%
\draw (2.4,-1.5) node{$\displaystyle = \ \ {1 \over s_{1234}} \Big( {T_{1234}\over s_{12}s_{123} } 
+ {T_{3214}\over s_{23}s_{123} }+ {T_{3421} \over s_{34}s_{234} } + {T_{3241} \over s_{23}s_{234} } 
+ {2T_{12[34]}\over s_{12}s_{34} }\Big)$};
\endtikzpicture
}
\tikzcaption\figMs{Diagrammatic construction of the Berends--Giele current $M_{1234}$ in terms of the cubic graphs
of the five-point amplitude with one leg off-shell.}
\endinsert
%%%%%
\noindent
where their signs can be fixed by requiring that they form a solution of \recBG. To see this one uses the BRST variations \powerQTs\
to obtain
\eqnn\QM
$$
\eqalignno{
Q M_{12} &= V_1 V_2 = M_1 M_2 , \cr 
Q M_{123} &= {V_1 T_{23}\over s_{23}} + {T_{12}V_3\over s_{12}} = M_1 M_{23} + M_{12} M_3, \cr
QM_{1234} &= {V_1 \over s_{234}} \Big( {T_{234} \over s_{23}} + {T_{432} \over s_{34}} \Big) 
+ {T_{12} T_{34} \over s_{12} s_{34}} + \Big( {T_{123} \over s_{12}} + {T_{321} \over s_{23}} \Big) {V_4 \over s_{123}} \cr
&= M_1 M_{234} + M_{12}M_{34} + M_{123}M_4 & \QM \cr
}
$$
and therefore the expressions for $M_{12}$, $M_{123}$ and $M_{1234}$ given above
form a solution of the system \recBG\ up to this order. Using this method it is straightforward to
obtain higher-point currents, and the explicit expressions of currents up to $M_{1234567}$ will be given in the Appendix~\AppendixMs.

Therefore by using the diagrammatic interpretation of $M_{123\ldots  p}$ in terms of the $p+1$ amplitude with one leg off-shell
one is able to efficiently construct any higher-order current in terms of building blocks.
However, in the later section \Mfromstring\ we will derive a formula for $M_{123\ldots  p}$ in terms of the field-theory 
limit $\a'\to 0$ of hypergeometric integrals occurring in a $p+2$ point string theory amplitude. This allows for a direct computation 
of $M_{12\ldots p}$, therefore bypassing the need to draw the cubic diagrams of the $(p+1)$-point SYM amplitude to find their
corresponding building blocks.

Note that \recBG\ can be written as
\eqn \QMp{
Q M_{12\ldots p} = \sum_{j=1}^{p-1} M_{12\ldots j} M_{j+1\ldots p}
}
and therefore one can interpret the action of $Q$ as
cutting $M_{12\ldots p}$ in each way compatible with the color ordering, see \figQM.
Furthermore, equation \QMp\ is the supersymmetric pure spinor analogue of the recursive construction of 
the Berends--Giele gluon currents in \BG, whose schematic form is
\eqn \BGJ{
J_n \sim {1 \over s_{12\ldots n}} \( \sum_{m=1}^{n-1}  J_m , J_{n-m} 
+ \sum_{m=1}^{n-2} \sum_{k=m+1}^{n-1} J_m J_{k-m} J_{n-k} \).
}
The cubic term in the lower order currents represents the four gluon vertex in the QCD action.
It does not enter into our supersymmetric version \QMp\ which encompasses diagrams 
with cubic vertices only. After multiplying the external propagator $1/s_{12\ldots n}$ to the 
left hand side of \BGJ\ one could symbolically reproduce \QMp\ by identifying $s_{12\ldots n} \equiv Q$.

\topinsert
\centerline{\hskip\parindent
\tikzpicture [scale=1,line width=0.30mm]
\draw (-1.5,0) node{ $Q$};
\draw (0,0) -- (1.5,0) node[right]{$\displaystyle \ldots \ \ \ = \ \ \sum_{j=1}^{p-1}$};
\draw (0,0) -- (0,-1.25) node[below]{$2$};
\draw (0,0) -- (0,1.25) node[above]{$p-1$};
\draw (0,0) -- (0.88,-0.88) node[below]{$1$}; % for 1
\draw (0,0) -- (-0.88,-0.88) node[below]{$3$}; % for 3
\draw (0,0) -- (0.88,0.88) node[above]{$p$}; % for p
\draw[dashed] (0,1) arc (90:225:1cm);
\draw[fill=white] (0,0) circle (0.5cm);
\draw (0,0) node {$M^p$};
\scope[xshift=0.2cm]
\draw (5,0) -- (5,-1.25) node[below]{$1$};
\draw (5,0) -- (4.3,-0.88) node[below]{$2$};
\draw (5,0) -- (5,1.25) node[above]{$j$};
\draw[dashed] (5,1) arc (90:225:1cm);
\draw[fill=white] (5,0) circle (0.5cm);
\draw (5,0) node {$M^j$};
\draw (5.5,0) -- (6.2,0) node[right]{$\displaystyle \ldots$};
\draw (7.25,0) node {$\times$};
\draw (7.25,1) node {$E_{12\ldots p}$};
\draw (9.5,0) -- (8.3,0) node[left]{$\displaystyle \ldots$};
\draw (9.5,0) -- (9.5,-1.25) node[below]{$p$};
\draw (9.5,0) -- (9.5,1.25) node[above]{$j+1$};
\draw (9.5,0) -- (10.2,0.88) node[above]{$j+2$};
\draw[dashed] (9.5,-1) arc (-90:45:1cm);
\draw[fill=white] (9.5,0) circle (0.5cm);
\draw (9.5,0) node {$M^{p-j}$};
\endscope
\endtikzpicture
}
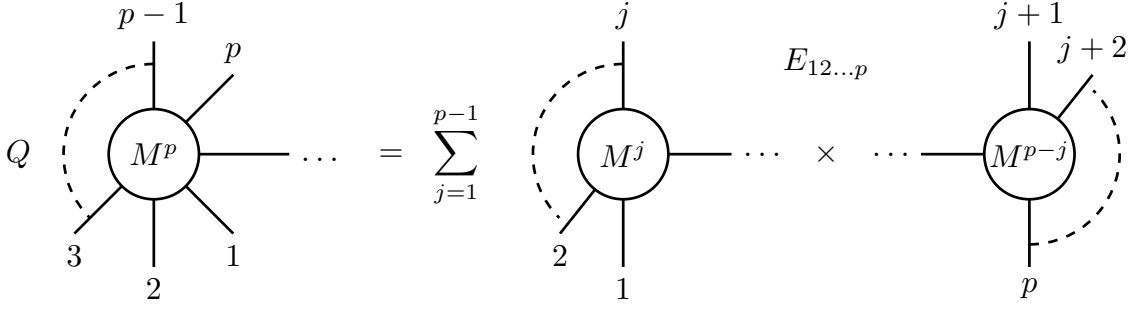
\tikzcaption\figQM{Decomposition of $M_{12\ldots p}$ into its factorization channels under the action of
the pure spinor BRST charge; $QM_{12\ldots  p} =\sum_{j=1}^{p-1} M_{12\ldots j} M_{j+1\ldots p}$.}
\endinsert
\noindent

%*****************************************************************
\subsec{Symmetry properties of $M_{12\ldots p}$}
%*****************************************************************

As a further motivation for identifying $M_{12\ldots p}$ with supersymmetric Berends--Giele currents, 
we discuss their symmetry properties in this subsection. First of all, $M_{12}$ trivially 
satisfies $M_{12}+M_{21} = 0$ because the building block $T_{ij}$ is antisymmetric. Similar 
identities hold for $M_{123}$
\eqn\Mtc{
M_{123} + M_{231} + M_{312} = 0, \qquad M_{123} - M_{321} = 0,
}
as one can easily check by plugging in the expression for $M_{ijk}$ given in \Mgraphs.

At higher $n \geq 4$, this generalizes as follows:
\eqnn\MJs
$$\eqalignno{
M_{12\ldots n} &= (-1)^{n-1}M_{n\ldots 21}\cr
\sum_{\sigma \in {\rm cyclic}} \!\!\!\! M_{\sigma(1,2,\ldots ,n)} &=0. & \MJs \cr
}$$
The proof of these identities is most conveniently carried out on the level of the 
corresponding $E_{12\ldots n} = QM_{12\ldots n} = \sum_{p=1}^{n-1} M_{12\ldots p} M_{p+1\ldots n}$. 
Since all the BRST closed components of the $M_{12\ldots n}$ have been removed by construction of 
its $T_{12\ldots n}$ constituents, the BRST variation $E_{12\ldots n}$ contains all information 
on the symmetry properties of its $M_{12\ldots n}$ ancestor. The reflection identity can be 
easily checked by induction, and the vanishing cyclic sum follows from
\eqn \Mcyc{ \eqalign{
\sum_{\sigma \in {\rm cyclic}}&E_{\sigma(1,2,\ldots ,n)}=\sum_{\sigma \in {\rm cyclic}} 
\sum_{p=1}^{n-1}M_{\sigma(1,2,\ldots ,p)} M_{\sigma(p+1,\ldots ,n)} \cr
&=\sum_{\sigma \in {\rm cyclic}} \sum_{p=1}^{n-1} {1\over 2} (M_{\sigma(1,2,\ldots ,p)} M_{\sigma(p+1,\ldots ,n)} 
+ M_{\sigma(p+1,\ldots ,n)}M_{\sigma(1,2,\ldots ,p)}) = 0
}}
where the last step exploits the overall cyclic sum to shift all labels of the second term by~$p$
and that the $M_{12\ldots p}$ anticommute.

The properties \MJs\ are shared by the $n$-gluon Berends--Giele currents $J_n$ of \BG\ and can
be naturally explained by the construction of currents $M_{123\ldots  n}$ as $(n+1)$-point 
amplitudes with one off-shell leg. Inspired by this explanation, we explicitly checked using the expressions
of Appendix~\AppendixMs\ that $M_{12\ldots  n}$ for $n\le 7$ also satisfy an additional relation -- 
obtained by removing the $(n+1)$-th leg from the $(n+1)$-point Kleiss-Kuijf identity \KK:
\eqn\KKMs{
M_{\{ \beta\} ,1,\{\alpha \}} =\, (-1)^{n_{\beta}}\hskip-0.7cm 
\sum_{\sigma\, \in\, {\rm OP} ( \{\alpha \}, \{\beta^T \})} \hskip-0.4cm M_{1,\{\sigma \}}.
}
The summation range ${\rm OP} ( \{\alpha \}, \{\beta^T \})$ denotes the set of all the permutations 
of $\{\alpha \} \bigcup  \{\beta^T \}$ that maintain the order of the individual elements of both 
sets $\{\alpha \}$ and $\{\beta^T \}$. The notation $\{\beta^T \}$ represents the set $\{\beta \}$ 
with reversed ordering of its $n_\beta$ elements. The Kleiss-Kuijf identity is well known to reduce 
the number of independent color ordered $n+1$ point amplitudes down to $(n-1)!\,$. 

The specialization of \KKMs\ to sets $\{ \beta \}$ with one element only, say $\{ \beta \}  = \{ n \}$, 
reproduces the second property of \MJs. However, this so-called dual Ward identity or photon decoupling 
identity by itself is not sufficient for a reduction to $(n-1)!$ independent $M_{i_1i_2\ldots i_n}$ at 
$n \geq 6$ \KK. Since there are only $(n-1)!$ independent $T_{i_1i_2\ldots i_n}$ which constitute the 
$M_{i_1i_2\ldots i_n}$, also the latter must have a basis of no more than $(n-1)!$ elements. This 
suggests the Kleiss-Kuijf identity \KKMs\ to hold beyond our checks for $n \leq 7$.

The reflection- and Kleiss-Kuijf identity for the $M_{12\ldots n}$ are inherited from their 
associated $n+1$ point amplitudes with one leg off-shell. The off-shellness of one leg is no 
obstruction for the aforementioned identities to hold because they do not involve any kinematic 
factors. However, the field theory version of the monodromy relations \refs{\monodVanhove,\monodStie}
\eqn \Ftmono{ \eqalign{
s_{12}\, {\cal A}_{YM}(2,1,3,\ldots,N) &+( s_{12}+s_{13}) {\cal A}_{YM}(2,3,1,\ldots,N) + \cdots \cr
&+ (s_{12} + \cdots + s_{1,N-1})  {\cal A}_{YM}(2,3,\ldots,N-1,1,N)=0
}}
rely on having on-shell momenta, so the $M_{12\ldots n}$ do not obey any analogue of \Ftmono\ 
and cannot be reduced to $(n-2)!$ independent permutations.

%*************************************************
\subsec{The $N$--point field-theory tree amplitude}
%*************************************************

The expressions found for $Q M_{12\ldots p} = E_{12\ldots p}$ might look familiar from lower order field theory amplitudes such as
\eqn \Ft{ \eqalign {
{\cal A}_{YM}(1,2,3) &= \langle V_1V_{2} V_3 \rangle = \langle E_{12} V_3 \rangle \cr
{\cal A}_{YM}(1,2,3,4) &= \langle \( {V_1 T_{23}\over s_{23}} + {T_{12}V_3\over s_{12}}  \) V_4 \rangle = \langle E_{123} V_4 \rangle
}}
From $QV=0$, one might naively expect that the three-point amplitude would be BRST-exact, ${\cal A}(1,2,3) = \langle Q(T_{12}V_3/s_{12})\rangle$, 
and thus doomed to vanish. However, all Mandelstam invariants $s_{ij}$ vanish in the momentum phase space of three 
massless particles -- therefore writing $V_1 V_2 = Q(T_{12}/s_{12})$ is not allowed and BRST triviality of the amplitude is avoided.

More generally, the prefactor $M_{12\ldots p} \sim 1/ s_{12\ldots p}$ in the $p$ point current is incompatible with putting 
the external state with $k_{p+1} = - \sum_{i=1}^p k_i$ on-shell $k_{p+1}^2 = 0$. Since $N$ particle kinematics forbids 
the existence of $M_{12\ldots N-1}$, the corresponding $E_{12\ldots N-1}$ is not BRST exact. Hence, the following
expression for the $N$--point field theory amplitude is in the 
cohomology of the pure spinor BRST charge\foot{It is interesting to note that the cohomology formula \SYMN\ together with the property 
of $E_{n(n-1)\ldots  1} = (-1)^{n-1}E_{12\ldots  n}$ (which follows from \MJs) imply that if
the amplitude satisfies the reflection property of
${\cal A}(n,n-1,\ldots , 1) = (-1)^n {\cal A}(1,2,\ldots ,n)$ then it is also {\it cyclically\/} symmetric,
${\cal A}(2,3,\ldots ,n,1) = {\cal A}(1,2,\ldots ,n)$.}\MSSTFT
\eqn\SYMN{
{\cal A}_{YM}(1,2,\ldots,N) = \langle E_{12\ldots N-1} V_N \rangle =  \sum_{j=1}^{N-2} \langle  M_{12\ldots j} M_{j+1\ldots N-1} V_N \rangle. 
}
The diagrammatic representation of $\sum_{j=1}^{p-1} M_{12\ldots j} M_{j+1\ldots p}$ in \figQM\ can be uplifted to 
the on-shell $N=p+1$ point amplitude ${\cal A}_{YM}(1,\ldots ,N)$ where an additional cubic vertex connects the $N^{\rm th}$ 
leg with the two currents of rank $j$ and $N-1-j$, respectively.

\topinsert
\centerline{
\tikzpicture [scale=1.0,line width=0.30mm]
\draw (0,0) node{$\displaystyle {\cal A}_{YM}(1,2,\ldots,N) \ \ = \ \ \sum_{j=1}^{N-2}$ };
\draw (4,0) -- (4,1.8) node[above]{$j$};
\draw (4,0) -- (4,-1.8) node[below]{$1$};
\draw (4,0) -- (2.75,-1.25) node[below]{$2$};
\draw[dashed] (4,1.4) arc (90:225:1.4cm);
\draw[fill=white] (4,0) circle (0.7cm);
\draw (4,0) node{$M^j$ };
\draw (4.7,0) -- (6.3,0);
\draw (5.5,0) -- (5.5,-1.4)node[below]{$V_N$};
\draw (7,0) -- (7,1.8) node[above]{$j+1$};
\draw (7,0) -- (8.25,1.25) node[above]{$j+2$};
\draw (7,0) -- (7,-1.8) node[below]{$N-1$};
\draw[dashed] (7,-1.4) arc (-90:45:1.4cm);
\draw[fill=white] (7,0) circle (0.7cm);
\draw (7,0) node{$M^{N-j-1}$ };
\endtikzpicture
}
\tikzcaption\figYsix{Berends--Giele decomposition of ${\cal A}_{YM}$ according to the pure spinor cohomology 
formula \SYMN.}
\endinsert

\noindent
The $N$--point formula \SYMN\ is analogous to the Berends--Giele formula for the color ordered $N$ gluon
amplitude of \BG. The latter is written as a product of a rank $N-1$ current $J_{N-1}$ and another $J_1$ for the $N^{\rm th}$ leg, multiplied by 
the Mandelstam factor $s_{12\ldots N-1}$ to cancel the divergent propagator;
${\cal A}_{YM} = s_{12\ldots  N-1}J(1,\ldots, N-1)\, J(N)$.
In our case, the somewhat artificial 
object $s_{12\ldots N-1} J_{N-1}$ is replaced by $E_{12\ldots N-1}$, which could be written as $QM_{12\ldots N-1}$ 
in a larger momentum phase space. Therefore this parallel also suggests the schematic identification $s_{12\ldots  N-1} \rightarrow Q$
mentioned after \BGJ.

%*****************************************************
\subsec{BRST integration by parts and cyclic symmetry}
\subseclab\secBRSTbyparts
%*****************************************************

The strength of our presentation \SYMN\ of the $N$--point field theory amplitude is the manifestation
of its factorization properties. But singling out a particular leg $V_N$ obscures the cyclic symmetry 
required for color stripped amplitudes. The essential tool to restore manifest cyclicity is BRST 
integration by parts,
\eqn \ME{
\langle M_{i_1 \ldots i_p} E_{j_1 \ldots j_q} \rangle = \langle E_{i_1 \ldots i_p} M_{j_1 \ldots j_q} \rangle.
}
Using the definition of $E_{123 \ldots p}$ in \recBG\ it follows that,
\eqn\Ecyc{
E_{12 \ldots N-1} V_N = E_{23 \ldots N} V_1 + \sum_{j=2}^{N-2}\bigl( M_{12 \ldots j} E_{j+1 \ldots N} - E_{12 \ldots j}M_{j+1 \ldots N}\bigr),
}
therefore $\langle E_{12 \ldots N-1}V_N \rangle = \langle E_{23 \ldots N}V_1 \rangle$ and the $N$--point subamplitude \SYMN\
is cyclically invariant. However, to obtain a formula with {\it manifest\/} cyclic symmetry one needs to explicitly use BRST integration
by parts in \SYMN. And as a byproduct of that, the maximum rank of the Berends--Giele currents needed for the $N$--point amplitude is reduced.
To see this, note that the term containing the 
maximum rank of $M_{i_1\ldots i_p}$ appearing in the $N$--point amplitude \SYMN\ is $p=N-2$ and has
the form $\langle M_{i_1 \ldots i_{N-2}} V_{i_{N-1}} V_N \rangle$, therefore the use of \Ecyc\ leads to
\eqn \MVV{
\langle M_{i_1 \ldots i_{N-2}} V_{i_{N-1}} V_N \rangle =  \langle M_{i_1 \ldots i_{N-2}} QM_{i_{N-1} N} \rangle 
=  \langle E_{i_1 \ldots i_{N-2}} M_{i_{N-1} N} \rangle,
}
so the BRST integration reduced the maximum rank to $p=N-3$ (because $E_{12 \ldots (N-2)}$ contains at most $M_{12 \ldots N-3}$).
It turns out that the cohomology formula \SYMN\ allows enough BRST integration by parts
as to reduce the maximum rank of the currents to $p=[N/2]$,
leading to manifestly cyclic-symmetric amplitudes
\eqn \cyc{ \eqalign{
{\cal A}_{YM}(1,2,\ldots,5) &= \langle M_{12} V_3 M_{45} \rangle \ + \ {\rm cyclic}(12345) \cr
{\cal A}_{YM}(1,2,\ldots,6) &= {1 \over 3}\langle M_{12} M_{34} M_{56} \rangle + {1 \over 2} \langle M_{123} E_{456} \rangle  \ + \ {\rm cyclic}(123456) \cr
{\cal A}_{YM}(1,2,\ldots,7) &= \langle M_{123}  M_{45} M_{67} \rangle + \langle V_1 M_{234} M_{567} \rangle \ + \ {\rm cyclic}(1234567) \cr
{\cal A}_{YM}(1,2,\ldots,8) &= \langle M_{123} M_{456} M_{78} \rangle + {1 \over 2} \langle M_{1234} E_{5678} \rangle  \ + \ {\rm cyclic}(12345678) 
}}
The fractional prefactors ${1 \over 2}$ or ${1\over 3}$ compensate for the fact that cyclic orbits for 
particularly symmetric superfield kinematics are shorter than the number $N$ of legs. At $N=6$, for instance,
$M_{12}M_{34} M_{56}$ has just one distinct cyclic image $M_{23}M_{45} M_{61}$, hence the full ${\rm cyclic}(123456)$
overcounts the occurring diagrams by a factor of three.

%**************************************************
\subsec{Factorization in cyclically symmetric form}
%**************************************************

In this subsection, we introduce a cyclically symmetric presentation of SYM amplitudes where their factorization 
into two Berends--Giele currents becomes even more obvious. 

One can check by evaluating the BRST variations that the amplitudes in \cyc\ can be equivalently written as
\eqnn\cycfactor 
$$\eqalignno{
{\cal A}_{YM}(1,2,\ldots,4) &= {1 \over 2} \langle M_{12} Q M_{34} \rangle + {\rm cyclic}(1234) \cr 
{\cal A}_{YM}(1,2,\ldots,5) &= {1 \over 4} \Big( \langle M_{12} Q M_{345} \rangle + \langle M_{123} Q M_{45} \rangle \Big)+{\rm cyclic}(12345) \cr 
{\cal A}_{YM}(1,2,\ldots,6) &= {1 \over 6} \Big( \langle M_{12} Q M_{3456}  \rangle +  \langle M_{123} Q M_{456} \rangle  +  \langle M_{1234} Q M_{56} \rangle \Big) + {\rm cyclic}(123456) \cr  
{\cal A}_{YM}(1,2,\ldots,7) &= {1 \over 8} \Big( \langle M_{12} Q M_{34567}  \rangle +  \langle M_{123} Q M_{4567} \rangle + \langle M_{1234} Q M_{567}  \rangle \cr
& \ \ \ \ \ \  +  \langle M_{12345} Q M_{67} \rangle \Big) \ + \ {\rm cyclic}(1234567) & \cycfactor \cr  
{\cal A}_{YM}(1,2,\ldots,8) &= {1 \over 10} \Big( \langle M_{12} Q M_{345678}  \rangle + \langle M_{123} Q M_{45678}  \rangle + \langle M_{1234} Q M_{5678} \rangle \cr 
& \ \ \ \ \ \ + \langle M_{12345} Q M_{678}  \rangle + \langle M_{123456} Q M_{78} \rangle \Big) + {\rm cyclic}(12345678)  
}$$
Note that some terms in the formul{\ae} are naively overcounted by a factor of 2 because
the cyclic orbits of $\langle M_{12\ldots j} Q M_{j+1\ldots N} \rangle$ and $\langle M_{12\ldots N-j} Q M_{N-j+1\ldots N} \rangle$
are the same. The purpose of including both of them is to obtain a uniform overall coefficient in \cycfactor\
and to simplify the transition to the general $N$--point formula,
\eqn \Ncycfactor{
{\cal A}_{YM}(1,2,\ldots,N) = {1 \over 2(N-3)} \sum_{j=2}^{N-2} \langle M_{12\ldots j} Q M_{j+1\ldots N} \rangle \ + \ {\rm cyclic}(1\ldots N) 
}
whose graphical representation is shown in \figfactorN.
We have explicitly checked up to $N=10$ points that the formula \Ncycfactor\ exactly reproduces the 
expression ${\cal A}_{YM} = \langle E_{12\ldots N-1} V_N \rangle$ of \MSSTFT, including prefactors.

\topinsert
\centerline{
\tikzpicture [line width=0.30mm]
\draw (0.9,0) node{$\displaystyle {\cal A}_{YM}(1,2,\ldots,N) = {1 \over 2(N-3)} \sum_{j=2}^{N-2}$};
\draw (10.1,0) node{$\displaystyle +\ \ {\rm cyclic}(1\ldots N)$};
\draw (5,0) -- (5,-1.25) node[below]{$1$};
\draw (5,0) -- (4.3,-0.88) node[below]{$2$};
\draw (5,0) -- (5,1.25) node[above]{$j$};
\draw[dashed] (5,1) arc (90:225:1cm);
\draw[fill=white] (5,0) circle (0.5);
\draw (5.5,0) -- (6,0) ;
\draw (6.25,0) node{$Q$};
\draw (6.5,0) -- (7,0) ;
\draw (5,0) node{$M^j$};
\scope[xshift=0.5cm]
\draw (7,0) -- (7,-1.25) node[below]{$N$};
\draw (7,0) -- (7,1.25) node[above]{$j+1$};
\draw (7,0) -- (7.7,0.88) node[above]{$j+2$};
\draw[fill=white] (7,0) circle (0.5);
\draw[dashed] (7,-1) arc (-90:45:1cm);
\draw (7,0) node{$M^{N-j}$};
\endscope
\endtikzpicture
}
\tikzcaption\figfactorN{Cyclic factorization of the $N$--point field-theory 
amplitude ${\cal A}_{YM}(1,2,\ldots,N)$ into different Berends--Giele partitions
according to equation \Ncycfactor.}
\endinsert

The factorization formula \Ncycfactor\ can also be interpreted as coming from the factorization channels of
two amplitudes with one leg $x$ off-shell each with the form $\langle E_{12{\dots}j}V_{x}\rangle$ and $\langle V_{x}E_{j+1\ldots N}\rangle$ 
that are connected by a pure spinor propagator
which effectively replaces\foot{CM thanks Nathan Berkovits for suggesting back in 2006 how one could view an operation like
$V_x \; V_x \rightarrow {1\over Q}$ as possibly being related to a massless propagator in pure spinor superspace.} 
$V_x \; V_x \rightarrow {1\over Q}$, resulting in 
$$\eqalignno{
{\cal A}_{YM}(1,2,\dots, N) &= {1\over 2(N-3)}\sum_{j=2}^{N-2} \langle E_{12{\dots}j \mathstrut} {1\over Q} \; E_{\mathstrut j+1\ldots N}\rangle
+ {\rm cyclic}(1 \ldots N)\cr
 &= {1\over 2(N-3)}\sum_{j=2}^{N-2} \langle M_{12{\dots}j \mathstrut} Q M_{\mathstrut j+1 \ldots N}\rangle + {\rm cyclic}(1\ldots N)
}$$
which reproduces the formula \Ncycfactor.

%****************************************************************
\newsec{The superstring tree amplitude in pure spinor superspace}
%****************************************************************

This section derives our central result \nptstring\ for the superstring $N$ point tree amplitude of the massless gauge 
multiplet. The BRST building blocks $T_{12\ldots p}$ and their combinations to form supersymmetric Berends--Giele 
currents $M_{12\ldots p}$ turn out to be very efficient bookkeeping devices to handle the kinematic structures 
of a superstring amplitude in a universal way, i.e. for any number $N$ of external legs.

According to the tree level prescription \treepresc, the task in computing superstring amplitudes in the canonical color ordering $(1,2,\ldots,N)$ is 
to evaluate the CFT correlator
\eqn \treecorr{ \prod_{j=2}^{N-2} \int dz_j
\langle V^1(0)V^{(N-1)}(1)V^N(\infty)  U^2(z_2) 
U^3(z_3) \ldots   U^{(N-2)}(z_{(N-2)})\rangle
}
integrated over $z_1=0 \leq z_2 \leq \cdots \leq z_{N-2} \leq z_{N-1} = 1$. We will first 
of all give a representation of \treecorr\ in terms of $(N-2)!$ different $z_i$ polynomials in the integrand. 
Then, performing manipulations on the level of both the building blocks and the associated integrals reduces 
the number of distinct integrals to $(N-3)!$ each of which multiplies a full-fledged SYM amplitude \SYMN\ 
in a color ordering specific to the integral.

%**************************
\subsec{The CFT correlator}
%**************************
 
Since the conformal $h=1$ primaries $[\p \theta^\alpha,\Pi^m,d_\alpha , N^{mn}]$ within the integrated vertex 
do not have zero modes at tree level, the correlator \treecorr\ can be computed by summing all their OPE 
singularities. Generically, this gives rise to a set of $(N-2)!$ worldsheet functions where all the $z_{ij}$ 
appear as single poles, and additionally to a set of double pole integrands $\sim z_{ij}^{-2}$. It has been 
observed in \MSST\ that the role of the double pole integrals is to correct the numerators of the $(N-2)!$ 
single pole integrals such that any OPE residue $L_{jiki\ldots li}$ is transformed to the associated BRST 
building block $T_{ijk\ldots l}$. This is the consequence of a subtle interplay between the integrals along the lines of subsection 5.4, in 
particular the tachyon poles due to double pole integrals are cancelled by the superfield kinematics in a 
highly nontrivial way.

A bit of care is needed to reduce the single pole residue among two integrated vertices $U^i(z_i) U^j(z_j)$ 
to the more basic $L_{jiki\ldots li}$ superfields which appear when $U^j U^k \ldots U^l$ successively approach 
an unintegrated vertex $V^i$. The required manipulations are based on the independence of correlation functions
on the order of integrating out the $h=1$ fields~\FivePt. The relations up to the six point case can be found in \refs{\FivePt,\MSST},
\eqn \UU{ \eqalign{
V^1(z_1) U^2(z_2) U^3(z_3) &\sim { L_{3121} - L_{2131} \over z_{23} z_{31} } =: { 2L_{[31,21]}  \over z_{23} z_{31} } \cr
 V^1(z_1) U^2(z_2) U^3(z_3) U^4(z_4)  &\sim { L_{413121} - L_{412131} + L_{213141} - L_{312141} \over z_{23} z_{34}  z_{41} }
 =: { 4L_{[41,[31,21]]} \over z_{23} z_{34}  z_{41}},
}}
we are picking out one particular residue here when the arguments approach each other in the order 
$z_2 \rightarrow z_3 \rightarrow z_1$ and $z_2 \rightarrow z_3 \rightarrow z_4 \rightarrow z_1$, respectively. 
This order is reflected in the specific $z_{ij}$ in the denominator. 

Higher order analogues of \UU\ involve nested antisymmetrizations:
\eqn \UUU{ \eqalign{
V^1(z_1) U^2(z_2) U^3(z_3) U^4(z_4) U^5(z_5) &\sim { 8 L_{[51,[41,[31,21]]]}\over z_{23} z_{34} z_{45} z_{51} } \cr
V^1(z_1) U^2(z_2) U^3(z_3) \cdots  U^p(z_p) &\sim {2^{p-2} L_{[p1,[(p-1)1,[ \ldots ,[41,[31,21]]\ldots]]]}\over z_{23} z_{34} \cdots z_{p-1,p} z_{p1} } 
}}
When all the single pole numerators are reduced to $L_{jiki\ldots li}$ and the double pole corrections are 
absorbed into $L_{jiki\ldots li} \mapsto T_{ijk\ldots l}$, the integrated correlator \treecorr\ assumes a manifestly 
symmetric form in the labels $2,3,\ldots, N-2$ of the $U^j$ vertices
\eqnn\corr 
$$\eqalignno{
&\prod_{j=2}^{N-2} \int dz_j \langle V^1(0)V^{(N-1)}(1)V^N(\infty)\,  U^2(z_2)\,
U^3(z_3) \cdots  U^{(N-2)}(z_{(N-2)})\rangle   & \corr \cr
& \quad = \prod_{j=2}^{N-2} \int dz_j \, \prod_{i<j} |z_{ij}|^{-s_{ij}} 
 \sum_{p=1}^{N-2} \bigg \langle { T_{12\ldots p}\, T_{N-1,N-2,\ldots, p+1} V_N  \over (z_{12}  z_{23} \cdots z_{p-1,p} )
( z_{N-1,N-2}  z_{N-2,N-3} \cdots z_{p+2,p+1})} \cr
& \hskip5cm  \ + \ {\cal P}(2,3,\ldots,N-2) \, \bigg \rangle ,
}$$
where ${\cal P}(2,3,\ldots,N-2)$ denotes a symmetric sum over the $(N-3)!$ permutations of the labels $(2,3,\ldots,N-2)$.
The $z_{ij}$ polynomials associated with a specific BRST building block $T_{ij_1j_2\ldots j_p}$ follow an intriguing 
pattern (where the first label $i$ belongs to an unintegrated vertex $V^1$ or $V^{N-1}$ and the remaining ones to the 
integrated vertices $j_k \in \{ 2,3,\ldots, N-2 \}$):
\eqn \TZ{
T_{ij_1j_2\ldots j_p} \ \ \leftrightarrow \ \ {1 \over z_{ij_1} z_{j_1 j_2} z_{j_2 j_3} \cdots z_{j_{p-1},j_p}}
}
%The $N$--point superstring amplitude ${\cal A}(1,\ldots, N)$ is obtained by integrating over $z_1=0 \leq z_2 \leq \cdots \leq z_{N-2} \leq z_{N-1} = 1$. 
Since there are $(N-3)!$ permutations 
of the $(2,3,\ldots,N-2)$ labels and the $p$ sum collects $(N-2)$ distinct permutation orbits, \corr\ yields
an expression for the $N$--point superstring amplitude \treepresc\
in terms of $(N-2)!$ kinematic numerators and hypergeometric integrals,
\eqnn\stringWithTs
$$\eqalignno{
{\cal A}_N \ \equiv \ &{\cal A}(1,2,\ldots,N) \ = \ \prod_{j=2}^{N-2} \int dz_j \prod_{i<j} |z_{ij}|^{-s_{ij}}  \cr
\sum_{p=1}^{N-2} & \bigg \langle {  T_{12 \ldots p} T_{N-1,N-2, \ldots ,p+1} V_N  
\over (z_{12}  z_{23} \cdots z_{p-1,p} )( z_{N-1,N-2} \cdots z_{p+2,p+1})}  \ + \; {\cal P}(2,\ldots,N-2) \bigg \rangle. & \stringWithTs \cr
}$$
\vskip-0.2cm
\noindent
The cases $N=5$ and $N=6$ of \stringWithTs\ reproduce the formul{\ae} obtained in
\refs{\MSST, \FTAmps} and \stringWithTs\ has also been used in \KITP\ to obtain (via the field-theory limit $\a' \to 0$)
local expressions for all $(2N-5)!!$ kinematic numerators entering the field-theory $N$--point amplitude
which manifestly satisfy all BCJ numerator identities \BCJ.

%*****************************************************************
\subsec{A closed formula for $M_{12\ldots p}$ from the superstring}
\subseclab\Mfromstring
%*****************************************************************

In this subsection we will show that the result \stringWithTs\ for the $N$--point superstring amplitude
allows to extract a closed formula for the Berends--Giele current $M_{12\ldots p}$. The $p$~sum in \stringWithTs\
partitions the legs $2,3,\ldots, N-2$ into two groups -- one of them gets connected to leg $1$,
the other to leg $N-1$. The same structure is also present in the cohomology formula \SYMN\ for the 
field-theory amplitude;
${\cal A}_{YM}^N = \sum_{p=1}^{N-2} \langle M_{12\ldots p} M_{p+1\ldots N-1} V_N \rangle$.

Since the kinematic factors within individual terms of the $p$ sum are linearly independent, we can directly compare the $p=N-2$ term on both sides of
${\cal A}_N \buildrel{\a'\to 0}\over\longrightarrow {\cal A}_{YM}^N$ --  with the string- and field-theory
amplitudes given respectively by \stringWithTs\ and \SYMN:
$$
{\cal A}_N = (2\alpha')^{N-3} \prod_{j=2}^{N-2} \int dz_j \prod_{i<j} |z_{ij}|^{-2\alpha' s_{ij}} 
\bigg \langle {T_{12\ldots N-2} V_{N-1} V_N \over z_{12} z_{23} \cdots z_{N-3,N-2} } \ 
+ \ {\cal P}(2,\ldots ,N-2) + \cdots \bigg \rangle 
$$
\eqn \Ftlim{
\buildrel{\alpha' \rightarrow 0}\over\longrightarrow \; \langle M_{12\ldots N-2} V_{N-1} V_N \rangle \ + \ \cdots
}
This yields a closed-formula solution for the rank $p=N-2$ current $M_{12\ldots  p}$,
\eqn \Mstring{
M_{12...p} = \lim_{\alpha' \rightarrow 0} (2\alpha')^{p-1} \prod_{j=2}^{p} \int_{z_{j-1}}^{1} dz_j 
\prod_{i<j}^{p+1} |z_{ij}|^{-2 \alpha' s_{ij}} \( {T_{12...p} \over z_{12} z_{23} \cdots z_{p-1,p} }
+ {\cal P}(2,3,\ldots, p) \),
}
where $z_1 = 0$ and $z_{p+1}=1$ as customary for a $(p+2)-$point amplitude. For example, using the momentum 
expansion of the five-point superstring integrals \Medinas\ and the BRST symmetry $T_{123} + T_{231} + T_{312} = 0$ 
of \BRSTexact\ the following $M_{123}$ is generated
\eqnn\MthreeS
$$\eqalignno{
M_{123} & = \lim_{\alpha' \rightarrow 0} (2\alpha')^{2} \int_{0}^{1}dz_2 \int_{z_2}^{1} dz_3
\prod_{i<j}^4 |z_{ij}|^{-2 \alpha' s_{ij}} \( {T_{123} \over z_{12} z_{23} } + {T_{132} \over z_{13} z_{32} }\) \cr
& = {T_{123}\over s_{12}s_{123}} + {T_{123}\over s_{23}s_{123}} - {T_{132}\over s_{23}s_{123}} 
= {T_{123}\over s_{12}s_{123}} + {T_{321}\over s_{23}s_{123}}, & \MthreeS \cr
}$$
which is easily shown to satisfy $QM_{123} = E_{123}$. Similarly, we checked that the formula \Mstring\ correctly generates 
solutions of \QMp\ up to and including $M_{1234567}$.

%*****************************************************************
\subsec{Trading $T_{12\ldots p}$ for $M_{12\ldots p}$}
%*****************************************************************

As will be shown in the next subsections, in order to simplify even further the expression \stringWithTs\ 
of the superstring $N$--point amplitude
it will be convenient to trade the BRST building blocks 
$T_{12\ldots p}$ for the Berends--Giele currents $M_{12\ldots p}$. 

This exchange will be possible because of the particular pattern \TZ\ of $z_{ij}$ dependence 
along with the $T_{12\ldots p}$.
The lowest order example of $T \leftrightarrow M$ conversion is a triviality 
${ T_{12} \over z_{12}} = { s_{12} \over z_{12} } M_{12}$, but already the simplest generalization is a result 
of partial fraction relations and the symmetry properties of $T_{ijk}$:
\eqn \TMc{
{ T_{123} \over z_{12} z_{23}} \ + \ {\cal P}(2,3) = { s_{12} \over z_{12} }  
\left( { s_{13} \over z_{13}} + { s_{23} \over z_{23}} \right)  M_{123} \ + \ {\cal P}(2,3). }
Similar identities have been checked at $p=4$ and $p=5$ level:
\eqnn\TMde
$$
\eqalignno{
{ T_{1234} \over z_{12}  z_{23} z_{34}} \ + \ {\cal P}(2,3,4)  &= { s_{12} \over z_{12} } 
\left( { s_{13} \over z_{13}} + { s_{23} \over z_{23}} \right)  \left( { s_{14} \over z_{14}} 
+ { s_{24}\over z_{24}} + { s_{34} \over z_{34}} \right) M_{1234} + \ {\cal P}(2,3,4) \cr %\cr & \hskip4.5cm + \ {\cal P}(2,3,4) \cr
{ T_{12345} \over z_{12} z_{23} z_{34} z_{45}} \ + \ {\cal P}(2,3,4,5) &= { s_{12} \over z_{12} } 
\left(  { s_{13} \over z_{13}} + { s_{23} \over z_{23}} \right) 
\left( { s_{14} \over z_{14}} + { s_{24} \over z_{24}} + { s_{34} \over z_{34}} \right) \cr
 &\hskip-1.5cm \times \ \left( { s_{15} \over z_{15}} + { s_{25} \over z_{25}} 
 +  { s_{35} \over z_{35}} + { s_{45} \over z_{45}} \right)  M_{12345} \ + \ {\cal P}(2,3,4,5). & \TMde
}
$$
These identities heavily rely on the interplay of different terms in the permutation sum and on 
the symmetry properties \TrankN\ of the BRST building blocks which leave no more than $(p-1)!$ independent 
permutations of $T_{i_1 \ldots i_p}$ at level $p$. 

The natural $n$ point generalization of \TMc\ and \TMde\ reads as follows:
\eqnn\TMN 
$$\eqalignno{
{ T_{12\ldots p} \over z_{12} z_{23} \cdots z_{p-1,p}} \ + \ {\cal P}(2,\ldots,p) &= \prod_{k=2}^p 
\sum_{m=1}^{k-1} { s_{mk} \over z_{mk} } M_{12\ldots p} \ + \ {\cal P}(2,\ldots,p) \cr
{ T_{N-1,N-2,\ldots, p+1} \over z_{N-1,N-2} \cdots z_{p+2,p+1} } \ + \ {\cal P}(2,\ldots,p) &= \!
\prod_{k=p+1}^{N-2} \sum_{n=k+1}^{N-1} {s_{nk} \over z_{nk}} M_{N-1,N-2,\ldots ,p+1} & \TMN \cr
\ + \ {\cal P}(2,\ldots,p) = \prod_{k=p+1}^{N-2} \sum_{n=k+1}^{N-1} &{s_{kn} \over z_{kn}} M_{p+1,p+2,\ldots,N-1} \ + \ {\cal P}(2,\ldots,p),
}$$
where in the last line the rank $N-1-p$ Berends--Giele current with leg $N-1$ involved was 
reflected via \MJs; $M_{N-1, \ldots ,p+1} = (-1)^{N-p-2} M_{p+1,\ldots ,N-1}$.

%*****************************************************************
\subsec{Worldsheet integration by parts}
\subseclab\sectotderiv
%*****************************************************************

This subsection focuses on the integrals rather than the kinematic factors in the superstring amplitude. 
The chain of ${s_{mk} \over z_{mk}}$ sums which appears as a result of \TMN\ when all the $T_{12\ldots p}$ 
are converted to $M_{12\ldots p}$ is particularly suitable to perform integration by parts with respect 
to $z_j$ variables. Further details on the structure and manipulations of the integrals can be found in \PARTTWO.

The key idea is the vanishing of boundary terms in the worldsheet integrals:
\eqn \tot{
\int d z_j \cdots \int d z_{N-2}\; {\partial \over \partial z_k} { \prod_{i<j} |z_{ij}|^{-s_{ij}} \over z_{i_1 j_1} \cdots z_{i_{N-4} j_{N-4}} } = 0.
}
This identity provides relations between the integrals in an $N$--point superstring amplitude with $N-3$ 
powers of $z_{ij}$ in the denominator. They become particularly easy if the differentiation variable 
$z_k$ does not appear in the denominator (i.e. if $k \notin \{ i_l,j_l \}$) because ${\partial \over \partial z_k}$ 
only hits the $\prod_{m \neq k} |z_{mk}|^{-s_{mk}}$ factor in that case:
\eqn \total{
\int d z_2 \cdots \int d z_{N-2} { \prod_{i<j} |z_{ij}|^{-s_{ij}} \over z_{i_1 j_1} \cdots z_{i_{N-4} j_{N-4}} } 
\sum_{m=1 \atop{ m \neq k}}^{N-1} {s_{mk} \over z_{mk}} = 0.
}
This can be directly applied to the integrands on the right hand side of \TMc, \TMde\ and \TMN, namely:
\eqnn\byparts 
$$\eqalignno{
&\prod_{j=2}^{3} \int dz_j \prod_{i<j} |z_{ij}|^{-s_{ij}} { s_{12} \over z_{12} }  
\left( { s_{13} \over z_{13}} + { s_{23} \over z_{23}} \right)  = 
\prod_{j=2}^{3} \int dz_j \prod_{i<j} |z_{ij}|^{-s_{ij}} { s_{12} \over z_{12} }  { s_{34} \over z_{34}}   \cr
&\prod_{j=2}^{4} \int dz_j \prod_{i<j} |z_{ij}|^{-s_{ij}} { s_{12} \over z_{12} }  
\left( { s_{13} \over z_{13}} + { s_{23} \over z_{23}} \right) \left( { s_{14} \over z_{14}} + { s_{24}\over z_{24}} + { s_{34} \over z_{34}} \right)  \cr
& \ \ \ \ \ \ = \prod_{j=2}^{4} \int dz_j \prod_{i<j} |z_{ij}|^{-s_{ij}} { s_{12} \over z_{12} }  { s_{45} \over z_{45}}  
\left\{ \eqalign {& \left( { s_{13} \over z_{13}} + { s_{23} \over z_{23}} \right) \cr 
&\left( { s_{34} \over z_{34}} + { s_{35} \over z_{35}} \right)  } \right. & \byparts \cr
&\prod_{j=2}^{5} \int dz_j \prod_{i<j} |z_{ij}|^{-s_{ij}} { s_{12} \over z_{12} }  
\left( { s_{13} \over z_{13}} + { s_{23} \over z_{23}} \right) \left( { s_{14} \over z_{14}} + { s_{24}\over z_{24}} + { s_{34} \over z_{34}} \right) 
\left( { s_{15} \over z_{15}} + { s_{25} \over z_{25}} +  { s_{35} \over z_{35}} + { s_{45} \over z_{45}} \right) \cr
& \ \ \ \ \ \ = \prod_{j=2}^{5} \int dz_j \prod_{i<j} |z_{ij}|^{-s_{ij}} { s_{12} \over z_{12} }  { s_{56} \over z_{56}} 
\left( { s_{13} \over z_{13}} + { s_{23} \over z_{23}} \right) \left( { s_{45} \over z_{45}} + { s_{46} \over z_{46}} \right) 
}$$
In the general $N$ point case, it is most economic to leave the first $[N/2]-1$ factors of $\sum_{m=1}^{k-1} {s_{mk} \over z_{mk}}$ 
as they are, and to integrate the remaining $[(N-3)/2]$ such factors by parts:
\eqnn\bypartsN 
$$\eqalignno{
&\prod_{j=2}^{N-2} \int dz_j \prod_{i<j} |z_{ij}|^{-s_{ij}} { s_{12} \over z_{12} }  
\left( { s_{13} \over z_{13}} + { s_{23} \over z_{23}} \right) \cdots \left( { s_{1,N-2} \over z_{1,N-2}} + \cdots + { s_{N-1,N-2} \over z_{N-1,N-2}} \right) \cr
& \ \ = \prod_{j=2}^{N-2} \int dz_j \prod_{i<j} |z_{ij}|^{-s_{ij}} { s_{12} \over z_{12} } \left( { s_{13} \over z_{13}} + { s_{23} \over z_{23}} \right) \cdots 
\left( { s_{1,[N/2]} \over z_{1,[N/2]}} + \cdots + { s_{[N/2]-1,[N/2]} \over z_{[N/2]-1,[N/2]}} \right) \cr
& \ \ \ \ \ \ \times \left( { s_{[N/2]+1,[N/2]+2} \over z_{[N/2]+1,[N/2]+2}} + \cdots + { s_{[N/2]+1,N-1} \over z_{[N/2]+1,N-1}} \right) \cdots 
\left( { s_{N-3,N-2} \over z_{N-3,N-2}} + { s_{N-3,N-1} \over z_{N-3,N-1}} \right) {s_{N-2,N-1} \over z_{N-2,N-1} } \cr
& \ \ = \prod_{j=2}^{N-2} \int dz_j \prod_{i<j} |z_{ij}|^{-s_{ij}} \left( \prod_{k=2}^{[N/2]}  \sum_{m=1}^{k-1}  { s_{mk} \over z_{mk}} \right) 
\left( \prod_{k=[N/2]+1}^{N-2} \sum_{n=k+1}^{N-1}  { s_{kn} \over z_{kn} } \right) & \bypartsN
}$$
In contrast to the $T_{12\ldots p} \rightarrow M_{12\ldots p}$ reshuffling identities from the previous subsection, 
\byparts\ and \bypartsN\ are valid before summing over permutations of $(2,3,\ldots,N-2)$.

%*********************************************************
\subsec{The complete $N$--point superstring disk amplitude}
%*********************************************************

This subsection completes the derivation of the striking result \nptstring\ for the superstring $N$--point amplitude 
${\cal A}_N \equiv {\cal A}(1,2,\ldots,N)$ by combining the results of the previous subsections. Let us first look 
at the four-, five- and six-point examples to get a better feeling of the mechanisms at work.

After using $T_{ij} = s_{ij}M_{ij}$, the total derivative relation ${s_{23} \over z_{23}} \mapsto {s_{12} \over z_{12}}$ 
as well as $E_{123} = M_{12}V_3 + V_1 M_{23}$, the four-point open string disk amplitude is easily seen to be 
\eqnn \fourSYM
$$ \eqalignno{
{\cal A}_4 &= \int dz_2 \prod_{i<j} |z_{ij}|^{-s_{ij}} \left \langle { T_{12} V_3 V_4 \over z_{12}} 
+ { V_{1} T_{32} V_4 \over z_{32}} \right \rangle \cr &= \int dz_2 \prod_{i<j} |z_{ij}|^{-s_{ij}} 
\left \langle { s_{12} \over z_{12}}  M_{12} V_3 V_4 + { s_{23} \over z_{23}}  V_{1} M_{23} V_4\right \rangle \cr
&= \int dz_2 \prod_{i<j} |z_{ij}|^{-s_{ij}}\; {s_{12} \over z_{12}} \, \bigl\langle (M_{12} V_3 + V_1 M_{23} ) V_4 \bigr\rangle  \cr &
= \int dz_2 \prod_{i<j} |z_{ij}|^{-s_{ij}}\; {s_{12}\over z_{21}} {\cal A}_{\rm YM}(1,2,3,4).
} $$
Similarly, the five-point superstring amplitude \stringWithTs\ contains six different integrands and kinematic terms.
After applying \TMc, the $T_{ij}$ and $T_{ijk}$ conspire to give $M_{ij}$ and $M_{ijk}$ with modified 
integrals, then we use integration by parts according to \byparts\ on the way to the third equality of \fivePT. Remarkably, many 
of the initially $(N-2)!=6$ distinct integrals now coincide: The three kinematic terms $M_{123} V_4 V_5$, $M_{12} M_{34} V_5$
and $V_1 M_{234} V_5$ are multiplied by the same integral after partial integration, the same is true for the 
$(2 \leftrightarrow 3)$ permutation. That is why we can identify color ordered field-theory amplitudes \SYMN\ in the last line:
\eqnn\fivePT 
$$\eqalignno{
{\cal A}_5 &= \int dz_2 dz_3 \prod_{i<j} |z_{ij}|^{-s_{ij}} \bigg 
\langle { T_{123} V_4 V_5 \over z_{12} z_{23}} +  {  T_{12} T_{43} V_5   \over z_{12} z_{43}} + {V_1 T_{432} V_5 \over z_{43} z_{32} } 
+ (2 \leftrightarrow 3) \bigg \rangle \cr
&= \int dz_2 dz_3 \prod_{i<j} |z_{ij}|^{-s_{ij}} \left\langle {s_{12} \over z_{12}} 
\left( {s_{13} \over z_{13} } + {s_{23} \over z_{23}} \right) M_{123} V_4 V_5 +  {s_{12} s_{34} \over z_{12} z_{34}} M_{12} M_{34} V_5 \right. \cr
& \hskip3.5cm \left. + {s_{43} \over z_{43}} \left( {s_{42} \over z_{42}} + {s_{32} \over z_{32} } \right) 
V_1 M_{432} V_5 + (2 \leftrightarrow 3) \right \rangle & \fivePT\cr
&= \int dz_2 dz_3 \prod_{i<j} |z_{ij}|^{-s_{ij}} \left\{ { s_{12} s_{34} \over z_{12} z_{34}} 
\langle M_{123} V_4 V_5 + M_{12} M_{34} V_5 + V_1 M_{234} V_5 \rangle +  (2 \leftrightarrow 3) \right\} \cr
&= \int dz_2 dz_3 \prod_{i<j} |z_{ij}|^{-s_{ij}} 
\left\{ { s_{12} s_{34} \over z_{12} z_{34}} {\cal A}_{\rm YM}(1,2,3,4,5) + { s_{13} s_{24} \over z_{13} z_{24}} {\cal A}_{YM}(1,3,2,4,5)  \right\}
}$$
Simplifying the six-point amplitudes ${\cal A}_6$ follows similar steps. In this case, \TMde\ takes care of the conversion 
of $T_{ijkl}$ into $M_{ijkl}$, then integration by parts makes the four integrals within a given $(2,3,4)$ permutation coincide:
\eqnn\sixPT 
$$\eqalignno{
{\cal A}_6 &= \prod_{j=2}^4 \int dz_j \prod_{i<j} |z_{ij}|^{-s_{ij}} \bigg 
\langle {  T_{1234} V_5 V_6  \over z_{12} z_{23} z_{34}} +  { T_{123} T_{54} V_6   \over z_{12} z_{23}z_{54}} 
+ {T_{12} T_{543} V_6 \over z_{12} z_{54} z_{43} }  %\cr
%& \hskip6.5cm   
+  {V_1 T_{5432} V_6 \over z_{54} z_{43} z_{32} }  +  {\cal P}(2,3,4) \bigg \rangle \cr
&= \prod_{j=2}^4 \int dz_j \prod_{i<j} |z_{ij}|^{-s_{ij}} \left\langle {s_{12} \over z_{12}} 
\left( {s_{13} \over z_{13} } + {s_{23} \over z_{23}} \right) \left( {s_{14} \over z_{14} } + {s_{24} \over z_{24}} 
+ {s_{34} \over z_{34} } \right) M_{1234} V_5 V_6 \right. \cr
& \hskip2cm \left. + {s_{12} \over z_{12}} \left( {s_{13} \over z_{13} } + {s_{23} \over z_{23}} \right) {s_{45} \over z_{45}} M_{123} M_{45} V_6 
+{s_{12} \over z_{12}} {s_{45} \over z_{45}} \left( {s_{34} \over z_{34}} + {s_{35} \over z_{35} } \right)M_{12} M_{543} V_6  \right. \cr
& \hskip2cm \left. + {s_{45} \over z_{45}} \left( {s_{34} \over z_{34}} + {s_{35} \over z_{35} } \right) 
\left( {s_{52} \over z_{52}} + {s_{42} \over z_{42} }+ {s_{32} \over z_{32} } \right) V_1 M_{5432} V_6 \ +
\ {\cal P}(2,3,4) \right \rangle \cr
&= \prod_{j=2}^4 \int dz_j \prod_{i<j} |z_{ij}|^{-s_{ij}} \biggl\{ { s_{12} s_{45} \over z_{12} z_{45}}  
\left( {s_{13} \over z_{13} } + {s_{23} \over z_{23}} \right) \langle M_{1234} V_5 V_6 + M_{123} M_{45} V_6 \cr
& \hskip 4.5cm  + M_{12} M_{345} V_6 + V_1 M_{2345} V_6 \rangle \ +
\ {\cal P}(2,3,4) \biggr\} & \sixPT \cr
&= \prod_{j=2}^4 \int dz_j \prod_{i<j} |z_{ij}|^{-s_{ij}} \biggl\{ { s_{12} s_{45} \over z_{12} z_{45}}  
\left( {s_{13} \over z_{13} } + {s_{23} \over z_{23}} \right) {\cal A}_{\rm YM}(1,2,3,4,5,6)  \ +
\ {\cal P}(2,3,4)  \biggr\}
}$$
The identities \TMde\ and \byparts\ are sufficient to also reduce the superstring seven-point amplitude ${\cal A}_{7}$ to 
its field-theory constituents:
\eqnn\sevenPT 
$$\eqalignno{
{\cal A}_7 &= \prod_{j=2}^5 \int dz_j \prod_{i<j} |z_{ij}|^{-s_{ij}} \bigg 
\langle {  T_{12345} V_6 V_7  \over z_{12} z_{23} z_{34} z_{45}} +  { T_{1234} T_{65} V_7   \over z_{12} z_{23}z_{34} z_{65}} 
+ {T_{123} T_{654} V_7 \over z_{12} z_{23} z_{65} z_{54} }  \cr
& \hskip3.5cm  + {T_{12} T_{6543} V_7 \over z_{12}  z_{65} z_{54} z_{43} }  +  {V_1 T_{65432} V_7 \over z_{65} z_{54} z_{43} z_{32} } \ 
+ \ {\cal P}(2,3,4,5) \bigg \rangle \cr
&= \prod_{j=2}^5 \int dz_j \prod_{i<j} |z_{ij}|^{-s_{ij}} \biggl\{ { s_{12} s_{56} \over z_{12} z_{56}}  
\left( {s_{13} \over z_{13} } + {s_{23} \over z_{23}} \right) \left( {s_{45} \over z_{45} } 
+ {s_{46} \over z_{46}} \right) {\cal A}_{\rm YM}(1,2,3,4,5,6,7) \cr
& \hskip3.5cm  + \ {\cal P}(2,3,4,5)  \biggr\}. &\sevenPT \cr
}$$
The $N$--point generalization is based on introducing currents $M_{i_1i_2\ldots i_p}$ via \TMN\ followed by integration 
by parts using \bypartsN. The latter makes the integral independent on $p$ such that the $z_{ij}$ can be placed outside 
the $p$~sum and SYM amplitudes emerge from the kinematics.
\eqnn\NPTstring 
$$\eqalignno{
{\cal A}_N &= \prod_{j=2}^{N-2} \int dz_j \prod_{i<j} |z_{ij}|^{-s_{ij}} \bigg \langle
\sum_{p=1}^{N-2} {  T_{12 \ldots p}\; T_{N-1,N-2, \ldots, p+1} V_N   \over (z_{12}  z_{23} \cdots z_{p-1,p} )( z_{N-1,N-2}   \cdots z_{p+2,p+1})}  \cr
& \hskip4.5cm + \ {\cal P}(2,3,\ldots,N-2) \bigg \rangle \cr
&= \prod_{j=2}^{N-2} \int dz_j \prod_{i<j} |z_{ij}|^{-s_{ij}} \biggl\langle
\sum_{p=1}^{N-2} \left( \prod_{k=2}^{p}  \sum_{m=1}^{k-1} {s_{mk} \over z_{mk}} M_{12\ldots p} \right) \cr
&\hskip 2.5cm \times \left( \prod_{k=p+1}^{N-2} \sum_{n=k+1}^{N-1} {s_{kn} \over z_{kn}} M_{ p+1,\ldots , N-2,N-1} \right) V_N \ 
+ \ {\cal P}(2,3,\ldots,N-2) \biggr\rangle\cr
&= \prod_{j=2}^{N-2} \int dz_j \prod_{i<j} |z_{ij}|^{-s_{ij}} \bigg\{ \left( \prod_{k=2}^{[N/2]}  
\sum_{m=1}^{k-1} {s_{mk} \over z_{mk}}  \right) \left( \prod_{k=[N/2]+1}^{N-2} \sum_{n=k+1}^{N-1} {s_{kn} \over z_{kn}}  \right) \cr
&\hskip 2.5cm \times \sum_{p=1}^{N-2} \langle M_{12\ldots p} M_{p+1\ldots N-2,N-1} V_N \rangle \ + \ {\cal P}(2,3,\ldots,N-2) \bigg\} \cr
&= \prod_{j=2}^{N-2} \int dz_j \prod_{i<j} |z_{ij}|^{-s_{ij}} \bigg\{ \left( \prod_{k=2}^{[N/2]}  
\sum_{m=1}^{k-1} {s_{mk} \over z_{mk}}  \right) \left( \prod_{k=[N/2]+1}^{N-2} \sum_{n=k+1}^{N-1} {s_{kn} \over z_{kn}}  \right) \cr
&\hskip 2.5cm \times {\cal A}_{YM}(1,2,3,\ldots,N-1,N) \ + \ {\cal P}(2,3,\ldots,N-2) \bigg\}. & \NPTstring \cr
}$$
Equivalently, by undoing the total derivative relation used in \NPTstring\ the full $N$--point superstring amplitude becomes
\eqn\nptstring{
{\cal A}_N = \int \limits _{z_i < z_{i+1}} \prod_{i<j} |z_{ij}|^{-s_{ij}} \bigg[ \prod_{k=2}^{N-2}  
\sum_{m=1}^{k-1} {s_{mk} \over z_{mk}}\; {\cal A}_{YM}(1,2,\ldots,N)  +  {\cal P}(2,\dots,N-2) \bigg],
}
where the integration region $\int _{z_i < z_{i+1}} \equiv \prod_{j=2}^{N-2} \int^1 _{z_{j-1}} dz_j$ is responsible for dictating which color-ordered
string subamplitude is being computed. Therefore the end result of all these pure spinor superspace manipulations is that
the $N$--point superstring disk amplitude is written in terms of the
explicit sum of $(N-3)!$ basis of field-theory amplitudes multiplied by an equal number of hypergeometric integrals,
as mentioned in the Introduction and further elaborated in \PARTTWO.

\bigskip
\noindent
{\bf Acknowledgements:} We thank Dimitrios Tsimpis for his contributions at earlier stages of this project and for
useful discussions. We also thank Nathan Berkovits for pointing out the general argument about the
BRST-exactness of the sums in subsection~\BBB. Furthermore, we thank the Kavli Institute for Theoretical Physics in
Santa Barbara for hospitality and partial financial support. C.M. also thanks the Werner-Heisenberg-Institut in M\"unchen 
for hospitality and partial financial support during preparation of this work and acknowledges support by
the Deutsch-Israelische Projektkooperation (DIP H52). O.S. is indebted to  UCLA and in particular to Martin Ammon for hospitality during preparation of this work. St.St. is grateful to UCLA, Caltech, and SLAC for hospitality and financial support during completion of this work. This research was supported in part by the National Science Foundation under Grant No. NSF PHY05Ð51164.

%*****************************************************
\appendix{A}{The explicit construction of $T_{12345}$}
\applab\AppendixTs
%*****************************************************

In order to find the appropriate redefinition of $L_{21314151}$ leading to ${\tilde T}_{12345}$
one simply uses the known redefinitions of $[L_{21}, L_{2131}, L_{213141}] \rightarrow [T_{12}, T_{123}, T_{1234}]$
in the right-hand side of \QLs. Even though it is not obvious, all terms from these lower-order redefinitions 
group together into a BRST-exact combination which can be moved to the left-hand side of \QLs. Doing that finally leads
to the definition of ${\tilde T}_{12345}$, given by
\eqnn\TtildeFivedef
$$\eqalignno{
{\tilde T}_{12345} = \; &L_{21314151}\cr
 &- {1\over 4}(s_{13} + s_{23})\big[ D_{12} D_{34} V_{5} (s_{35} + s_{45})
       + D_{12} D_{35} V_{4}  s_{34}
       - D_{12} D_{45} V_{3}  s_{34}
       \big]\cr
  & - {1\over 4}s_{12}\Big[ 
         D_{13} D_{24} V_{5} (s_{25} + s_{45})
       + D_{14} D_{23} V_{5} (s_{25} + s_{35})
       + D_{15} D_{23} V_{4} (s_{24} + s_{34})\cr
&+ s_{24}( D_{13} D_{25} V_{4} - D_{13} D_{45} V_{2} )
+ s_{13}(D_{34} D_{25} V_{1} + D_{35} D_{24} V_{1})\cr
 &  + s_{23}( D_{14} D_{25} V_{3} - D_{14} D_{35} V_{2} + D_{15} D_{24} V_{3} - D_{15} D_{34} V_{2})     
       + s_{14} D_{45} D_{23} V_{1} 
       \Big]\cr
 &  -( s_{15} + s_{25} + s_{35} + s_{45}) S^{(2)}_{1234} V_{5}
 - (s_{14}+s_{24}+s_{34}) \(S^{(1)}_{123} L_{54} + S^{(2)}_{1235} V_{4}\)\cr
 &      - (s_{13}+s_{23})\(S^{(1)}_{124}L_{53} + S^{(1)}_{125} L_{43} - S^{(1)}_{345} L_{21} + S^{(2)}_{1245} V_{3}\) \cr
  & - s_{12}\Big[ 
         S^{(1)}_{134} L_{52} 
       + S^{(1)}_{135} L_{42}
       + S^{(1)}_{145} L_{32}
       + S^{(2)}_{1345} V_{2}
       - (1\leftrightarrow 2)
       \Big]\cr
  &     - \half\Big[ T_{123} D_{45} (s_{14}+s_{24}+s_{34})       
  + (T_{125} D_{34} - T_{345} D_{12} + T_{124} D_{35}) (s_{13} + s_{23})\cr
%\eqn\TtildeF{
 &      + s_{12}(  
         T_{134} D_{25} 
       + T_{135} D_{24} 
       + T_{145} D_{23} 
       - (1\leftrightarrow 2)
       )
       \Big] &\TtildeFivedef
}$$
which, by construction, is guaranteed to satisfy
\eqnn\satis
$$\eqalignno{
Q{\tilde T}_{12345} = & + (s_{15} + s_{25} + s_{35} + s_{45}) T_{1234} V_{5} + (s_{14} + s_{24} + s_{34}) (T_{1235} V_{4} + T_{123}T_{45}) \cr
 & + (s_{13} + s_{23}) (T_{1245} V_{3} + T_{124} T_{35} + T_{125} T_{34} + T_{12} T_{345}) \cr
 & + s_{12} (T_{1345} V_{2} + V_{1} T_{2345} + T_{134} T_{25} + T_{135} T_{24} + T_{145} T_{23} \cr
 & + T_{13} T_{245} + T_{14} T_{235} + T_{15} T_{234}). & \satis \cr
}$$
One can also show that\foot{The tedious algebra was handled using {\tt FORM} \FORM.}
\eqnn\Rfours
$$\eqalignno{
{\tilde T}_{12345} + {\tilde T}_{21345} = &QR^{(1)}_{12345},\cr
{\tilde T}_{12345} + {\tilde T}_{23145} + {\tilde T}_{31245} = &QR^{(2)}_{12345}, & \Rfours \cr
{\tilde T}_{12345} - {\tilde T}_{12435} + {\tilde T}_{34125} - {\tilde T}_{34215} = &QR^{(3)}_{12345},\cr
         {\tilde T}_{12345}
       - {\tilde T}_{12354}
       + {\tilde T}_{45123}
       - {\tilde T}_{45213}
       - {\tilde T}_{45312}
       + {\tilde T}_{45321}
       = &QR^{(4)}_{12345}
}$$
\noindent
where the BRST-exact parts are given by
\eqnn\TfiveExact
$$
\eqalignno{
R^{(1)}_{12345} = &D_{12} (k^{12}\cdot A^3)(k^{123}\cdot A^4) (k^{1234}\cdot A^5)
+ {1\over 6} (s_{13}+s_{23})D_{12}\Big[
D_{45}\( (k^{4}\cdot A^3) - (k^5\cdot A^3)\) \cr
      & +   D_{35}\( (k^{5}\cdot A^4) - (k^3\cdot A^4)\)
       - 2  D_{34}\( (k^3\cdot A^5) + 2 (k^4\cdot A^5)\)
    \Big],\cr
%%%%    
R^{(2)}_{12345} = &D_{12} (k^2\cdot A^3)(k^{123}\cdot A^4)(k^{1234}\cdot A^5)
+ {1\over 6}\Big[ s_{12}D_{13} \(
D_{45}((k^4\cdot A^2)-(k^5\cdot A^2)) \right.\cr
&\left.
       + D_{25}((k^5\cdot A^4) - (k^2\cdot A^4))
       - 2 D_{24}((k^2\cdot A^5) + 2 (k^4\cdot A^5))\)
+{\rm cyclic}(123)       \Big],\cr
%%%%
R^{(3)}_{12345} =  &-  (W^1 \g^{m} W^2) (W^3 \g^{m} W^4) (k^{1234}\cdot A^5) \cr
    &+  \big[ D_{12} (k^3\cdot A^4)(k^2\cdot A^3)(k^{1234}\cdot A^5)
    +  {1 \over 3}(s_{24} - 2 s_{23})  D_{34} D_{12} (k^4\cdot A^5)
      - (3\leftrightarrow 4)
      \big]\cr
   &+  {1 \over 6} (s_{14} + s_{24}) \big[ 
           D_{25} D_{34} ((k^2\cdot A^1) -  (k^5\cdot A^1))
       +   D_{15} D_{34} ((k^5\cdot A^2) -  (k^1\cdot A^2))
       \big]\cr       
   &+  {1 \over 6}(s_{23} + s_{24})\big[
           D_{45} D_{12} ((k^4\cdot A^3) - (k^5\cdot A^3))
       +   D_{35} D_{12} ((k^5\cdot A^4) - (k^3\cdot A^4))\big]\cr
   & +  \Big[ \(D_{13} (k^1\cdot A^2)(k^3\cdot A^4)
    +  D_{24} (k^2\cdot A^1)(k^4\cdot A^3)
    +  D_{34} (k^1\cdot A^2)(k^4\cdot A^1)\)(k^{1234}\cdot A^5)\cr
   & +  {1 \over 3} (s_{24} - 2 s_{14}) D_{34} D_{12}  (k^2\cdot A^5)
    - (1\leftrightarrow 2)
    \Big]\cr
%%%%
R^{(4)}_{12345} = 
 &(W^1 \g^{m} W^2)\big[ (W^4 \g^{n} W^5) \cF^3_{mn} - (W^4 \g^{m} W^5) (k^{12} \cdot A^3) \big]\cr
       & + \Big[  (W^1 \g^{m} W^2) (W^3 \g^{m} W^5) (k^5 \cdot A^4)  +  {1 \over 4} (W^1 \g^{m} W^2) (W^5 \g^{np} \g^{m} W^3) \cF^4_{np}\cr
       & + D_{12} (k^2 \cdot A^3) (k^{23} \cdot A^4) (k^4 \cdot A^5) +  D_{12} (k^1 \cdot A^3) (k^2 \cdot A^4) (k^4 \cdot A^5)\cr 
       &+  {1 \over 6} D_{12} D_{35} (k^3 \cdot A^4) s_{23} 
        +  {5 \over 6} D_{12} D_{35} (k^5 \cdot A^4) s_{23}
        +  {1 \over 3} D_{12} D_{45} (k^4 \cdot A^3) s_{23} \cr
       &+  D_{14} (k^1 \cdot A^2) (k^{12} \cdot A^3) (k^4 \cdot A^5) 
        +  D_{25} (k^2 \cdot A^1) (k^{12} \cdot A^3) (k^5 \cdot A^4)\cr
       & +  D_{34} (k^2 \cdot A^1) (k^3 \cdot A^2) (k^4 \cdot A^5) 
        +  D_{35} (k^3 \cdot A^1) (k^1 \cdot A^2) (k^5 \cdot A^4) - (4\leftrightarrow 5) \Big]
	\cr
       & + \Big[ (W^2 \g^{m} W^3) (W^4 \g^{m} W^5) (k^2 \cdot A^1) + {1 \over 4} (W^4 \g^{m} W^5) (W^1 \g^{np} \g^{m} W^3) \cF^2_{np} \cr
      &+  D_{13} (k^1 \cdot A^2) (k^3 \cdot A^4) (k^4 \cdot A^5) 
        -  D_{13} (k^5 \cdot A^4) (k^1 \cdot A^2) (k^3 \cdot A^5)\cr
       & +  D_{45} (k^2 \cdot A^1) (k^3 \cdot A^2) (k^5 \cdot A^3) 
        +  D_{45} (k^5 \cdot A^1) (k^1 \cdot A^2) (k^{12} \cdot A^3)\cr
       & +  {1 \over 3} D_{12} D_{45} (k^2 \cdot A^3) (-2s_{15} + s_{25} + s_{35})
        +  {1 \over 6} D_{13} D_{45} (k^3 \cdot A^2) (s_{15} + s_{25} + s_{35}) \cr
       & -  {1 \over 6} D_{13} D_{45} (k^1 \cdot A^2) (s_{15} + s_{25} - 5s_{35}) - (1\leftrightarrow 2)\Big] & \TfiveExact \cr
}       
$$

Removing these BRST-exact parts is accomplished by the second redefinition
${\tilde T}_{12345} \longrightarrow T_{12345}$, leading to the rank-five BRST building block
\eqn\Tfivedef{
T_{12345} = {\tilde T}_{12345} - QS_{12345}^{(3)},
}
where the expression for $S^{(3)}_{12345}$ can be written recursively as
\eqnn\StwoFdef
$$\eqalignno{
S^{(3)}_{12345} = &  {4\over 5} S^{(2)}_{12345}
       + {1\over 5} \(
          S^{(2)}_{12354}
       -  S^{(2)}_{45123}
       +  S^{(2)}_{45213}
       +  S^{(2)}_{45312}
       -  S^{(2)}_{45321}
       \)
       + {1\over 5} R^{(4)}_{12345},\cr
S^{(2)}_{12345} = &{3\over 4}S^{(1)}_{12345} + {1\over 4}( S^{(1)}_{12435} - S^{(1)}_{34125} + S^{(1)}_{34215}) 
+ {1\over 4}R^{(3)}_{12345}, & \StwoFdef \cr
S^{(1)}_{12345} = &{1\over 2}R^{(1)}_{12345} + {1\over 3}R^{(2)}_{[12]345}.
}$$
To see that \Tfivedef\ and \StwoFdef\ imply all the BRST-symmetries of $T_{12345}$
\eqnn\Tfiman
$$\eqalignno{
0 & = T_{12345} + T_{21345} \cr
0 & = T_{12345} + T_{31245} +T_{23145} \cr
0 & = T_{12345} - T_{12435} + T_{34125} - T_{34215} \cr
0 & = T_{12345}
    - T_{12354}
    + T_{45123}
    - T_{45213}
    - T_{45312}
    + T_{45321} & \Tfiman \cr
}$$
it suffices to check that the following identities hold,
\vskip-0.4cm
\eqnn\boring
$$\eqalignno{
S^{(3)}_{12345} + S^{(3)}_{21345} = &R^{(1)}_{12345}\cr
S^{(3)}_{12345} + S^{(3)}_{31245} + S^{(3)}_{23145} = &R^{(2)}_{12345} &\boring \cr
S^{(3)}_{12345} - S^{(3)}_{12435} + S^{(3)}_{34125} - S^{(3)}_{34215} = &R^{(3)}_{12345}\cr
  S^{(3)}_{12345} 
- S^{(3)}_{12354} 
+ S^{(3)}_{45123} 
- S^{(3)}_{45213}
- S^{(3)}_{45312}
+ S^{(3)}_{45321}
= &R^{(4)}_{12345}.
}$$
Having the explicit superfield expressions for the building blocks up to $T_{12345}$ allows
all component amplitudes up to $N=11$ to be evaluated.

%****************************************************************************************************************
\appendix{B}{The solutions for $M_{i_1i_2\ldots  i_n}$ in terms of BRST building blocks}
\applab\AppendixMs
%****************************************************************************************************************

From the relation between $M_{123\ldots n}$ and the cubic diagrams of the $(n+1)$-point amplitude 
discussed in subsection~\secBGJs, it follows that the solutions for $M_{123}$, $M_{1234}$, 
$M_{12345}$, $M_{123456}$ and $M_{1234567}$ which satisfy \recBG\
contain 2, 5, 14, 42 and 132 different kinematic pole configurations, which are represented by
the cubic-graph expansion of the tree amplitudes. Their explicit
expressions  can then be read off from the dictionary between those
cubic graphs and the BRST building blocks; as discussed in subsection~\secgraphs. Furthermore, using 
the antisymmetry on the first two labels of $T_{ijk \dots }$, one can always choose an 
ordering such that all terms in$M_{123\ldots n}$ have a positive coefficient, leading to:
\eqnn\Mtwoformula
\eqnn\Mthreeformula
\eqnn\Mfourformula
\eqnn\Mfiveformula
\eqnn\Msixformula
\eqnn\Msevenformula
$$\eqalignno{
M_{12} &= {T_{12}\over s_{12}}\, , & \Mtwoformula \cr
M_{123} &={1\over s_{123}}\({T_{123}\over s_{12} } + {T_{321}\over s_{23} }\), & \Mthreeformula \cr
 M_{1234} &=
{1\over s_{1234}}  \(
  {T_{1234} \over s_{12}s_{123}}
+ {T_{3214} \over s_{23}s_{123}}
+ {T_{3241} \over s_{23}s_{234}}
+ {T_{3421} \over s_{34}s_{234}}
+ {2T_{12[34]} \over s_{12} s_{34}}
\), & \Mfourformula \cr
\noalign{\vskip0.3cm}
M_{12345} &={1\over s_{12345}}\biggl[
        {1\over s_{1234}}\(       
         {T_{12345} \over s_{12} s_{123}}
       + {T_{32145} \over s_{23} s_{123} }
       + {T_{32415} \over s_{23} s_{234} }
       + {T_{34215} \over s_{34} s_{234} }
       + {2 T_{12[34]5} \over s_{12} s_{34} }
       \) \cr
      & \phantom{{1\over s_{12345}}\Big[} + {1\over s_{2345}} \(
         {T_{34251} \over s_{34} s_{234} }
       + {T_{32451} \over s_{23} s_{234} }
       + {T_{34521} \over s_{34} s_{345} }
       + {T_{54321} \over s_{45} s_{345} }
       + {2T_{45[23]1} \over s_{23} s_{45} }
       \)\cr
       &\phantom{{1\over s_{12345}}\Big[} + 
         {2T_{123[45]}\over s_{12} s_{123} s_{45} }
       + {2T_{321[45]}\over s_{23} s_{123} s_{45} }
       + 
         {2T_{453[12]}\over s_{45} s_{345} s_{12}}
       + {2T_{435[12]}\over s_{34} s_{345} s_{12}}\biggr], & \Mfiveformula \cr
}$$       
\smallskip
$$\eqalignno{
%\noalign{\vskip0.3cm}
M_{123456} &={1\over s_{123456}}\Big[
        {
            4T_{12[34][56]}
          \over s_{12} s_{34} s_{56} s_{1234}}
       + { 
           4 T_{34[56][21]}
          \over s_{12} s_{34} s_{56} s_{3456}}
       + {
           4 T_{123[[45]6]}
          \over s_{12} s_{45} s_{123} s_{456}} 
       + {
            4T_{123[4[56]]}
          \over s_{12} s_{56} s_{123} s_{456}}   \cr
       & + {
            4T_{231[[54]6]}
          \over s_{23} s_{45} s_{123} s_{456}}   
       + {
           4 T_{231[4[65]]}
          \over s_{23} s_{56} s_{123} s_{456}}
       + {
            2T_{345[21]6}
          \over s_{12} s_{34} s_{345} s_{12345}}
       + {
            2T_{3456[21]}
          \over s_{12} s_{34} s_{345} s_{3456}} \cr
      & + {
            2T_{12[34]56}
                 \over s_{12} s_{34} s_{1234} s_{12345}} 
       + {
           2 T_{123[45]6}
          \over s_{12} s_{45} s_{123} s_{12345}}
       + {
           2 T_{543[21]6}
          \over s_{12} s_{45} s_{345} s_{12345}}   
       + {
            2T_{5436[21]}
          \over s_{12} s_{45} s_{345} s_{3456}} \cr
      & + {
           2 T_{4563[12]}
          \over s_{12} s_{45} s_{456} s_{3456}}   
       + {
           2 T_{1234[56]}
          \over s_{12} s_{56} s_{123} s_{1234}}
       + {
           2 T_{5643[21]}
          \over s_{12} s_{56} s_{456} s_{3456}}   
       + {
           2 T_{231[54]6}
          \over s_{23} s_{45} s_{123} s_{12345}} \cr
      & + {
            2T_{456[23]1}
          \over s_{23} s_{45} s_{456} s_{23456}}
	+ {
            2T_{34[56]21}
          \over s_{34} s_{56} s_{23456} s_{3456}}
       + {
            2T_{23[54]16}
          \over s_{23} s_{45} s_{12345} s_{2345}}   
       + {
           2 T_{23[54]61}
          \over s_{23} s_{45} s_{2345} s_{23456}} \cr
      & + {
           2 T_{2314[65]}
          \over s_{23} s_{56} s_{123} s_{1234}}   
       + {
           2 T_{2341[65]}
          \over s_{23} s_{56} s_{234} s_{1234}}   
       + {
           2 T_{234[65]1}
          \over s_{23} s_{56} s_{234} s_{23456}}   
       + {
           2 T_{564[32]1}
          \over s_{23} s_{56} s_{456} s_{23456}} \cr
      & + {
            2T_{3421[56]}
          \over s_{34} s_{56} s_{234} s_{1234}}   
       + {
           2 T_{342[56]1}
          \over s_{34} s_{56} s_{234} s_{23456}}   
       + {
           T_{321456}
          \over s_{23} s_{123} s_{1234} s_{12345}}   
       + {
           T_{324156}
          \over s_{23} s_{234} s_{1234} s_{12345}} \cr
      & + {
           T_{324516}
          \over s_{23} s_{234} s_{12345} s_{2345}}   
       + {
           T_{324561}
          \over s_{23} s_{234} s_{2345} s_{23456}}   
       + {
           T_{342156}
          \over s_{34} s_{234} s_{1234} s_{12345}}
       + {
           T_{342516}
          \over s_{34} s_{234} s_{12345} s_{2345}}  \cr
      & + {
           T_{342561}
          \over s_{34} s_{234} s_{2345} s_{23456}}   
       + {
           T_{345216}
          \over s_{34} s_{345} s_{12345} s_{2345}}   
       + {
           T_{345261}
          \over s_{34} s_{345} s_{2345} s_{23456}}   
       + {
           T_{345621}	   
          \over s_{34} s_{345} s_{23456} s_{3456}} \cr
      & + {
           T_{543216}
          \over s_{45} s_{345} s_{12345} s_{2345}}   
       + {
           T_{543261}
          \over s_{45} s_{345} s_{2345} s_{23456}}
       + {
           T_{123456}
          \over s_{12} s_{123} s_{1234} s_{12345}}
       + {
           T_{543621}
          \over s_{45} s_{345} s_{23456} s_{3456}} \cr
      & + {
           T_{546321}
          \over s_{45} s_{456} s_{23456} s_{3456}}   
       + {
           T_{564321}
          \over s_{56} s_{456} s_{23456} s_{3456}}
	  \Big], & \Msixformula \cr
}	  
$$
\smallskip
\noindent
\line{$ s_{1234567} M_{1234567} =$ \hfill}
\vskip0.1cm
\openup6pt
\halign{ & $\displaystyle #$ \hfill & $\displaystyle #$ \hfill & $\displaystyle #$ \hfill & $\displaystyle #$ \hfill \cr
   & \hskip-0.5cm   + \, {
           8T_{12[34][[56]7]}
           \over \;\; s_{12} s_{34} s_{56} s_{567} s_{1234} \;\;}
  & \hskip-0.1cm    +\, {
           8T_{54[67][1[23]]}
           \over \;\; s_{23} s_{45} s_{67} s_{123} s_{4567} \;\;  }
  & \hskip-0.1cm    +\, {
           8T_{12[34][5[67]]}
           \over \;\; s_{12} s_{34} s_{67} s_{567} s_{1234} \;\;  }
  & \hskip-0.1cm    +\, { 
           8T_{45[67][3[12]]}
           \over \;\; s_{12} s_{45} s_{67} s_{123} s_{4567}  }\cr
 & \hskip-0.5cm      +\, {
           4T_{567[34][12]}
           \over \;\; s_{12} s_{34} s_{56} s_{567} s_{34567} \; }
   & \hskip-0.1cm    +\, { 
           4T_{12[34][56]7}          
           \over s_{12} s_{34} s_{56} s_{1234} s_{123456}  }
    & \hskip-0.1cm   +\, { 4T_{43[56][12]7}
           \over s_{12} s_{34} s_{56} s_{123456} s_{3456}   }
    & \hskip-0.1cm   +\, {
           4T_{43[56]7[12]}
           \over s_{12} s_{34} s_{56} s_{3456} s_{34567}   } \cr
 & \hskip-0.5cm      +\, {
           4T_{435[12][67]}
           \over \; s_{12} s_{34} s_{67} s_{345} s_{12345}\;\; }
 & \hskip-0.1cm      +\, {
           4T_{435[67][12]}
           \over \; s_{12} s_{34} s_{67} s_{345} s_{34567}\;\;   }
 & \hskip-0.1cm      +\, {
           4T_{765[34][12]}
           \over \;\; s_{12} s_{34} s_{67} s_{567} s_{34567} \;\,  }
 & \hskip-0.1cm      +\, {
           4T_{12[34]5[67]}
           \over \; s_{12} s_{34} s_{67} s_{1234} s_{12345}   }\cr
 & \hskip-0.5cm      +\, {
           4T_{123[45][67]}          
          \over \;\; s_{12} s_{45} s_{67} s_{123} s_{12345} \;  }
 & \hskip-0.1cm      +\, {
           4T_{453[12][67]}
          \over \;\; s_{12} s_{45} s_{67} s_{345} s_{12345} \; }
 & \hskip-0.1cm      +\, {
           4T_{453[67][12]}         
          \over \;\; s_{12} s_{45} s_{67} s_{345} s_{34567} \;  }
 & \hskip-0.1cm      +\, {
           4T_{45[67]3[12]}         
          \over s_{12} s_{45} s_{67} s_{34567} s_{4567}   } \cr
%$$
%$$
 & \hskip-0.5cm      +\, {
           4T_{123[[45]6]7}
          \over s_{12} s_{45} s_{123} s_{456} s_{123456}   }
 & \hskip-0.1cm      +\, {
          4T_{5467[[12]3]}
          \over \;\; s_{12} s_{45} s_{123} s_{456} s_{4567} \;  }
  & \hskip-0.1cm     +\, {
           4T_{7654[[12]3]}
          \over \;\; s_{12} s_{67} s_{123} s_{567} s_{4567} \;  }
  & \hskip-0.1cm     +\, {
           4T_{1234[5[67]]}
          \over \;\; s_{12} s_{67} s_{123} s_{567} s_{1234}   } \cr
%$$
%$$       
 & \hskip-0.5cm      +\, {
           4T_{321[45][67]}
          \over \;\; s_{23} s_{45} s_{67} s_{123} s_{12345} \;  }
  & \hskip-0.1cm     +\, {
           4T_{32[45]1[67]}
          \over \; s_{23} s_{45} s_{67} s_{12345} s_{2345} \;  }
  & \hskip-0.1cm     +\, {
           4T_{32[45][67]1}
          \over \; s_{23} s_{45} s_{67} s_{2345} s_{234567}   }
  & \hskip-0.1cm     +\, {
           4T_{5647[[12]3]}
          \over \;\;s_{12} s_{56} s_{123} s_{456} s_{4567}   } \cr
%$$
%$$
 & \hskip-0.5cm      +\, {
           4T_{1234[[56]7]}
          \over \;\; s_{12} s_{56} s_{123} s_{567} s_{1234} \;  }
  & \hskip-0.1cm     +\, {
           4T_{5674[[12]3]}
          \over \;\; s_{12} s_{56} s_{123} s_{567} s_{4567} \;  }
  & \hskip-0.1cm     +\, {
           4T_{123[4[56]]7}
          \over \; s_{12} s_{56} s_{123} s_{456} s_{123456}   }
  & \hskip-0.1cm     +\, {
           4T_{5647[1[23]]}
          \over \;\; s_{23} s_{56} s_{123} s_{456} s_{4567}   }\cr	  
%$$
%$$
 & \hskip-0.5cm      +\, {
           4T_{3214[[56]7]}
          \over \;\; s_{23} s_{56} s_{123} s_{567} s_{1234} \;  }
 & \hskip-0.1cm      +\, {
           4T_{5674[1[23]]}
          \over \;\; s_{23} s_{56} s_{123} s_{567} s_{4567} \;  }
 & \hskip-0.1cm      +\, {
           4T_{321[4[56]]7}
          \over \;\; s_{23} s_{56} s_{123} s_{456} s_{123456}  }       
 & \hskip-0.1cm      +\, {
           4T_{3214[5[67]]}
          \over \;\; s_{23} s_{67} s_{123} s_{567} s_{1234}   } \cr
%$$
%$$
 & \hskip-0.5cm      +\, {
           4T_{7654[1[23]]}
          \over \;\; s_{23} s_{67} s_{123} s_{567} s_{4567} \;  }
& \hskip-0.1cm       +\, {
           4T_{3241[5[67]]}
          \over \;\; s_{23} s_{67} s_{234} s_{567} s_{1234} \;  }
 & \hskip-0.1cm      +\, {
           4T_{324[5[67]]1}
          \over \;\; s_{23} s_{67} s_{234} s_{567} s_{234567}   }
 & \hskip-0.1cm      +\, {
           4T_{45[67][23]1}
          \over  s_{23} s_{45} s_{67} s_{234567} s_{4567}   } \cr
%$$
%$$       
 & \hskip-0.5cm      +\, {
           4T_{321[[45]6]7}
          \over s_{23} s_{45} s_{123} s_{456} s_{123456}   }
 & \hskip-0.1cm      +\, {
           4T_{5467[1[23]]}
          \over \;\; s_{23} s_{45} s_{123} s_{456} s_{4567} \;  }
 & \hskip-0.1cm      +\, {
           4T_{3241[[56]7]}
          \over \;\; s_{23} s_{56} s_{234} s_{567} s_{1234} \;\,  }
 & \hskip-0.1cm      +\, {
           4T_{324[[56]7]1}
          \over \; s_{23} s_{56} s_{234} s_{567} s_{234567}   }\cr
%$$
%$$
 & \hskip-0.5cm      + {
           4T_{3421[[56]7]}
          \over \;\; s_{34} s_{56} s_{234} s_{567} s_{1234} \;  }
& \hskip-0.1cm       + {
           4T_{342[[56]7]1}
          \over s_{34} s_{56} s_{234} s_{567} s_{234567}   }
 & \hskip-0.1cm      + {
           4T_{3421[5[67]]}
          \over\;\; s_{34} s_{67} s_{234} s_{567} s_{1234} \;\;  }
 & \hskip-0.1cm      + {
           4T_{342[5[67]]1}
          \over s_{34} s_{67} s_{234} s_{567} s_{234567}   }\cr
}
\halign{ & $\displaystyle #$ \hfill &  $\displaystyle #$ \hfill & $\displaystyle #$ \hfill \cr
 &       + \; {
           2T_{435[12]67}          
          \over \, s_{12} s_{34} s_{345} s_{12345} s_{123456}   }
  &      + \; {
           2T_{4356127}
          \over s_{12} s_{34} s_{345} s_{123456} s_{3456} \;  }
  &      + \; {
           2T_{43567[12]}
          \over \; s_{12} s_{34} s_{345} s_{3456} s_{34567} \;  } \cr
%$$
%$$
&        + \; {
           2T_{12[34]567}
          \over s_{12} s_{34} s_{1234} s_{12345} s_{123456}   }
   &     + \; {
           2T_{453[12]67}
          \over s_{12} s_{45} s_{345} s_{12345} s_{123456}   }
  &      +\; {
           2T_{4536[12]7}
          \over s_{12} s_{45} s_{345} s_{123456} s_{3456}  \; } \cr
%$$
%$$
 &       +\; {
           2T_{45367[12]}
          \over \;\;\; s_{12} s_{45} s_{345} s_{3456} s_{34567} \;  }
  &      +\; {
           2T_{4563[12]7}
          \over \; s_{12} s_{45} s_{456} s_{123456} s_{3456} \;  }
  &      +\; {
           2T_{45637[12]}
          \over \; s_{12} s_{45} s_{456} s_{3456} s_{34567} \;\,  }\cr
%$$
%$$
  &      +\; {
           2T_{45673[12]}
          \over \;\;\; s_{12} s_{45} s_{456} s_{34567} s_{4567}  }
  &      +\; {
           2T_{1234[56]7}
          \over \; s_{12} s_{56} s_{123} s_{1234} s_{123456} \;  }
   &     +\; {
           2T_{123[45]67}
         \over s_{12} s_{45} s_{123} s_{12345} s_{123456}   }\cr
%$$
%$$
  &      +\; {
           2T_{6543[12]7}          
          \over \;\; s_{12} s_{56} s_{456} s_{123456} s_{3456}  }
  &      +\; {
           2T_{65437[12]}          
          \over \;\; s_{12} s_{56} s_{456} s_{3456} s_{34567} \;\,  }
  &      +\; {
           2T_{65473[12]}
          \over \;\; s_{12} s_{56} s_{456} s_{34567} s_{4567} \;  }\cr
%$$
%$$
&      +\; {
           2T_{65743[12]}         
          \over \;\;\; s_{12} s_{56} s_{567} s_{34567} s_{4567}  }
&       +\; {
           2T_{12345[67]}
          \over \;\; s_{12} s_{67} s_{123} s_{1234} s_{12345} \;\,  }
&       +\; {
           2T_{67543[12]}
          \over \;\; s_{12} s_{67} s_{567} s_{34567} s_{4567} \;  }\cr
%$$
%$$
 &      +\; {
           2T_{321[45]67}
          \over s_{23} s_{45} s_{123} s_{12345} s_{123456}   }
 &      +\; {
           2T_{456[23]17}
          \over s_{23} s_{45} s_{456} s_{123456} s_{23456}\,   }
  &     +\; {
           2T_{456[23]71}
          \over s_{23} s_{45} s_{456} s_{23456} s_{234567}   }\cr
%$$
%$$
  &     +\; {
           2T_{4567[23]1}
          \over \; s_{23} s_{45} s_{456} s_{234567} s_{4567}   }
   &    +\; {
           2T_{32[45]167}
          \over s_{23} s_{45} s_{12345} s_{123456} s_{2345}   }
& 	+\; {
           2T_{3241[56]7}
          \over s_{23} s_{56} s_{234} s_{1234} s_{123456}   }\cr
%$$
%$$
   &    +\; {
           2T_{32[45]617}
          \over s_{23} s_{45} s_{123456} s_{2345} s_{23456}   }
 &      +\; {
           2T_{32[45]671}
          \over s_{23} s_{45} s_{2345} s_{23456} s_{234567}   }
  &     +\; {
           2T_{3214[56]7}
          \over s_{23} s_{56} s_{123} s_{1234} s_{123456}   }\cr
%$$
%$$
    &   +\; {
           2T_{324[56]17}
          \over s_{23} s_{56} s_{234} s_{123456} s_{23456}   }
   &    +\; {
           2T_{324[56]71}
          \over \; s_{23} s_{56} s_{234} s_{23456} s_{234567}   }
   &    +\; {
           2T_{654[23]17}
          \over s_{23} s_{56} s_{456} s_{123456} s_{23456}   }\cr
%$$
%$$
&       +\; {
           2T_{654[23]71}
          \over s_{23} s_{56} s_{456} s_{23456} s_{234567}   }
&       +\; {
           2T_{6547[23]1}
          \over \; s_{23} s_{56} s_{456} s_{234567} s_{4567}   }
 &      +\; {
           2T_{6574[23]1}
          \over s_{23} s_{56} s_{567} s_{234567} s_{4567}   }\cr
%$$
%$$
 &      +\; {
           2T_{32145[67]}
          \over \;\; s_{23} s_{67} s_{123} s_{1234} s_{12345}\;   }
  &     +\; {
           2T_{32415[67]}
          \over \;\, s_{23} s_{67} s_{234} s_{1234} s_{12345}\;   }
   &    +\; {
           2T_{32451[67]}
          \over \, s_{23} s_{67} s_{234} s_{12345} s_{2345}   }\cr
%$$
%$$
  &     +\; {
           2T_{3245[67]1}
          \over \; s_{23} s_{67} s_{234} s_{2345} s_{234567} \;  }
  &     +\; {
           2T_{6754[23]1}
          \over \; s_{23} s_{67} s_{567} s_{234567} s_{4567}   }
  &     +\; {
           2T_{3421[56]7}
          \over s_{34} s_{56} s_{234} s_{1234} s_{123456}   }\cr
%$$
%$$       
   &    +\; {
           2T_{342[56]17}
          \over s_{34} s_{56} s_{234} s_{123456} s_{23456}   }
   &    +\; {
           2T_{342[56]71}
          \over s_{34} s_{56} s_{234} s_{23456} s_{234567}   }
   &    +\; {
           2T_{657[34]21}
          \over \, s_{34} s_{56} s_{567} s_{234567} s_{34567}   }\cr
%$$
%$$
    &   +\; {
           2T_{34[56]217}
          \over s_{34} s_{56} s_{123456} s_{23456} s_{3456}   }
   &    +\; {
           2T_{34[56]271}
          \over s_{34} s_{56} s_{23456} s_{234567} s_{3456}   }
   &    +\; {
           2T_{34[56]721}
          \over s_{34} s_{56} s_{234567} s_{3456} s_{34567}   }\cr
%$$
%$$
&       +\; {
           2T_{34215[67]}
          \over \;\; s_{34} s_{67} s_{234} s_{1234} s_{12345} \;\;  }
&       +\; {
           2T_{34251[67]}
          \over \;\; s_{34} s_{67} s_{234} s_{12345} s_{2345} \;\;  }
 &      +\; {
           2T_{3425[67]1}
          \over \;\; s_{34} s_{67} s_{234} s_{2345} s_{234567}   }\cr
%$$
%$$
 &      +\; {
           2T_{5432[67]1}
          \over \;\, s_{45} s_{67} s_{345} s_{2345} s_{234567}\;\;   }
  &     +\; {
           2T_{543[67]21}
          \over \;\; s_{45} s_{67} s_{345} s_{234567} s_{34567}   }
   &    +\; {
           2T_{54[67]321}
          \over s_{45} s_{67} s_{234567} s_{34567} s_{4567}   }\cr
%$$
%$$
  &     +\; {
           2T_{34521[67]}
          \over \;\;\; s_{34} s_{67} s_{345} s_{12345} s_{2345}\;\;   }
  &     +\; {
           2T_{3452[67]1}
          \over \;\;\; s_{34} s_{67} s_{345} s_{2345} s_{234567}\;   }
  &     +\; {
           2T_{345[67]21}
          \over \;\; s_{34} s_{67} s_{345} s_{234567} s_{34567}   }\cr
%$$
%$$
   &    +\; {
           2T_{675[34]21}
          \over \;\, s_{34} s_{67} s_{567} s_{234567} s_{34567}\;   }
   &    +\; {
           2T_{54321[67]}
          \over \;\;\; s_{45} s_{67} s_{345} s_{12345} s_{2345} \;\;  }
   &    +\; {
           T_{1234567}
          \over s_{12} s_{123} s_{1234} s_{12345} s_{123456}   }\cr
%$$
%$$
    &   +\; {
           T_{3214567}
          \over s_{23} s_{123} s_{1234} s_{12345} s_{123456}   }
  &     +\; {
           T_{3241567}
          \over s_{23} s_{234} s_{1234} s_{12345} s_{123456}   }
  &     +\; {
           T_{3245167}
          \over s_{23} s_{234} s_{12345} s_{123456} s_{2345}   }\cr
%$$
%$$
&       +\; {
           T_{3245617}
          \over s_{23} s_{234} s_{123456} s_{2345} s_{23456}   }
&       +\; {
           T_{3245671}
          \over s_{23} s_{234} s_{2345} s_{23456} s_{234567}   }
 &      +\; {
           T_{3456217}
          \over s_{34} s_{345} s_{123456} s_{23456} s_{3456}   }\cr
%$$
%$$
 &      +\; {
           T_{3421567}
          \over s_{34} s_{234} s_{1234} s_{12345} s_{123456}   }
  &     +\; {
           T_{3425167}
          \over s_{34} s_{234} s_{12345} s_{123456} s_{2345}   }
   &    +\; {
           T_{3452617}
          \over s_{34} s_{345} s_{123456} s_{2345} s_{23456}   }\cr
%$$
%$$
  &     +\; {
           T_{3425617}
          )\over s_{34} s_{234} s_{123456} s_{2345} s_{23456}   }
  &     +\; {
           T_{3425671}
          )\over s_{34} s_{234} s_{2345} s_{23456} s_{234567}   }
  &     +\; {
           T_{3452167}
          \over s_{34} s_{345} s_{12345} s_{123456} s_{2345}   }\cr
%$$
%$$
   &    +\; {
           T_{3452671}
          \over s_{34} s_{345} s_{2345} s_{23456} s_{234567}   }
   &    +\; {
           T_{3456271}
          \over s_{34} s_{345} s_{23456} s_{234567} s_{3456}   }
   &    +\; {
           T_{3456721}
          \over s_{34} s_{345} s_{234567} s_{3456} s_{34567}   }\cr
%$$
%$$
    &   +\; {
           T_{5432167}
          \over s_{45} s_{345} s_{12345} s_{123456} s_{2345}   }
   &    +\; {
           T_{5432617}
          \over s_{45} s_{345} s_{123456} s_{2345} s_{23456}   }
    &   +\; {
           T_{5436217}
          \over s_{45} s_{345} s_{123456} s_{23456} s_{3456}   }\cr
%$$
%$$
&       +\; {
           T_{5432671}
          \over s_{45} s_{345} s_{2345} s_{23456} s_{234567}   }
&       +\; {
           T_{5436271}
          \over s_{45} s_{345} s_{23456} s_{234567} s_{3456}   }
 &      +\; {
           T_{5436721}
          \over s_{45} s_{345} s_{234567} s_{3456} s_{34567}   }\cr
%$$
%$$
 &      +\; {
           T_{5463217}
          \over s_{45} s_{456} s_{123456} s_{23456} s_{3456}   }
  &     +\; {
           T_{5463271}
          \over s_{45} s_{456} s_{23456} s_{234567} s_{3456}   }
   &    +\; {
           T_{5463721}
          \over s_{45} s_{456} s_{234567} s_{3456} s_{34567}   }\cr
%$$
%$$
  &     +\; {
           T_{5467321}
          \over s_{45} s_{456} s_{234567} s_{34567} s_{4567}   }
  &     +\; {
           T_{5643217}
          \over s_{56} s_{456} s_{123456} s_{23456} s_{3456}   }
  &     +\; {
           T_{5643271}
          \over s_{56} s_{456} s_{23456} s_{234567} s_{3456}   }\cr
%$$
%$$
   &    +\; {
           T_{5643721}
          \over s_{56} s_{456} s_{234567} s_{3456} s_{34567}   }
   &    +\; {
           T_{5647321}
          \over s_{56} s_{456} s_{234567} s_{34567} s_{4567}   }
   &    +\; {
           T_{5674321}
          \over s_{56} s_{567} s_{234567} s_{34567} s_{4567}   }\cr
%$$
%$$
&       +\; {
           T_{7654321}
          \over s_{67} s_{567} s_{234567} s_{34567} s_{4567}   }. & & \hskip3.3cm \Msevenformula \cr
%$$
}
\bigskip

%\end

%********************************************************************
\appendix{C}{The cubic graphs of $M_{123{\ldots} n}$}
\applab\appgraphs
%********************************************************************
\raggedbottom

As discussed in section \secBGJs, the expressions for $M_{123{\ldots} n}$ of Appendix~\AppendixMs\ were
found using the dictionary between the cubic diagrams of the $(n+1)-$point amplitude with one leg off-shell
and BRST building blocks. The graphs which compose the expressions for $M_{123},\dots, M_{123456}$ are given 
below (the 132 graphs used in the derivation of $M_{1234567}$ would occupy to much space and were omitted).
\medskip
%****************
% The M3 diagrams
%****************
\centerline{
\tikzpicture [scale=0.8,line width=0.30mm]
\scope[xshift=-1cm]
\draw (0,0) -- (-1,1) node[above]{$2$};
\draw (0,0) -- (-1,-1) node[below]{$1$};
\draw (0,0) -- (1.8,0);
\draw (0.5,-0.2) node{$s_{12}$};
\draw (1,0) -- (1,1) node[above]{$3$};
\draw (1.5,0.2) node{$s_{123}$};
\draw (2.3, 0) node{${\ldots} $};
\draw (3.7,0) node{$=\displaystyle {T_{123} \over s_{12}s_{123}}$};
\endscope
\draw (5.5,0) -- (4.5,1) node[above]{$3$};
\draw (5.5,0) -- (4.5,-1) node[below]{$2$};
\draw (5.5,0) -- (7.3,0);
\draw (6,-0.2) node{$s_{23}$};
\draw (6.5,0) -- (6.5,-1) node[below]{$1$};
\draw (7,0.2) node{$s_{123}$};
\draw (7.8, 0) node{${\ldots} $};
\draw (9.1,0) node{$=\displaystyle {T_{321}\over s_{23}s_{123}}$};
\endtikzpicture
}
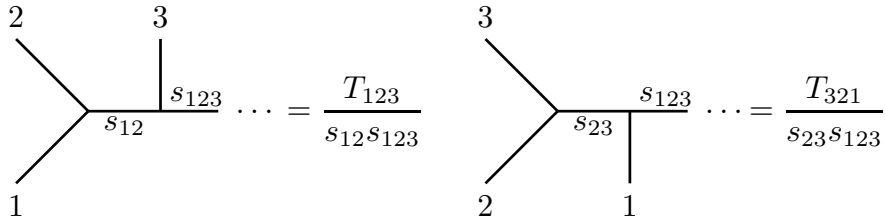
\tikzcaption\Mtdiags{The two cubic diagrams which constitute $M_{123}$.}

%****************
% The M4 diagrams
%****************
\centerline{\hskip\parindent
\tikzpicture [scale=1.1,line width=0.30mm]
\scope[yshift=2cm,xshift=-1cm]
\draw (0,0) -- (-0.5,0.5) node[above]{$2$};
\draw (0,0) -- (-0.5,-0.5) node[below]{$1$};
\draw (0,0) -- (1.3,0);
\draw (0.25,0.2) node{$s_{12}$};
\draw (0.5,0) -- (0.5,0.5) node[above]{$3$};
\draw (0.75,-0.2) node{$s_{123}$};
\draw (1,0) -- (1,0.5) node[above]{$4$};
\draw (1.45,0.2) node{$s_{1234}$};
\draw (1.6, 0) node{$\ldots  $};
\draw (2.9, 0) node{ $=\displaystyle {T_{1234}\over s_{12}s_{123}s_{1234}}$};
\endscope
\scope[yshift=-2cm,xshift=-1cm]
\draw (0,2) -- (-0.5,2.5) node[above]{$3$};
\draw (0,2) -- (-0.5,1.5) node[below]{$2$};
\draw (0,2) -- (1.3,2);
\draw (0.25,2.2) node{$s_{23}$};
\draw (0.5,2) -- (0.5,1.5) node[below]{$1$};
\draw (0.85,1.8) node{$s_{123}$};
\draw (1,2) -- (1,2.5) node[above]{$4$};
\draw (1.45,2.2) node{$s_{1234}$};
\draw (1.6,2) node{$\ldots  $};
\draw (2.9, 2) node{ $=\displaystyle {T_{3214}\over s_{23}s_{123}s_{1234}}$};
\endscope
\scope[xshift=0.3cm]
\draw (4.2, 2) -- (3.7, 2.5) node[above]{$4$};
\draw (4.2, 2) -- (3.7, 1.5) node[below]{$3$};
\draw (4.2, 2) -- (5.5, 2);
\draw (4.45, 1.8) node{$s_{34}$};
\draw (4.7, 2) -- (4.7, 1.5) node[below]{$2$};
\draw (4.95, 2.2) node{$s_{234}$};
\draw (5.2, 2) -- (5.2, 1.5) node[below]{$1$};
\draw (5.65, 1.8) node{$s_{1234}$};
\draw (5.8, 2) node{$\ldots  $};
\draw (7.1, 2) node{$=\displaystyle {T_{3421} \over s_{34}s_{234}s_{1234}}$};
\endscope
\scope[xshift=0.3cm]
\draw (4.2, 0) -- (3.7, 0.5) node[above]{$3$};
\draw (4.2, 0) -- (3.7, -0.5) node[below]{$2$};
\draw (4.2, 0) -- (5.5, 0);
\draw (4.45,-0.2) node{$s_{23}$};
\draw (4.7, 0) -- (4.7, 0.5) node[above]{$4$};
\draw (5.05, 0.2) node{$s_{234}$};
\draw (5.2, 0) -- (5.2, -0.5) node[below]{$1$};
\draw (5.65, -0.2) node{$s_{1234}$};
\draw (5.8, 0) node{$\ldots  $};
\draw (7.1, 0) node{$=\displaystyle {T_{3241} \over s_{23}s_{234}s_{1234}}$};
\endscope
\scope[xshift=0.0cm,yshift=-0.4cm]
  \draw (1, -1.5) -- (4.0, -1.5);
  \draw (1, -1.5) -- (0.5, -1.0) node[above]{$2$};
  \draw (1, -1.5) -- (0.5, -2.0) node[below]{$1$};
  \draw (4.0, -1.5) -- (4.5, -1.0) node[above]{$3$};
  \draw (4.0, -1.5) -- (4.5, -2.0) node[below]{$4$};
  %off-shell leg
  \draw (2.5, -1.5) -- (2.5, -1.75);
  \draw (2.5, -1.85) node{$\vdots$};
  \draw (1.75, -1.3) node{$s_{12}$};
  \draw (3.25, -1.3) node{$s_{34}$};
  \draw (2.9, -1.7) node{$s_{1234}$};
  \draw (5.5, -1.5) node{$= \displaystyle {2T_{12[34]}\over s_{12}s_{34}s_{1234}}$};
\endscope
\endtikzpicture
}
\tikzcaption\Mfourtikz{The five cubic diagrams which constitute $M_{1234}$. The signs match the corresponding terms
given in the formula \Mfourformula.}

%****************
% The M5 diagrams
%****************
\vskip0.8cm
\centerline{
\tikzpicture[scale=1.3,line width=0.30mm]
% +++
\draw  (2,2) -- (1.5,2.5);
\draw  (2,2) -- (1.5,1.5);
\draw (2.25,1.8) node{$s_{12}$};
\draw (2.75,2.2) node{$s_{123}$};
\draw (3.25,1.8) node{$s_{1234}$};
\draw (3.95,2.2) node{$s_{12345}$};
\draw  (2,2) -- (3.8,2); 
\draw (4,2) node{$\ldots  $};
\draw  (2.5,2) -- (2.5,2.5); 
\draw  (3,2) -- (3,2.5);
\draw  (3.5,2) -- (3.5,2.5);
\draw (1.5,2.7) node{$2$} ; 
\draw (1.5,1.3) node{$1$} ;
\draw (2.5,2.7) node{$3$} ;
\draw (3,2.7) node{$4$} ;
\draw (3.5,2.7) node{$5$} ;
\draw (5.2,2)node{{$ \displaystyle = \ \frac{ T_{12345} / s_{12345}}{s_{12} \, s_{123} \, s_{1234} }$}};
% ++-
\draw  (7,2) -- (6.5,2.5);
\draw  (7,2) -- (6.5,1.5);
\draw (7.25,1.8) node{$s_{23}$};
\draw (7.75,2.2) node{$s_{234}$};
\draw (8.25,1.8) node{$s_{2345}$};
\draw (8.95,2.2) node{$s_{12345}$};
\draw  (7,2) -- (8.8,2); 
\draw (9,2) node{$\ldots  $};
\draw  (7.5,2) -- (7.5,2.5); 
\draw  (8,2) -- (8,2.5);
\draw  (8.5,2) -- (8.5,1.5);
\draw (6.5,2.7) node{$3$} ; 
\draw (6.5,1.3) node{$2$} ;
\draw (7.5,2.7) node{$4$} ;
\draw (8,2.7) node{$5$} ;
\draw (8.5,1.3) node{$1$} ;
\draw (10.2,2)node{{$ \displaystyle = \ \frac{ T_{32451} / s_{12345}}{s_{23} \, s_{234} \, s_{2345} }$}};
% +-+
\endtikzpicture
}
\centerline{
\tikzpicture[scale=1.3,line width=0.30mm]
\draw  (2,0) -- (1.5,0.5);
\draw  (2,0) -- (1.5,-0.5);
\draw (2.25,-0.2) node{$s_{23}$};
\draw (2.75,0.2) node{$s_{234}$};
\draw (3.3,-0.2) node{$s_{1234}$};
\draw (3.95,0.2) node{$s_{12345}$};
\draw  (2,0) -- (3.8,0); 
\draw (4,0) node{$\ldots  $};
\draw  (2.5,0) -- (2.5,0.5); 
\draw  (3,0) -- (3,-0.5);
\draw  (3.5,0) -- (3.5,0.5);
\draw (1.5,0.7) node{$3$} ; 
\draw (1.5,-0.7) node{$2$} ;
\draw (2.5,0.7) node{$4$} ;
\draw (3,-0.7) node{$1$} ;
\draw (3.5,0.7) node{$5$} ;
\draw (5.2,0)node{{$ \displaystyle = \ \frac{ T_{32415} / s_{12345}}{s_{23} \, s_{234} \, s_{1234} }$}};
% +--
\draw  (7,0) -- (6.5,0.5);
\draw  (7,0) -- (6.5,-0.5);
\draw (7.25,0.2) node{$s_{34}$};
\draw (7.75,-0.2) node{$s_{345}$};
\draw (8.25,0.2) node{$s_{2345}$};
\draw (8.95,-0.2) node{$s_{12345}$};
\draw  (7,0) -- (8.8,0); 
\draw (9,0) node{$\ldots  $};
\draw  (7.5,0) -- (7.5,0.5); 
\draw  (8,0) -- (8,-0.5);
\draw  (8.5,0) -- (8.5,-0.5);
\draw (6.5,0.7) node{$4$} ; 
\draw (6.5,-0.7) node{$3$} ;
\draw (7.5,0.7) node{$5$} ;
\draw (8,-0.7) node{$2$} ;
\draw (8.5,-0.7) node{$1$} ;
\draw (10.2,0)node{{$ \displaystyle = \ \frac{ T_{34521} / s_{12345}}{s_{34} \, s_{345} \, s_{2345} }$}};
\endtikzpicture
}
%%%%%%%%%%%%%
\centerline{
\tikzpicture[scale=1.3,line width=0.30mm]
% -++
\draw  (2,-2) -- (1.5,-1.5);
\draw  (2,-2) -- (1.5,-2.5);
\draw (2.25,-2.2) node{$s_{23}$};
\draw (2.75,-1.8) node{$s_{123}$};
\draw (3.25,-2.2) node{$s_{1234}$};
\draw (3.95,-1.8) node{$s_{12345}$};
\draw  (2,-2) -- (3.8,-2); 
\draw (4,-2) node{$\ldots  $};
\draw  (2.5,-2) -- (2.5,-2.5); 
\draw  (3,-2) -- (3,-1.5);
\draw  (3.5,-2) -- (3.5,-1.5);
\draw (1.5,-1.3) node{$3$} ; 
\draw (1.5,-2.7) node{$2$} ;
\draw (2.5,-2.7) node{$1$} ;
\draw (3,-1.3) node{$4$} ;
\draw (3.5,-1.3) node{$5$} ;
\draw (5.2,-2)node{{$ \displaystyle = \ \frac{ T_{32145} / s_{12345}}{s_{23} \, s_{123} \, s_{1234} }$}};
% -+-
\draw  (7,-2) -- (6.5,-1.5);
\draw  (7,-2) -- (6.5,-2.5);
\draw (7.25,-2.2) node{$s_{34}$};
\draw (7.75,-1.8) node{$s_{234}$};
\draw (8.15,-2.2) node{$s_{2345}$};
\draw (8.85,-1.8) node{$s_{12345}$};
\draw  (7,-2) -- (8.8,-2); 
\draw (9,-2) node{$\ldots  $};
\draw  (7.5,-2) -- (7.5,-2.5); 
\draw  (8,-2) -- (8,-1.5);
\draw  (8.5,-2) -- (8.5,-2.5);
\draw (6.5,-1.3) node{$4$} ; 
\draw (6.5,-2.7) node{$3$} ;
\draw (7.5,-2.7) node{$2$} ;
\draw (8,-1.3) node{$5$} ;
\draw (8.5,-2.7) node{$1$} ;
\draw (10.2,-2)node{{$ \displaystyle = \ \frac{ T_{34251} / s_{12345}}{s_{34} \, s_{234} \, s_{2345} }$}};
\endtikzpicture
}
\centerline{
\tikzpicture[scale=1.3,line width=0.30mm]
% --+
\draw  (2,-4) -- (1.5,-3.5);
\draw  (2,-4) -- (1.5,-4.5);
\draw (2.25,-4.2) node{$s_{34}$};
\draw (2.75,-3.8) node{$s_{234}$};
\draw (3.4,-4.2) node{$s_{1234}$};
\draw (4.05,-3.8) node{$s_{12345}$};
\draw  (2,-4) -- (3.8,-4); 
\draw (4,-4) node{$\ldots  $};
\draw  (2.5,-4) -- (2.5,-4.5); 
\draw  (3,-4) -- (3,-4.5);
\draw  (3.5,-4) -- (3.5,-3.5);
\draw (1.5,-3.3) node{$4$} ; 
\draw (1.5,-4.7) node{$3$} ;
\draw (2.5,-4.7) node{$2$} ;
\draw (3,-4.7) node{$1$} ;
\draw (3.5,-3.3) node{$5$} ;
\draw (5.2,-4)node{{$ \displaystyle = \ \frac{ T_{34215} / s_{12345}}{s_{34} \, s_{234} \, s_{1234} }$}};
% ---
\draw  (7,-4) -- (6.5,-3.5);
\draw  (7,-4) -- (6.5,-4.5);
\draw (7.25,-3.8) node{$s_{45}$};
\draw (7.75,-4.2) node{$s_{345}$};
\draw (8.35,-3.8) node{$s_{2345}$};
\draw (8.95,-4.2) node{$s_{12345}$};
\draw  (7,-4) -- (8.8,-4); 
\draw (9,-4) node{$\ldots  $};
\draw  (7.5,-4) -- (7.5,-4.5); 
\draw  (8,-4) -- (8,-4.5);
\draw  (8.5,-4) -- (8.5,-4.5);
\draw (6.5,-3.3) node{$5$} ; 
\draw (6.5,-4.7) node{$4$} ;
\draw (7.5,-4.7) node{$3$} ;
\draw (8,-4.7) node{$2$} ;
\draw (8.5,-4.7) node{$1$} ;
\draw (10.2,-4)node{{$ \displaystyle = \ \frac{ T_{54321} / s_{12345}}{s_{45} \, s_{345} \, s_{2345} }$}};
\endtikzpicture
}
%%%%
\centerline{
\tikzpicture[scale=1.2,line width=0.30mm]
% And now, the completely different ones
\draw (2,-6) -- (1.5, -5.5) node[above] {$2$};
\draw (2,-6) -- (2.5, -5.5) node[above] {$3$};
\draw (2,-6) -- (3.5,-6);
\draw  (3.5,-6) -- (3, -5.5) node[above] {$4$};
\draw  (3.5,-6) -- (4, -5.5) node[above] {$5$};
\draw (2.75, -6) -- (2.75,-6.8);
\draw (2.75, -6.9) node{$\vdots$};
\draw (2.75, -6.5) -- (2.25, -6.5) node[left] {$1$};
\draw (2.5,-5.8) node{$s_{23}$} ;
\draw (3.0,-5.8) node{$s_{45}$} ;
\draw (3.15,-6.25) node{$s_{2345}$} ;
\draw (2.3,-6.8) node{$s_{12345}$} ;
\draw (5.0,-6) node{$\displaystyle = \ \frac{ 2 T_{54[32]1} / s_{12345}}{s_{23} \, s_{45} \, s_{2345} }$};
\draw (7,-6) -- (6.5, -5.5) node[above] {$1$};
\draw (7,-6) -- (7.5, -5.5) node[above] {$2$};
\draw (7,-6) -- (8.5,-6);
\draw  (8.5,-6) -- (8, -5.5) node[above] {$3$};
\draw  (8.5,-6) -- (9, -5.5) node[above] {$4$};
\draw (7.75, -6) -- (7.75,-6.8);
\draw (7.75, -6.9) node{$\vdots$};
\draw (7.75, -6.5) -- (8.25, -6.5) node[right] {$5$};
\draw (7.5,-5.8) node{$s_{12}$} ;
\draw (8.0,-5.8) node{$s_{34}$} ;
\draw (7.35,-6.25) node{$s_{1234}$} ;
\draw (8.2,-6.8) node{$s_{12345}$} ;
\draw (10.0,-6) node{$\displaystyle = \ \frac{ 2 T_{12[34]5} / s_{12345}}{s_{12} \, s_{34} \, s_{1234} }$};
\endtikzpicture
}
\vskip-0.4cm
% Now even more different ones
\centerline{
\tikzpicture[scale=1.3,line width=0.30mm]
\draw (2,-8) -- (1.5, -7.5) node[above] {$2$};
\draw (2,-8) -- (2.5, -7.5) node[above] {$3$};
\draw (2,-8.5) -- (3.5,-8.5);
\draw (2,-8.5) -- (2,-8);
\draw  (3.5,-8.5) -- (3, -8) node[above] {$4$};
\draw  (3.5,-8.5) -- (4, -8) node[above] {$5$};
\draw (2, -8.5) -- (1.5,-9) node[above] {$1$};
\draw (2.75,-8.5) -- (2.75,-8.8);
\draw (2.75, -8.9) node{$\vdots$};
\draw (3.2,-8.8) node{$s_{12345}$} ;
\draw (2.4,-8.3) node{$s_{123}$} ;
\draw (3.0,-8.3) node{$s_{45}$} ;
\draw (1.8,-8.25) node{$s_{23}$} ;
\draw (5.0,-8.5) node{$\displaystyle = \ \frac{ 2 T_{321[45]} / s_{12345}}{s_{23} \, s_{123} \, s_{45} }$};
\draw (7,-8.5) -- (6.5, -8) node[above] {$1$};
\draw (7,-8.5) -- (7.5, -8) node[above] {$2$};
\draw (7,-8.5) -- (8.5,-8.5);
\draw (8.5,-8.5) -- (8.5,-8);
\draw  (8.5,-8) -- (8, -7.5) node[above] {$3$};
\draw  (8.5,-8) -- (9, -7.5) node[above] {$4$};
\draw  (8.5,-8.5) -- (9, -9) node[above] {$5$};
\draw (7.75,-8.5) -- (7.75,-8.8);
\draw (7.75, -8.9) node{$\vdots$};
\draw (7.3,-8.8) node{$s_{12345}$} ;
\draw (7.5,-8.3) node{$s_{12}$} ;
\draw (8.1,-8.3) node{$s_{345}$} ;
\draw (8.7,-8.25) node{$s_{34}$} ;
\draw (10.0,-8.5) node{$\displaystyle = \ \frac{ 2 T_{435[12]} / s_{12345}}{s_{12} \, s_{34} \, s_{345} }$};
\endtikzpicture
}
\vskip-0.4cm
\centerline{
\tikzpicture[scale=1.3,line width=0.30mm]
% Two more of this strange type
\draw (2,-10) -- (1.5, -9.5) node[above] {$1$};
\draw (2,-10) -- (2.5, -9.5) node[above] {$2$};
\draw (2,-10.5) -- (3.5,-10.5);
\draw (2,-10.5) -- (2,-10);
\draw  (3.5,-10.5) -- (3, -10) node[above] {$4$};
\draw  (3.5,-10.5) -- (4, -10) node[above] {$5$};
\draw (2, -10.5) -- (2.5,-10) node[above] {$3$};
\draw (2.75,-10.5) -- (2.75,-10.8);
\draw (2.75, -10.9) node{$\vdots$};
\draw (3.2,-10.8) node{$s_{12345}$} ;
\draw (2.4,-10.7) node{$s_{123}$} ;
\draw (3.0,-10.3) node{$s_{45}$} ;
\draw (1.8,-10.25) node{$s_{12}$} ;
\draw (5.0,-10.5) node{$\displaystyle = \ \frac{ 2 T_{123[45]} / s_{12345}}{s_{12} \, s_{123} \, s_{45} }$};
\draw (7,-10.5) -- (6.5, -10) node[above] {$1$};
\draw (7,-10.5) -- (7.5, -10) node[above] {$2$};
\draw (7,-10.5) -- (8.5,-10.5);
\draw (8.5,-10.5) -- (8.5,-10);
\draw  (8.5,-10) -- (8, -9.5) node[above] {$4$};
\draw  (8.5,-10) -- (9, -9.5) node[above] {$5$};
\draw  (8.5,-10.5) -- (8, -10) node[above] {$3$};
\draw (7.75,-10.5) -- (7.75,-10.8);
\draw (7.75, -10.9) node{$\vdots$};
\draw (7.3,-10.8) node{$s_{12345}$} ;
\draw (7.5,-10.3) node{$s_{12}$} ;
\draw (8.1,-10.7) node{$s_{345}$} ;
\draw (8.7,-10.25) node{$s_{45}$} ;
\draw (10.0,-10.5) node{$\displaystyle = \ \frac{ 2 T_{453[12]} / s_{12345}}{s_{12} \, s_{45} \, s_{345} }$};
\endtikzpicture
}
\vskip-0.3cm
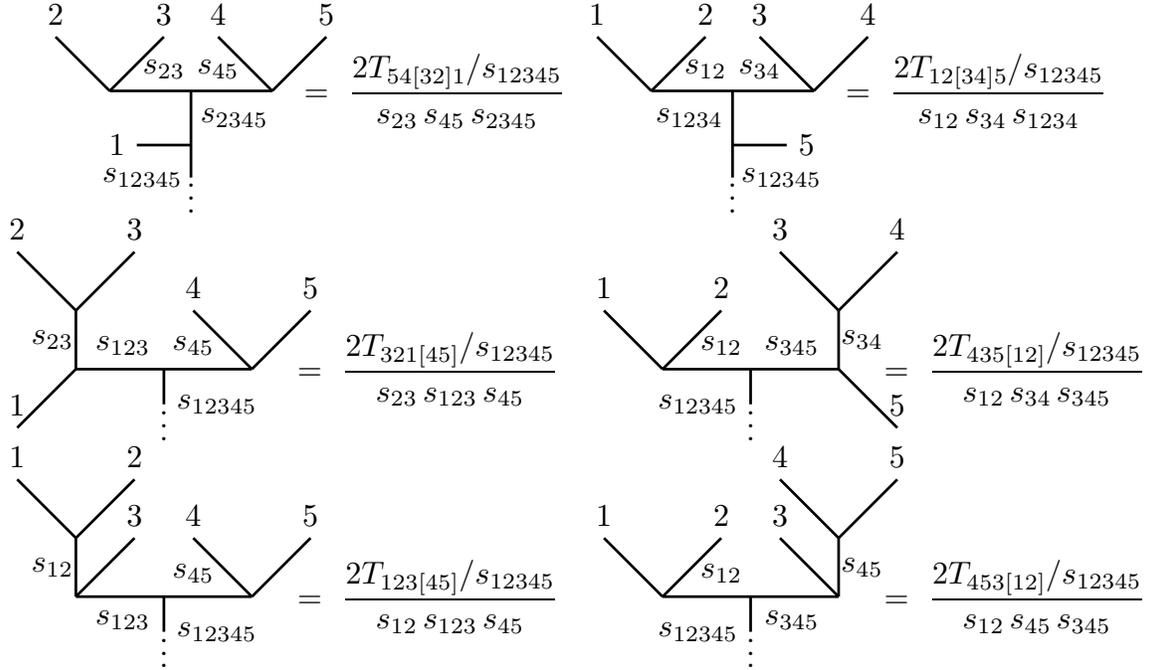
\tikzcaption\Mfivetikz{The 14 cubic diagrams which constitute $M_{12345}$. The signs of their corresponding formul{\ae}
are in one-to-one agreement with the terms in expression for $M_{12345}$ given by \Mfiveformula, which is
reproduced by summing all 14 graphs displayed here.}

%****************
% The M6 diagrams
%****************

\tikzpicture[scale=1.3,line width=0.30mm]
% ++++
\draw  (2,2) -- (1.5,2.5) node[left]{$2$};
\draw  (2,2) -- (1.5,1.5) node[left]{$1$};
\draw (2.35,1.8) node{$s_{12}$};
\draw (3.05,2.2) node{$s_{123}$};
\draw (3.75,1.8) node{$s_{1234}$};
\draw (4.45,2.2) node{$s_{12345}$};
\draw (5.25,1.8) node{$s_{123456}$};
\draw  (2,2) -- (5.1,2); 
\draw (5.3,2) node{$\ldots  $};
\draw  (2.7,2) -- (2.7,2.5) node[above]{$3$}; 
\draw  (3.4,2) -- (3.4,2.5) node[above]{$4$};
\draw  (4.1,2) -- (4.1,2.5) node[above]{$5$};
\draw  (4.8,2) -- (4.8,2.5) node[above]{$6$};
\draw (3.5,0.8)node{{$ \displaystyle = \ \ \frac{ T_{123456} \, / \, s_{123456}}{s_{12} \, s_{123} \, s_{1234} \, s_{12345} }$}};
% +++-
\scope[xshift=5cm]
\draw  (2,2) -- (1.5,2.5) node[left]{$3$};
\draw  (2,2) -- (1.5,1.5) node[left]{$2$};
\draw (2.35,1.8) node{$s_{23}$};
\draw (3.05,2.2) node{$s_{234}$};
\draw (3.75,1.8) node{$s_{2345}$};
\draw (4.45,2.2) node{$s_{23456}$};
\draw (5.25,1.8) node{$s_{123456}$};
\draw  (2,2) -- (5.1,2); 
\draw (5.3,2) node{$\ldots  $};
\draw  (2.7,2) -- (2.7,2.5) node[above]{$4$}; 
\draw  (3.4,2) -- (3.4,2.5) node[above]{$5$};
\draw  (4.1,2) -- (4.1,2.5) node[above]{$6$};
\draw  (4.8,2) -- (4.8,1.5) node[below]{$1$};
\draw (3.5,0.8)node{{$ \displaystyle = \ \ \frac{ T_{324561} \, / \, s_{123456}}{s_{23} \, s_{234} \, s_{2345} \, s_{23456} }$}};
\endscope
\endtikzpicture

\tikzpicture [scale=1.3, line width=0.30mm]
% ++-+
\scope
\draw  (2,2) -- (1.5,2.5) node[left]{$3$};
\draw  (2,2) -- (1.5,1.5) node[left]{$2$};
\draw (2.35,1.8) node{$s_{23}$};
\draw (3.05,2.2) node{$s_{234}$};
\draw (3.75,1.8) node{$s_{2345}$};
\draw (4.45,2.2) node{$s_{12345}$};
\draw (5.25,1.8) node{$s_{123456}$};
\draw  (2,2) -- (5.1,2); 
\draw (5.3,2) node{$\ldots  $};
\draw  (2.7,2) -- (2.7,2.5) node[above]{$4$}; 
\draw  (3.4,2) -- (3.4,2.5) node[above]{$5$};
\draw  (4.1,2) -- (4.1,1.5) node[below]{$1$};
\draw  (4.8,2) -- (4.8,2.5) node[above]{$6$};
\draw (3.5,0.8)node{{$ \displaystyle = \ \ \frac{ T_{324516} \, / \, s_{123456}}{s_{23} \, s_{234} \, s_{2345} \, s_{12345} }$}};
\endscope
% ++--
\scope[xshift=5cm]
\draw  (2,2) -- (1.5,2.5) node[left]{$4$};
\draw  (2,2) -- (1.5,1.5) node[left]{$3$};
\draw (2.35,1.8) node{$s_{34}$};
\draw (3.05,2.2) node{$s_{345}$};
\draw (3.75,1.8) node{$s_{3456}$};
\draw (4.45,2.2) node{$s_{23456}$};
\draw (5.25,1.8) node{$s_{123456}$};
\draw  (2,2) -- (5.1,2); 
\draw (5.3,2) node{$\ldots  $};
\draw  (2.7,2) -- (2.7,2.5) node[above]{$5$}; 
\draw  (3.4,2) -- (3.4,2.5) node[above]{$6$};
\draw  (4.1,2) -- (4.1,1.5) node[below]{$2$};
\draw  (4.8,2) -- (4.8,1.5) node[below]{$1$};
\draw (3.5,0.8)node{{$ \displaystyle = \ \ \frac{ T_{345621} \, / \, s_{123456}}{s_{34} \, s_{345} \, s_{3456} \, s_{23456} }$}};
\endscope
\endtikzpicture

\tikzpicture [scale=1.3, line width=0.30mm]
%%%%%%%%%%%%%%%%%%%
% +-++
\scope
\draw  (2,2) -- (1.5,2.5) node[left]{$3$};
\draw  (2,2) -- (1.5,1.5) node[left]{$2$};
\draw (2.35,1.8) node{$s_{23}$};
\draw (3.05,2.2) node{$s_{234}$};
\draw (3.75,1.8) node{$s_{1234}$};
\draw (4.45,2.2) node{$s_{12345}$};
\draw (5.25,1.8) node{$s_{123456}$};
\draw  (2,2) -- (5.1,2); 
\draw (5.3,2) node{$\ldots  $};
\draw  (2.7,2) -- (2.7,2.5) node[above]{$4$}; 
\draw  (3.4,2) -- (3.4,1.5) node[below]{$1$};
\draw  (4.1,2) -- (4.1,2.5) node[above]{$5$};
\draw  (4.8,2) -- (4.8,2.5) node[above]{$6$};
\draw (3.5,0.8)node{{$ \displaystyle = \ \ \frac{ T_{324156} \, / \, s_{123456}}{s_{23} \, s_{234} \, s_{1234} \, s_{12345} }$}};
\endscope
% +-+-
\scope[xshift=5cm]
\draw  (2,2) -- (1.5,2.5) node[left]{$4$};
\draw  (2,2) -- (1.5,1.5) node[left]{$3$};
\draw (2.35,1.8) node{$s_{34}$};
\draw (3.05,2.2) node{$s_{345}$};
\draw (3.75,1.8) node{$s_{2345}$};
\draw (4.45,2.2) node{$s_{23456}$};
\draw (5.25,1.8) node{$s_{123456}$};
\draw  (2,2) -- (5.1,2); 
\draw (5.3,2) node{$\ldots  $};
\draw  (2.7,2) -- (2.7,2.5) node[above]{$5$}; 
\draw  (3.4,2) -- (3.4,1.5) node[below]{$2$};
\draw  (4.1,2) -- (4.1,2.5) node[above]{$6$};
\draw  (4.8,2) -- (4.8,1.5) node[below]{$1$};
\draw (3.5,0.8)node{{$ \displaystyle = \ \ \frac{ T_{345261} \, / \, s_{123456}}{s_{34} \, s_{345} \, s_{2345} \, s_{23456} }$}};
\endscope
\endtikzpicture

\tikzpicture [scale=1.3, line width=0.30mm]
% +--+
\scope
\draw  (2,2) -- (1.5,2.5) node[left]{$4$};
\draw  (2,2) -- (1.5,1.5) node[left]{$3$};
\draw (2.35,1.8) node{$s_{34}$};
\draw (3.05,2.2) node{$s_{345}$};
\draw (3.75,1.8) node{$s_{2345}$};
\draw (4.45,2.2) node{$s_{12345}$};
\draw (5.25,1.8) node{$s_{123456}$};
\draw  (2,2) -- (5.1,2); 
\draw (5.3,2) node{$\ldots  $};
\draw  (2.7,2) -- (2.7,2.5) node[above]{$5$}; 
\draw  (3.4,2) -- (3.4,1.5) node[below]{$2$};
\draw  (4.1,2) -- (4.1,1.5) node[below]{$1$};
\draw  (4.8,2) -- (4.8,2.5) node[above]{$6$};
\draw (3.5,0.8)node{{$ \displaystyle = \ \ \frac{ T_{345216} \, / \, s_{123456}}{s_{34} \, s_{345} \, s_{2345} \, s_{12345} }$}};
\endscope
% +---
\scope[xshift=5cm]
\draw  (2,2) -- (1.5,2.5) node[left]{$5$};
\draw  (2,2) -- (1.5,1.5) node[left]{$4$};
\draw (2.35,1.8) node{$s_{45}$};
\draw (3.05,2.2) node{$s_{456}$};
\draw (3.75,1.8) node{$s_{3456}$};
\draw (4.45,2.2) node{$s_{23456}$};
\draw (5.25,1.8) node{$s_{123456}$};
\draw  (2,2) -- (5.1,2); 
\draw (5.3,2) node{$\ldots  $};
\draw  (2.7,2) -- (2.7,2.5) node[above]{$6$}; 
\draw  (3.4,2) -- (3.4,1.5) node[below]{$3$};
\draw  (4.1,2) -- (4.1,1.5) node[below]{$2$};
\draw  (4.8,2) -- (4.8,1.5) node[below]{$1$};
\draw (3.5,0.8)node{{$ \displaystyle = \ \ \frac{ T_{546321} \, / \, s_{123456}}{s_{45} \, s_{456} \, s_{3456} \, s_{23456} }$}};
\endscope
\endtikzpicture

%%%%%%%%%%%%%%
%%%%%%%%%%%%%%%%%%%%%%%%%%%%%%%
%%%%%%%%%%%%%%

\tikzpicture[scale=1.3, line width=0.30mm]
% -+++
\draw  (2,2) -- (1.5,2.5) node[left]{$3$};
\draw  (2,2) -- (1.5,1.5) node[left]{$2$};
\draw (2.35,1.8) node{$s_{23}$};
\draw (3.05,2.2) node{$s_{123}$};
\draw (3.75,1.8) node{$s_{1234}$};
\draw (4.45,2.2) node{$s_{12345}$};
\draw (5.25,1.8) node{$s_{123456}$};
\draw  (2,2) -- (5.1,2); 
\draw (5.3,2) node{$\ldots  $};
\draw  (2.7,2) -- (2.7,1.5) node[below]{$1$}; 
\draw  (3.4,2) -- (3.4,2.5) node[above]{$4$};
\draw  (4.1,2) -- (4.1,2.5) node[above]{$5$};
\draw  (4.8,2) -- (4.8,2.5) node[above]{$6$};
\draw (3.5,0.8)node{{$ \displaystyle = \ \ \frac{ T_{321456} \, / \, s_{123456}}{s_{23} \, s_{123} \, s_{1234} \, s_{12345} }$}};
% -++-
\scope[xshift=5cm]
\draw  (2,2) -- (1.5,2.5) node[left]{$4$};
\draw  (2,2) -- (1.5,1.5) node[left]{$3$};
\draw (2.35,1.8) node{$s_{34}$};
\draw (3.05,2.2) node{$s_{234}$};
\draw (3.75,1.8) node{$s_{2345}$};
\draw (4.45,2.2) node{$s_{23456}$};
\draw (5.25,1.8) node{$s_{123456}$};
\draw  (2,2) -- (5.1,2); 
\draw (5.3,2) node{$\ldots  $};
\draw  (2.7,2) -- (2.7,1.5) node[below]{$2$}; 
\draw  (3.4,2) -- (3.4,2.5) node[above]{$5$};
\draw  (4.1,2) -- (4.1,2.5) node[above]{$6$};
\draw  (4.8,2) -- (4.8,1.5) node[below]{$1$};
\draw (3.5,0.8)node{{$ \displaystyle = \ \ \frac{ T_{342561} \, / \, s_{123456}}{s_{34} \, s_{234} \, s_{2345} \, s_{23456} }$}};
\endscope
\endtikzpicture

\tikzpicture [scale=1.3, line width=0.30mm]
% -+-+
\scope
\draw  (2,2) -- (1.5,2.5) node[left]{$4$};
\draw  (2,2) -- (1.5,1.5) node[left]{$3$};
\draw (2.35,1.8) node{$s_{34}$};
\draw (3.05,2.2) node{$s_{234}$};
\draw (3.75,1.8) node{$s_{2345}$};
\draw (4.45,2.2) node{$s_{12345}$};
\draw (5.25,1.8) node{$s_{123456}$};
\draw  (2,2) -- (5.1,2); 
\draw (5.3,2) node{$\ldots  $};
\draw  (2.7,2) -- (2.7,1.5) node[below]{$2$}; 
\draw  (3.4,2) -- (3.4,2.5) node[above]{$5$};
\draw  (4.1,2) -- (4.1,1.5) node[below]{$1$};
\draw  (4.8,2) -- (4.8,2.5) node[above]{$6$};
\draw (3.5,0.8)node{{$ \displaystyle = \ \ \frac{ T_{342516} \, / \, s_{123456}}{s_{34} \, s_{234} \, s_{2345} \, s_{12345} }$}};
\endscope
% -+--
\scope[xshift=5cm]
\draw  (2,2) -- (1.5,2.5) node[left]{$5$};
\draw  (2,2) -- (1.5,1.5) node[left]{$4$};
\draw (2.35,1.8) node{$s_{45}$};
\draw (3.05,2.2) node{$s_{345}$};
\draw (3.75,1.8) node{$s_{3456}$};
\draw (4.45,2.2) node{$s_{23456}$};
\draw (5.25,1.8) node{$s_{123456}$};
\draw  (2,2) -- (5.1,2); 
\draw (5.3,2) node{$\ldots  $};
\draw  (2.7,2) -- (2.7,1.5) node[below]{$3$}; 
\draw  (3.4,2) -- (3.4,2.5) node[above]{$6$};
\draw  (4.1,2) -- (4.1,1.5) node[below]{$2$};
\draw  (4.8,2) -- (4.8,1.5) node[below]{$1$};
\draw (3.5,0.8)node{{$ \displaystyle = \ \ \frac{ T_{543621} \, / \, s_{123456}}{s_{45} \, s_{345} \, s_{3456} \, s_{23456} }$}};
\endscope
\endtikzpicture

\tikzpicture [scale=1.3, line width=0.30mm]
%%%%%%%%%%%%%%%%%%%
% --++
\scope 
\draw  (2,2) -- (1.5,2.5) node[left]{$4$};
\draw  (2,2) -- (1.5,1.5) node[left]{$3$};
\draw (2.35,1.8) node{$s_{34}$};
\draw (3.05,2.2) node{$s_{234}$};
\draw (3.75,1.8) node{$s_{1234}$};
\draw (4.45,2.2) node{$s_{12345}$};
\draw (5.25,1.8) node{$s_{123456}$};
\draw  (2,2) -- (5.1,2); 
\draw (5.3,2) node{$\ldots  $};
\draw  (2.7,2) -- (2.7,1.5) node[below]{$2$}; 
\draw  (3.4,2) -- (3.4,1.5) node[below]{$1$};
\draw  (4.1,2) -- (4.1,2.5) node[above]{$5$};
\draw  (4.8,2) -- (4.8,2.5) node[above]{$6$};
\draw (3.5,0.8)node{{$ \displaystyle = \ \ \frac{ T_{342156} \, / \, s_{123456}}{s_{34} \, s_{234} \, s_{1234} \, s_{12345} }$}};
\endscope
% --+-
\scope[xshift=5cm]
\draw  (2,2) -- (1.5,2.5) node[left]{$5$};
\draw  (2,2) -- (1.5,1.5) node[left]{$4$};
\draw (2.35,1.8) node{$s_{45}$};
\draw (3.05,2.2) node{$s_{345}$};
\draw (3.75,1.8) node{$s_{2345}$};
\draw (4.45,2.2) node{$s_{23456}$};
\draw (5.25,1.8) node{$s_{123456}$};
\draw  (2,2) -- (5.1,2); 
\draw (5.3,2) node{$\ldots  $};
\draw  (2.7,2) -- (2.7,1.5) node[below]{$3$}; 
\draw  (3.4,2) -- (3.4,1.5) node[below]{$2$};
\draw  (4.1,2) -- (4.1,2.5) node[above]{$6$};
\draw  (4.8,2) -- (4.8,1.5) node[below]{$1$};
\draw (3.5,0.8)node{{$ \displaystyle = \ \ \frac{ T_{543261} \, / \, s_{123456}}{s_{45} \, s_{345} \, s_{2345} \, s_{23456} }$}};
\endscope
\endtikzpicture

\tikzpicture [scale=1.3, line width=0.30mm]
% ---+
\scope
\draw  (2,2) -- (1.5,2.5) node[left]{$5$};
\draw  (2,2) -- (1.5,1.5) node[left]{$4$};
\draw (2.35,1.8) node{$s_{45}$};
\draw (3.05,2.2) node{$s_{345}$};
\draw (3.75,1.8) node{$s_{2345}$};
\draw (4.45,2.2) node{$s_{12345}$};
\draw (5.25,1.8) node{$s_{123456}$};
\draw  (2,2) -- (5.1,2); 
\draw (5.3,2) node{$\ldots  $};
\draw  (2.7,2) -- (2.7,1.5) node[below]{$3$}; 
\draw  (3.4,2) -- (3.4,1.5) node[below]{$2$};
\draw  (4.1,2) -- (4.1,1.5) node[below]{$1$};
\draw  (4.8,2) -- (4.8,2.5) node[above]{$6$};
\draw (3.5,0.8)node{{$ \displaystyle = \ \ \frac{ T_{543216} \, / \, s_{123456}}{s_{45} \, s_{345} \, s_{2345} \, s_{12345} }$}};
\endscope
% ----
\scope[xshift=5cm]
\draw  (2,2) -- (1.5,2.5) node[left]{$6$};
\draw  (2,2) -- (1.5,1.5) node[left]{$5$};
\draw (2.35,1.8) node{$s_{56}$};
\draw (3.05,2.2) node{$s_{456}$};
\draw (3.75,1.8) node{$s_{3456}$};
\draw (4.45,2.2) node{$s_{23456}$};
\draw (5.25,1.8) node{$s_{123456}$};
\draw  (2,2) -- (5.1,2); 
\draw (5.3,2) node{$\ldots  $};
\draw  (2.7,2) -- (2.7,1.5) node[below]{$4$}; 
\draw  (3.4,2) -- (3.4,1.5) node[below]{$3$};
\draw  (4.1,2) -- (4.1,1.5) node[below]{$2$};
\draw  (4.8,2) -- (4.8,1.5) node[below]{$1$};
\draw (3.5,0.8)node{{$ \displaystyle = \ \ \frac{ T_{564321} \, / \, s_{123456}}{s_{56} \, s_{456} \, s_{3456} \, s_{23456} }$}};
\endscope
\endtikzpicture
%%%%%
%%%% Next class of diagrams now

\tikzpicture [scale=1.3, line width=0.30mm]
% PLUS PLUS
\scope
\draw  (2,2) -- (1.5,2.5) node[left]{$4$};
\draw  (2,2) node[right]{$s_{34}$} -- (1.5,2) node[left]{$3$};
\draw  (2,2) -- (2.5,1.5);
\draw  (2,1) -- (2.5,1.5);
\draw  (2,1) node[right]{$s_{12}$} -- (1.5,1) node[left]{$2$};
\draw  (2,1) -- (1.5,0.5) node[left]{$1$};
\draw  (2.5,1.5) -- (4.6,1.5);
\draw (4.8,1.5) node{$\ldots  $};
\draw (2.9,1.7) node{$s_{1234}$};
\draw (3.75,1.3) node{$s_{12345}$};
\draw (4.6,1.7) node{$s_{123456}$};
\draw  (3.4,1.5) -- (3.4,2) node[above]{$5$};
\draw  (4.1,1.5) -- (4.1,2) node[above]{$6$};
\draw (3.5,0.3)node{{$ \displaystyle = \ \ \frac{ 2 T_{12[34]56} \, / \, s_{123456}}{s_{12} \, s_{34} \, s_{1234} \, s_{12345} }$}};
\endscope
% PLUS MINUS
\scope[xshift=5cm]
\draw  (2,2) -- (1.5,2.5) node[left]{$5$};
\draw  (2,2) node[right]{$s_{45}$} -- (1.5,2) node[left]{$4$};
\draw  (2,2) -- (2.5,1.5);
\draw  (2,1) -- (2.5,1.5);
\draw  (2,1) node[right]{$s_{23}$} -- (1.5,1) node[left]{$3$};
\draw  (2,1) -- (1.5,0.5) node[left]{$2$};
\draw  (2.5,1.5) -- (4.6,1.5);
\draw (4.8,1.5) node{$\ldots  $};
\draw (2.9,1.7) node{$s_{2345}$};
\draw (3.75,1.3) node{$s_{23456}$};
\draw (4.6,1.7) node{$s_{123456}$};
\draw  (3.4,1.5) -- (3.4,2) node[above]{$6$};
\draw  (4.1,1.5) -- (4.1,1) node[below]{$1$};
\draw (3.5,0.3)node{{$ \displaystyle = \ \ \frac{ 2 T_{32[45]61} \, / \, s_{123456}}{s_{23} \, s_{45} \, s_{2345} \, s_{23456} }$}};
\endscope
\endtikzpicture

\tikzpicture [scale=1.3, line width=0.30mm]
% MINUS PLUS
\scope
\draw  (2,2) -- (1.5,2.5) node[left]{$5$};
\draw  (2,2) node[right]{$s_{45}$} -- (1.5,2) node[left]{$4$};
\draw  (2,2) -- (2.5,1.5);
\draw  (2,1) -- (2.5,1.5);
\draw  (2,1) node[right]{$s_{23}$} -- (1.5,1) node[left]{$3$};
\draw  (2,1) -- (1.5,0.5) node[left]{$2$};
\draw  (2.5,1.5) -- (4.6,1.5);
\draw (4.8,1.5) node{$\ldots  $};
\draw (2.9,1.7) node{$s_{2345}$};
\draw (3.75,1.3) node{$s_{12345}$};
\draw (4.6,1.7) node{$s_{123456}$};
\draw  (3.4,1.5) -- (3.4,1) node[below]{$1$};
\draw  (4.1,1.5) -- (4.1,2) node[above]{$6$};
\draw (3.5,0.3)node{{$ \displaystyle = \ \ \frac{ 2  T_{32[45]16} \, / \, s_{123456}}{s_{23} \, s_{45} \, s_{2345} \, s_{12345} }$}};
\endscope
% MINUS MINUS
\scope[xshift=5cm]
\draw  (2,2) -- (1.5,2.5) node[left]{$6$};
\draw  (2,2) node[right]{$s_{56}$} -- (1.5,2) node[left]{$5$};
\draw  (2,2) -- (2.5,1.5);
\draw  (2,1) -- (2.5,1.5);
\draw  (2,1) node[right]{$s_{34}$} -- (1.5,1) node[left]{$4$};
\draw  (2,1) -- (1.5,0.5) node[left]{$3$};
\draw  (2.5,1.5) -- (4.6,1.5);
\draw (4.8,1.5) node{$\ldots  $};
\draw (2.9,1.7) node{$s_{3456}$};
\draw (3.75,1.3) node{$s_{23456}$};
\draw (4.6,1.7) node{$s_{123456}$};
\draw  (3.4,1.5) -- (3.4,1) node[below]{$2$};
\draw  (4.1,1.5) -- (4.1,1) node[below]{$1$};
\draw (3.5,0.3)node{{$ \displaystyle = \ \ \frac{ 2 T_{34[56]21} \, / \, s_{123456}}{s_{34} \, s_{56} \, s_{3456} \, s_{23456} }$}};
\endscope
\endtikzpicture
%%%
%%% NEXT CLASS
%%% No 1,2

\tikzpicture [scale=1.3, line width=0.30mm]
\scope
\draw (2,-6) -- (1.5, -5.5) node[above] {$2$};
\draw (2,-6) -- (2.5, -5.5) node[above] {$3$};
\draw (2,-6) -- (4,-6);
\draw  (4,-6) -- (3.5, -5.5) node[above] {$4$};
\draw  (4,-6) -- (4.5, -5.5) node[above] {$5$};
\draw  (3.5,-6) -- (3.5, -6.5) node[below] {$6$};
\draw (2.75, -6) -- (2.75,-6.8);
\draw (2.75, -6.9) node{$\vdots$};
\draw (2.75, -6.5) -- (2.25, -6.5) node[left] {$1$};
\draw (2.5,-5.8) node{$s_{23}$} ;
\draw (3.15,-5.8) node{$s_{456}$} ;
\draw (3.75,-6.2) node{$s_{45}$} ;
\draw (2.3,-6.25) node{$s_{23456}$} ;
\draw (2.25,-6.8) node{$s_{123456}$} ;
\draw (5.0,-6.5) node{$\displaystyle = \ \frac{ 2 T_{546[32]1} / s_{123456}}{s_{23} \, s_{45} \, s_{456} \, s_{23456} }$};
\endscope
\scope[xshift=5.5cm]
\draw (2,-6) -- (1.5, -5.5) node[above] {$2$};
\draw (2,-6) -- (2.5, -5.5) node[above] {$3$};
\draw (2,-6) -- (4,-6);
\draw  (4,-6) -- (3.5, -5.5) node[above] {$5$};
\draw  (4,-6) -- (4.5, -5.5) node[above] {$6$};
\draw  (3.5,-6) -- (3, -5.5) node[above] {$4$};
\draw (2.75, -6) -- (2.75,-6.8);
\draw (2.75, -6.9) node{$\vdots$};
\draw (2.75, -6.5) -- (2.25, -6.5) node[left] {$1$};
\draw (2.5,-5.8) node{$s_{23}$} ;
\draw (3.15,-6.2) node{$s_{456}$} ;
\draw (3.75,-6.2) node{$s_{56}$} ;
\draw (2.3,-6.25) node{$s_{23456}$} ;
\draw (2.25,-6.8) node{$s_{123456}$} ;
\draw (5.0,-6.5) node{$\displaystyle = \ \frac{ 2 T_{564[32]1} / s_{123456}}{s_{23} \, s_{56} \, s_{456} \, s_{23456} }$};
\endscope
\endtikzpicture
%%% No 3,4

\tikzpicture [scale=1.3, line width=0.30mm]
\scope
\draw (2,-6) -- (1.5, -5.5) node[above] {$2$};
\draw (2,-6) -- (2.5, -5.5) node[above] {$3$};
\draw (2,-6) -- (4,-6);
\draw  (4,-6) -- (3.5, -5.5) node[above] {$4$};
\draw  (4,-6) -- (4.5, -5.5) node[above] {$5$};
\draw  (2.5,-6) -- (2.5, -6.5) node[below] {$1$};
\draw (3.25, -6) -- (3.25,-6.8);
\draw (3.25, -6.9) node{$\vdots$};
\draw (3.25, -6.5) -- (3.75, -6.5) node[right] {$6$};
\draw (3.5,-5.8) node{$s_{45}$} ;
\draw (2.85,-5.8) node{$s_{123}$} ;
\draw (2.25,-6.2) node{$s_{23}$} ;
\draw (3.7,-6.25) node{$s_{12345}$} ;
\draw (3.75,-6.8) node{$s_{123456}$} ;
\draw (5.0,-6.5) node{$\displaystyle \ \ \ \ \ \ = \ \frac{ 2 T_{321[45]6} / s_{123456}}{s_{23} \, s_{45} \, s_{123} \, s_{12345} }$};
\endscope
\scope[xshift=5.5cm]
\draw (2,-6) -- (1.5, -5.5) node[above] {$1$};
\draw (2,-6) -- (2.5, -5.5) node[above] {$2$};
\draw (2,-6) -- (4,-6);
\draw  (4,-6) -- (3.5, -5.5) node[above] {$4$};
\draw  (4,-6) -- (4.5, -5.5) node[above] {$5$};
\draw  (2.5,-6) -- (3, -5.5) node[above] {$3$};
\draw (3.25, -6) -- (3.25,-6.8);
\draw (3.25, -6.9) node{$\vdots$};
\draw (3.25, -6.5) -- (3.75, -6.5) node[right] {$6$};
\draw (3.5,-5.8) node{$s_{45}$} ;
\draw (2.85,-6.2) node{$s_{123}$} ;
\draw (2.25,-6.2) node{$s_{12}$} ;
\draw (3.7,-6.25) node{$s_{12345}$} ;
\draw (3.75,-6.8) node{$s_{123456}$} ;
\draw (5.0,-6.5) node{$\displaystyle \ \ \ \ \ \ = \ \frac{ 2 T_{123[45]6} / s_{123456}}{s_{12} \, s_{45} \, s_{123} \, s_{12345} }$};
\endscope
\endtikzpicture
%%% No 5,6

\tikzpicture [scale=1.3, line width=0.30mm]
\scope
\draw (2,-6) -- (1.5, -5.5) node[above] {$1$};
\draw (2,-6) -- (2.5, -5.5) node[above] {$2$};
\draw (2,-6) -- (4,-6);
\draw  (4,-6) -- (3.5, -5.5) node[above] {$3$};
\draw  (4,-6) -- (4.5, -5.5) node[above] {$4$};
\draw  (3.5,-6) -- (3.5, -6.5) node[below] {$5$};
\draw (2.75, -6) -- (2.75,-6.8);
\draw (2.75, -6.9) node{$\vdots$};
\draw (2.75, -6.5) -- (2.95, -6.5) node[right] {$6$};
\draw (2.5,-5.8) node{$s_{12}$} ;
\draw (3.15,-5.8) node{$s_{345}$} ;
\draw (3.75,-6.2) node{$s_{34}$} ;
\draw (2.3,-6.25) node{$s_{12345}$} ;
\draw (2.25,-6.8) node{$s_{123456}$} ;
\draw (5.0,-6.5) node{$\displaystyle = \ \frac{ 2 T_{345[21]6} / s_{123456}}{s_{12} \, s_{34} \, s_{345} \, s_{12345} }$};
\endscope
\scope[xshift=5.5cm]
\draw (2,-6) -- (1.5, -5.5) node[above] {$1$};
\draw (2,-6) -- (2.5, -5.5) node[above] {$2$};
\draw (2,-6) -- (4,-6);
\draw  (4,-6) -- (3.5, -5.5) node[above] {$4$};
\draw  (4,-6) -- (4.5, -5.5) node[above] {$5$};
\draw  (3.5,-6) -- (3, -5.5) node[above] {$3$};
\draw (2.75, -6) -- (2.75,-6.8);
\draw (2.75, -6.9) node{$\vdots$};
\draw (2.75, -6.5) -- (3.25, -6.5) node[right] {$6$};
\draw (2.5,-5.8) node{$s_{12}$} ;
\draw (3.15,-6.2) node{$s_{345}$} ;
\draw (3.75,-6.2) node{$s_{45}$} ;
\draw (2.3,-6.25) node{$s_{12345}$} ;
\draw (2.25,-6.8) node{$s_{123456}$} ;
\draw (5.0,-6.5) node{$\displaystyle = \ \frac{ 2 T_{543[21]6} / s_{123456}}{s_{12} \, s_{45} \, s_{345} \, s_{12345} }$};
\endscope
\endtikzpicture
%%% No 7,8

\tikzpicture [scale=1.3, line width=0.30mm]
\scope
\draw (2,-6) -- (1.5, -5.5) node[above] {$3$};
\draw (2,-6) -- (2.5, -5.5) node[above] {$4$};
\draw (2,-6) -- (4,-6);
\draw  (4,-6) -- (3.5, -5.5) node[above] {$5$};
\draw  (4,-6) -- (4.5, -5.5) node[above] {$6$};
\draw  (2.5,-6) -- (2.5, -6.5) node[below] {$2$};
\draw (3.25, -6) -- (3.25,-6.8);
\draw (3.25, -6.9) node{$\vdots$};
\draw (3.25, -6.5) -- (3.05, -6.5) node[left] {$1$};
\draw (3.5,-5.8) node{$s_{56}$} ;
\draw (2.85,-5.8) node{$s_{234}$} ;
\draw (2.25,-6.2) node{$s_{34}$} ;
\draw (3.7,-6.25) node{$s_{23456}$} ;
\draw (3.75,-6.8) node{$s_{123456}$} ;
\draw (5.0,-6.5) node{$\displaystyle  = \ \frac{ 2 T_{342[56]1} / s_{123456}}{s_{34} \, s_{56} \, s_{234} \, s_{23456} }$};
\endscope
\scope[xshift=5.5cm]
\draw (2,-6) -- (1.5, -5.5) node[above] {$2$};
\draw (2,-6) -- (2.5, -5.5) node[above] {$3$};
\draw (2,-6) -- (4,-6);
\draw  (4,-6) -- (3.5, -5.5) node[above] {$5$};
\draw  (4,-6) -- (4.5, -5.5) node[above] {$6$};
\draw  (2.5,-6) -- (3, -5.5) node[above] {$4$};
\draw (3.25, -6) -- (3.25,-6.8);
\draw (3.25, -6.9) node{$\vdots$};
\draw (3.25, -6.5) -- (2.75, -6.5) node[left] {$1$};
\draw (3.5,-5.8) node{$s_{56}$} ;
\draw (2.85,-6.2) node{$s_{234}$} ;
\draw (2.25,-6.2) node{$s_{23}$} ;
\draw (3.7,-6.25) node{$s_{23456}$} ;
\draw (3.75,-6.8) node{$s_{123456}$} ;
\draw (5.0,-6.5) node{$\displaystyle  = \ \frac{ 2 T_{324[56]1} / s_{123456}}{s_{23} \, s_{56} \, s_{234} \, s_{23456} }$};
\endscope
\endtikzpicture
%%%%%%%%%%%%%%%%%%%%%%%%%%%%%%%%
%%%% Next class of diagrams now

\tikzpicture [scale=1.3, line width=0.30mm]
% LEFT
\scope
\draw  (2,2) -- (1.5,2.5) node[left]{$4$};
\draw  (2,2) node[right]{$s_{34}$} -- (1.5,2) node[left]{$3$};
\draw  (2,2) -- (2.5,1.5);
\draw  (2,1) -- (2.5,1.5);
\draw  (2,1) node[right]{$s_{12}$} -- (1.5,1) node[left]{$2$};
\draw  (2,1) -- (1.5,0.5) node[left]{$1$};
\draw  (2.5,1.5) -- (4,1.5);
\draw  (3.25,1.5) -- (3.25,0.7);
\draw (3.25, 0.6) node{$\vdots$};
\draw (3.75,1.0) node{$s_{123456}$} ;
\draw (2.9, 1.7) node{$s_{1234}$};
\draw (3.6, 1.7) node{$s_{56}$};
\draw  (4,1.5) -- (4.5,1) node[right]{$6$};
\draw  (4,1.5) -- (4.5,2) node[right]{$5$};
\draw (4.7,0.4) node{$\displaystyle  = \ \frac{ 4 T_{12[34][56]} / s_{123456}}{s_{12} \, s_{34} \, s_{56} \, s_{1234} }$};
\endscope
% RIGHT
\scope [xshift = 5.9cm]
\draw  (4.5,2) -- (5,2.5) node[right]{$3$};
\draw  (4.5,2) node[left]{$s_{34}$} -- (5,2) node[right]{$4$};
\draw  (4.5,2) -- (4,1.5);
\draw  (4.5,1) -- (4,1.5);
\draw  (4.5,1) node[left]{$s_{56}$} -- (5,1) node[right]{$5$};
\draw  (4.5,1) -- (5,0.5) node[right]{$6$};
\draw  (2.5,1.5) -- (4,1.5);
\draw  (3.25,1.5) -- (3.25,0.7);
\draw (3.25, 0.6) node{$\vdots$};
\draw (2.75,1.0) node{$s_{123456}$} ;
\draw (2.9, 1.7) node{$s_{12}$};
\draw (3.6, 1.7) node{$s_{3456}$};
\draw  (2.5,1.5) -- (2,1) node[left]{$1$};
\draw  (2.5,1.5) -- (2,2) node[left]{$2$};
\draw (1.9, 0.4) node{$\displaystyle =  \frac{ 4 T_{56[43][21]} / s_{123456}}{s_{12} \, s_{34} \, s_{56} \, s_{3456} }$};
\endscope
\endtikzpicture
%%%%%%%%%%%%%%%%%%%%%%%%%%%%%%%%
%%%% Next class of diagrams now

\tikzpicture [scale=1.3, line width=0.30mm]
% PLUS PLUS
\scope
\draw (2,-6) -- (1.5, -5.5) node[above] {$1$};
\draw (2,-6) -- (2.5, -5.5) node[above] {$2$};
\draw (2,-6) -- (4.5,-6);
\draw (2.5,-6) -- (2.5,-6.8);
\draw (2.5, -6.9) node{$\vdots$};
\draw (2.0, -6.6) node{$s_{123456}$};
\draw (3,-6) -- (3,-5.5) node[above]{$3$};
\draw (3.5,-6) -- (3.5,-5.5) node[above]{$4$};
\draw (4.5,-6) -- (4, -5.5) node[above] {$5$};
\draw (4.5,-6) -- (5, -5.5) node[above] {$6$};
\draw (2.25, -6.2) node{$s_{12}$};
\draw (2.65, -5.8) node{$s_{3456}$};
\draw (3.25, -6.2) node{$s_{456}$};
\draw (3.75, -5.8) node{$s_{56}$};
\draw (4.8,-6.8) node{$\displaystyle  = \ \frac{ 2 T_{5643[21]} / s_{123456}}{s_{56} \, s_{456} \, s_{3456} \, s_{12} }$};
\endscope
% PLUS MINUS
\scope [xshift=5.5cm]
\draw (2,-6) -- (1.5, -5.5) node[above] {$1$};
\draw (2,-6) -- (2.5, -5.5) node[above] {$2$};
\draw (2,-6) -- (4.5,-6);
\draw (2.5,-6) -- (2.5,-6.8);
\draw (2.5, -6.9) node{$\vdots$};
\draw (2.0, -6.6) node{$s_{123456}$};
\draw (3,-6) -- (3,-5.5) node[above]{$3$};
\draw (3.5,-6) -- (3.5,-6.5) node[below]{$6$};
\draw (4.5,-6) -- (4, -5.5) node[above] {$4$};
\draw (4.5,-6) -- (5, -5.5) node[above] {$5$};
\draw (2.25, -6.2) node{$s_{12}$};
\draw (2.65, -5.8) node{$s_{3456}$};
\draw (3.15, -6.2) node{$s_{456}$};
\draw (3.75, -5.8) node{$s_{45}$};
\draw (4.5,-7.0) node{$\displaystyle  = \ \frac{ 2 T_{5463[21]} / s_{123456}}{s_{45} \, s_{456} \, s_{3456} \, s_{12} }$};
\endscope
\endtikzpicture
%%%%%%%%

\tikzpicture [scale=1.3, line width=0.30mm]
% MINUS PLUS
\scope
\draw (2,-6) -- (1.5, -5.5) node[above] {$1$};
\draw (2,-6) -- (2.5, -5.5) node[above] {$2$};
\draw (2,-6) -- (4.5,-6);
\draw (2.5,-6) -- (2.5,-6.8);
\draw (2.5, -6.9) node{$\vdots$};
\draw (2.0, -6.6) node{$s_{123456}$};
\draw (3,-6) -- (3,-6.5) node[below]{$6$};
\draw (3.5,-6) -- (3.5,-5.5) node[above]{$3$};
\draw (4.5,-6) -- (4, -5.5) node[above] {$4$};
\draw (4.5,-6) -- (5, -5.5) node[above] {$5$};
\draw (2.25, -6.2) node{$s_{12}$};
\draw (2.65, -5.8) node{$s_{3456}$};
\draw (3.35, -6.2) node{$s_{345}$};
\draw (3.75, -5.8) node{$s_{45}$};
\draw (4.8,-6.8) node{$\displaystyle  = \ \frac{ 2 T_{5436[21]} / s_{123456}}{s_{45} \, s_{345} \, s_{3456} \, s_{12} }$};
\endscope
% MINUS MINUS
\scope [xshift=5.5cm]
\draw (2,-6) -- (1.5, -5.5) node[above] {$1$};
\draw (2,-6) -- (2.5, -5.5) node[above] {$2$};
\draw (2,-6) -- (4.5,-6);
\draw (2.5,-6) -- (2.5,-6.8);
\draw (2.5, -6.9) node{$\vdots$};
\draw (2.0, -6.6) node{$s_{123456}$};
\draw (3,-6) -- (3,-6.5) node[below]{$6$};
\draw (3.5,-6) -- (3.5,-6.5) node[below]{$5$};
\draw (4.5,-6) -- (4, -5.5) node[above] {$3$};
\draw (4.5,-6) -- (5, -5.5) node[above] {$4$};
\draw (2.25, -6.2) node{$s_{12}$};
\draw (2.65, -5.8) node{$s_{3456}$};
\draw (3.25, -6.2) node{$s_{345}$};
\draw (3.75, -5.8) node{$s_{34}$};
\draw (4.8,-6.95) node{$\displaystyle  = \ \frac{ 2 T_{3456[21]} / s_{123456}}{s_{34} \, s_{345} \, s_{3456} \, s_{12} }$};
\endscope
\endtikzpicture
%%%%%%%%%%%%%%%%%%%%%%%%%%%%%%%%
%%%% Next class of diagrams now

\tikzpicture [scale=1.3, line width=0.30mm]
% PLUS PLUS
\scope
\draw (4.5,-6) -- (5, -5.5) node[above] {$6$};
\draw (4.5,-6) -- (4, -5.5) node[above] {$5$};
\draw (2,-6) -- (4.5,-6);
\draw (4,-6) -- (4,-6.8);
\draw (4, -6.9) node{$\vdots$};
\draw (4.5, -6.6) node{$s_{123456}$};
\draw (3.5,-6) -- (3.5,-5.5) node[above]{$4$};
\draw (3,-6) -- (3,-5.5) node[above]{$3$};
\draw (2,-6) -- (2.5, -5.5) node[above] {$2$};
\draw (2,-6) -- (1.5, -5.5) node[above] {$1$};
\draw (4.25, -6.2) node{$s_{56}$};
\draw (3.85, -5.8) node{$s_{1234}$};
\draw (3.25, -6.2) node{$s_{123}$};
\draw (2.75, -5.8) node{$s_{12}$};
\draw (1.6,-6.8) node{$\displaystyle  = \frac{ 2 T_{1234[56]} / s_{123456}}{s_{12} \, s_{123} \, s_{1234} \, s_{56} }$};
\endscope
% PLUS MINUS
\scope [xshift=5cm]
\draw (4.5,-6) -- (5, -5.5) node[above] {$6$};
\draw (4.5,-6) -- (4, -5.5) node[above] {$5$};
\draw (2,-6) -- (4.5,-6);
\draw (4,-6) -- (4,-6.8);
\draw (4, -6.9) node{$\vdots$};
\draw (4.5, -6.6) node{$s_{123456}$};
\draw (3.5,-6) -- (3.5,-6.5) node[below]{$1$};
\draw (3,-6) -- (3,-5.5) node[above]{$4$};
\draw (2,-6) -- (2.5, -5.5) node[above] {$3$};
\draw (2,-6) -- (1.5, -5.5) node[above] {$2$};
\draw (4.25, -6.2) node{$s_{56}$};
\draw (3.85, -5.8) node{$s_{1234}$};
\draw (3.25, -6.2) node{$s_{234}$};
\draw (2.75, -5.8) node{$s_{23}$};
\draw (1.7,-6.8) node{$\displaystyle  = \ \frac{ 2 T_{3241[56]} / s_{123456}}{s_{23} \, s_{234} \, s_{1234} \, s_{56} }$};
\endscope
\endtikzpicture
%%%%

\tikzpicture [scale=1.3, line width=0.30mm]
% MINUS PLUS
\scope
\draw (4.5,-6) -- (5, -5.5) node[above] {$6$};
\draw (4.5,-6) -- (4, -5.5) node[above] {$5$};
\draw (2,-6) -- (4.5,-6);
\draw (4,-6) -- (4,-6.8);
\draw (4, -6.9) node{$\vdots$};
\draw (4.5, -6.6) node{$s_{123456}$};
\draw (3.5,-6) -- (3.5,-5.5) node[above]{$4$};
\draw (3,-6) -- (3,-6.5) node[below]{$1$};
\draw (2,-6) -- (2.5, -5.5) node[above] {$3$};
\draw (2,-6) -- (1.5, -5.5) node[above] {$2$};
\draw (4.25, -6.2) node{$s_{56}$};
\draw (3.85, -5.8) node{$s_{1234}$};
\draw (3.25, -6.2) node{$s_{123}$};
\draw (2.75, -5.8) node{$s_{23}$};
\draw (1.6,-6.8) node{$\displaystyle  = \frac{ 2 T_{3214[56]} / s_{123456}}{s_{23} \, s_{123} \, s_{1234} \, s_{56} }$};
\endscope
% MINUS MINUS
\scope [xshift=5cm]
\draw (4.5,-6) -- (5, -5.5) node[above] {$6$};
\draw (4.5,-6) -- (4, -5.5) node[above] {$5$};
\draw (2,-6) -- (4.5,-6);
\draw (4,-6) -- (4,-6.8);
\draw (4, -6.9) node{$\vdots$};
\draw (4.5, -6.6) node{$s_{123456}$};
\draw (3.5,-6) -- (3.5,-6.5) node[below]{$1$};
\draw (3,-6) -- (3,-6.5) node[below]{$2$};
\draw (2,-6) -- (2.5, -5.5) node[above] {$4$};
\draw (2,-6) -- (1.5, -5.5) node[above] {$3$};
\draw (4.25, -6.2) node{$s_{56}$};
\draw (3.85, -5.8) node{$s_{1234}$};
\draw (3.25, -6.2) node{$s_{234}$};
\draw (2.75, -5.8) node{$s_{34}$};
\draw (1.6,-6.8) node{$\displaystyle  =  \frac{ 2 T_{3421[56]} / s_{123456}}{s_{34} \, s_{234} \, s_{1234} \, s_{56} }$};
\endscope
\endtikzpicture
%%%%%%%%%%%%%%%%%%%%%%%%%%%%%%%%
%%%% LAST class of diagrams now

\tikzpicture [scale=1.3, line width=0.30mm]
% PLUS PLUS
\scope
\draw (5,-6) -- (5.5, -5.5) node[above] {$6$};
\draw (5,-6) -- (4.5, -5.5) node[above] {$5$};
\draw (4.5, -6.2) node{$s_{56}$};
\draw (2,-6) -- (5,-6);
\draw (2,-6) -- (2.5, -5.5) node[above] {$2$};
\draw (2,-6) -- (1.5, -5.5) node[above] {$1$};
\draw (2.5, -5.8) node{$s_{12}$};
\draw (3,-6) -- (3,-5.5) node[above]{$3$};
\draw (4,-6) -- (4,-5.5) node[above]{$4$};
\draw (3.25, -6.2) node{$s_{123}$};
\draw (3.75, -5.8) node{$s_{456}$};
\draw (3.5,-6) -- (3.5,-6.8);
\draw (3.5, -6.9) node{$\vdots$};
\draw (3.0, -6.6) node{$s_{123456}$};
\draw (5.3,-6.8) node{$\displaystyle  =  \frac{ 4 T_{123[4[56]]} / s_{123456}}{s_{12} \, s_{123} \, s_{56} \, s_{456} }$};
\endscope
% PLUS MINUS
\scope [xshift=5cm]
\draw (5,-6) -- (5.5, -5.5) node[above] {$5$};
\draw (5,-6) -- (4.5, -5.5) node[above] {$4$};
\draw (4.5, -6.2) node{$s_{45}$};
\draw (2,-6) -- (5,-6);
\draw (2,-6) -- (2.5, -5.5) node[above] {$2$};
\draw (2,-6) -- (1.5, -5.5) node[above] {$1$};
\draw (2.5, -5.8) node{$s_{12}$};
\draw (3,-6) -- (3,-5.5) node[above]{$3$};
\draw (4,-6) -- (4,-6.5) node[below]{$6$};
\draw (3.25, -6.2) node{$s_{123}$};
\draw (3.75, -5.8) node{$s_{456}$};
\draw (3.5,-6) -- (3.5,-6.8);
\draw (3.5, -6.9) node{$\vdots$};
\draw (3.0, -6.6) node{$s_{123456}$};
\draw (5.3,-6.9) node{$\displaystyle  = \frac{ 4 T_{123[[45]6]} / s_{123456}}{s_{12} \, s_{123} \, s_{45} \, s_{456} }$};
\endscope
\endtikzpicture
%%%%%

\tikzpicture [scale=1.3,line width=0.30mm]
% MINUS PLUS
\scope
\draw (5,-6) -- (5.5, -5.5) node[above] {$6$};
\draw (5,-6) -- (4.5, -5.5) node[above] {$5$};
\draw (4.5, -6.2) node{$s_{56}$};
\draw (2,-6) -- (5,-6);
\draw (2,-6) -- (2.5, -5.5) node[above] {$3$};
\draw (2,-6) -- (1.5, -5.5) node[above] {$2$};
\draw (2.5, -5.8) node{$s_{23}$};
\draw (3,-6) -- (3,-6.5) node[below]{$1$};
\draw (4,-6) -- (4,-5.5) node[above]{$4$};
\draw (3.25, -6.2) node{$s_{123}$};
\draw (3.75, -5.8) node{$s_{456}$};
\draw (3.5,-6) -- (3.5,-6.8);
\draw (3.5, -6.9) node{$\vdots$};
\draw (3.0, -7) node{$s_{123456}$};
\draw (5.3,-6.8) node{$\displaystyle  = \ \frac{ 4 T_{321[4[56]]} / s_{123456}}{s_{23} \, s_{123} \, s_{56} \, s_{456} }$};
\endscope
% MINUS MINUS
\scope [xshift=5cm]
\draw (5,-6) -- (5.5, -5.5) node[above] {$5$};
\draw (5,-6) -- (4.5, -5.5) node[above] {$4$};
\draw (4.5, -6.2) node{$s_{45}$};
\draw (2,-6) -- (5,-6);
\draw (2,-6) -- (2.5, -5.5) node[above] {$3$};
\draw (2,-6) -- (1.5, -5.5) node[above] {$2$};
\draw (2.5, -5.8) node{$s_{23}$};
\draw (3,-6) -- (3,-6.5) node[below]{$1$};
\draw (4,-6) -- (4,-6.5) node[below]{$6$};
\draw (3.25, -6.2) node{$s_{123}$};
\draw (3.75, -5.8) node{$s_{456}$};
\draw (3.5,-6) -- (3.5,-6.8);
\draw (3.5, -6.9) node{$\vdots$};
\draw (3.0, -7) node{$s_{123456}$};
\draw (5.3,-6.9) node{$\displaystyle  = \ \frac{ 4 T_{321[[45]6]} / s_{123456}}{s_{23} \, s_{123} \, s_{45} \, s_{456} }$};
\endscope
\endtikzpicture
\tikzcaption\Msixgraphs{The 42 cubic diagrams which constitute $M_{123456}$. The signs of their corresponding formul{\ae}
are in one-to-one agreement with the terms in expression for $M_{123456}$ given by \Msixformula, which is
reproduced by summing all 42 graphs displayed here.}

\listrefs

\bye